\documentclass[11pt]{article}
\usepackage{slashed}
\usepackage{jheppub}
\usepackage{float, extarrows, tikz-cd}
\usepackage{graphicx}
\usepackage{orcidlink}
\usepackage{mathrsfs}
\usepackage{tabularx,ragged2e}
\usepackage{subcaption}
\usepackage{amsmath,amssymb}
\usepackage{caption}
\usepackage{braket}
\usepackage{dsfont}
\usepackage{hyperref}
\usepackage{verbatim}
\usepackage{mathtools, xcolor,ytableau, amsfonts}

\usepackage{physics}
\newcolumntype{C}{>{\Centering\arraybackslash}X}

\numberwithin{equation}{section}

\newcommand{\mO}{\mathcal{O}}
\newcommand{\pd}{\partial}
\newcommand{\mL}{\mathcal{L}}
\newcommand{\eps}{\varepsilon}

\author[a,b]{Gabriel Cuomo,} 
\author[c]{Simone Giombi,} 
\author[d]{Luigi Tizzano}

\affiliation[a]{SISSA, Via Bonomea 265, 34136 Trieste, Italy}
\affiliation[b]{INFN, Sezione di Trieste, Via Valerio 2, 34127 Trieste, Italy}
\affiliation[c]{Joseph Henry Laboratories, Princeton University, Princeton, NJ  08544, USA}
\affiliation[d]{CERN, Theoretical Physics Department, CH-1211 Geneva 23, Switzerland}

\emailAdd{gcuomo@sissa.it}
\emailAdd{sgiombi@princeton.edu}  
\emailAdd{luigi.tizzano@cern.ch}

\title{Impurities Near the Light Cone}
\preprint{CERN-TH-2026-186}
\abstract{The lightlike Wilson line cusp underlies the Sudakov double logarithm and is a central ingredient in gauge-theory factorization. We ask what replaces this behavior for general conformal line defects in Lorentzian signature. We argue that, when the defects admit local endpoint operators, the cusp anomalous dimension has a finite large-boost limit fixed by the scaling dimensions of the corresponding defect-creation operators. We further analyze the distinct analytic continuations of the cusp, elucidate their physical interpretations, and derive a positivity condition on the Lorentzian cusp anomalous dimension. We test these predictions in several perturbative examples, including pinning-field defects and spin impurities, and discuss their possible implications for gauge theories.}

\begin{document}

\maketitle

\section{Introduction}

The study of cusps of line operators in quantum field theory has its origin in the analysis of the infrared properties of gauge theory scattering amplitudes. In his seminal paper \cite{Sudakov:1954sw}, Sudakov found a characteristic double-logarithmic behavior whose resummation leads to an exponential suppression of the electromagnetic form factor. These logarithms were subsequently understood in terms of the ultraviolet divergences of Wilson loops with cusps \cite{Polyakov:1980ca,Brandt:1981kf}. Indeed, in the eikonal limit, hard particles are represented by semi-infinite Wilson lines along their classical trajectories. For a form factor, two such trajectories meet at the hard interaction and form a cusp. The IR singularities of the original amplitude are then encoded in the UV renormalization of this cusp \cite{Korchemsky:1985xj,Korchemsky:1987wg}.

\begin{figure}
    \centering
    \includegraphics[trim= 3cm 10cm 3cm 10cm, clip=true, width=\linewidth]{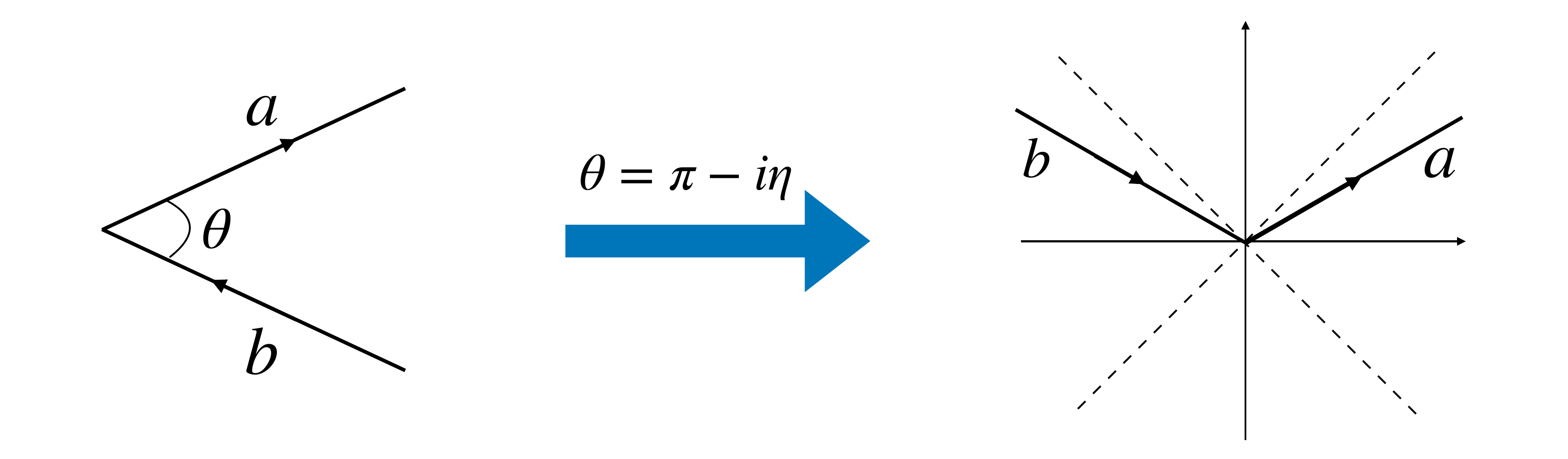}
    \caption{The Euclidean cusp (left) and its Wick rotation to a Lorentzian boosted configuration (right).}
    \label{fig:CuspIntro}
\end{figure}

In this work we explore cusp configurations in conformal field theories. Let $a$ and $b$ denote two conformal line defects meeting at a Euclidean angle $\theta$ as in fig.~\ref{fig:CuspIntro}. Joining the two line defects at a cusp preserves conformal dilations, under which the cusped configuration transforms as a local operator inserted at the junction, with scaling dimension $\Gamma_{ab}(\theta)$. The quantity $\Gamma_{ab}(\theta)$ is known as the cusp anomalous dimension.

The cusp anomalous dimension of Wilson lines in gauge theories has been studied extensively, see for instance~\cite{Drukker:1999zq,Correa:2012hh,Grozin:2015kna} and references therein. The notion of cusp anomalous dimension is, however, not intrinsically tied to gauge theories. For instance, the cusp anomalous dimension of replica twist defects determines the universal corner contribution to the Rényi entropies and, in the replica limit, to the ground-state entanglement entropy \cite{Casini:2006hu,Bueno:2015qya,Bianchi:2015liz}. 
More recently, cusp and related observables have been investigated for general conformal defects, see e.g. \cite{Wang:2021lmb,Cuomo:2024psk,Giombi:2025evu}.
The small-angle limit of the cusp is also closely related to the effective theory of defect fusion
\cite{Diatlyk:2024qpr,Kravchuk:2024qoh,Diatlyk:2024zkk,Cuomo:2024psk,Chernikov:2026lcv}.

Our main focus, however, will be on the \emph{Lorentzian} regime, obtained by analytically continuing the Euclidean angle:
\begin{equation}
    \theta=\pi-i\log y\,,
    \qquad
    \Gamma^L_{ab}(y)\equiv
    \Gamma_{ab}(\pi-i\log y)\,,
    \qquad
    \eta=\log y\,.
\end{equation}
For $y>1$, this continuation describes two everywhere spacelike-separated
lines boosted in real time with a rapidity parameter $\eta$ and $\Gamma^L_{ab}(y)$ is real, see fig.~\ref{fig:CuspIntro}.
% \footnote{Other Lorentzian configurations are obtained by further continuation in $y$. In particular, the ``quench'' configuration is described by the same real function $\Gamma_{ab}^L(y)$, whereas the physical Sudakov configuration is generically complex. See section \ref{sec:continuations} for further details.} 
As defects describe heavy particles traveling through a gapless bath, the Lorentzian cusp anomalous dimension characterizes the IR singularities generated by radiation emitted during quenches or production processes involving such massive particles in the CFT. Our goal will be to elucidate the general properties of the large-boost limit $\eta \to \infty$. For Wilson lines in gauge theories, this limit is familiar and has been studied extensively. In that context, one finds:
\begin{equation}\label{lightcuspgauge}
    \Gamma^L_{ab} (y) \overset{y\to \infty}{=} \log y~\gamma_{ab} +O(1)~,
\end{equation}
where $\gamma_{ab}$ governs the large boost behavior of the cusp anomalous dimension $\Gamma^L_{ab}(y)$. This quantity is often referred to as the lightlike cusp anomalous dimension  \cite{Korchemsky:1987wg,Korchemsky:1992xv} and also governs the $\log J$ growth of the anomalous dimension of single trace leading-twist large spin operators~\cite{Korchemsky:1988si,
Gubser:2002tv,Kruczenski:2002fb,Alday:2007mf}. Since $\Gamma_{ab}^L{(y)}$ appears as the coefficient of an infrared logarithm, its growth with $\log y$ is responsible for the usual double-logarithmic Sudakov suppression. The main question addressed in this work is what replaces this behavior for more general conformal line defects.

Our main result, which will be derived more systematically in the main text, shows that the Lorentzian cusp configuration $\Gamma^L_{ab}(y)$ for conformal field theories in $d\geq 3$ obeys a substantially different behavior than the one observed in \eqref{lightcuspgauge}. To argue for this, we assume that the line defects $a$ and $b$ admit local endpoint operators connecting them to the trivial line. We denote the dimensions of the lowest such defect-creation operators by $\Delta_{a0}$ and $\Delta_{b0}$. Assuming
also a twist gap above the identity in the sector that couples the
two defects, we can show that the two segments decouple in the large-boost
limit. The cusp anomalous dimension can therefore be written as
\begin{equation}\label{lorentzcusp}
    \Gamma^L_{ab}(y)
    =
    \Delta_{a0}+\Delta_{b0}-V_{\rm int}(y)\,,
    \qquad
    \lim_{y\to\infty}V_{\rm int}(y)=0\,.
\end{equation}
The origin of \eqref{lorentzcusp} becomes transparent after a Weyl transformation to an AdS frame, introduced in \cite{Alday:2007mf}. In the AdS frame, the rapidity becomes a large
spatial separation proportional to $\log y$, while the Hamiltonian
governing propagation between the defects is isomorphic to the twist
generator. Let $\mO_{\rm exch}$ be the lowest twist non-identity bulk primary
with nonzero overlap with both defects, of dimension $\Delta$, spin $J$ and
twist $\tau_{\rm exch} = \Delta-J$. We argue that the exchange of its conformal family controls the leading interaction between the two defects at large-boost. Within this description, we obtain
\begin{equation}\label{eq_int_intro}
    V_{\rm int}(y)
    =
    \frac{1}{y^{\tau_{\rm exch}}}
    \begin{cases}
        \#\log y+\#\,,
        & J(\mO_{\rm exch})=0\,,
        \\[0.4em]
        \#\,,
        & J(\mO_{\rm exch})>0\,.
    \end{cases}
\end{equation}
where each $\#$ denotes constants which are not determined by the general argument. 

We also relate the different analytic continuations of the Lorentzian cusp anomalous dimension to several physical Lorentzian configurations, including the ``quench'' process, in which a particle is suddenly kicked at $t=0$, and the Sudakov configuration describing the production of a particle--antiparticle pair. We prove that the spacelike Lorentzian cusp anomalous dimension satisfies
\begin{equation}
\Gamma_{ab}^L(y)\geq 0\,,
\end{equation}
and, for a broad class of defects, establish an analogous bound for the real part of the analytic continuation describing the Sudakov production process. See secs.~\ref{sec:continuations} and~\ref{subsec:rindler-positivity} for details.

To shed further light on the above discussion, we test our predictions in a broad range of conformal field theories, both with and without gauge fields where we are able to obtain explicit values for the unspecified constant terms in the equation above. Our examples include pinning-field defects in free scalar, $O(N)$ Wilson--Fisher and Yukawa
theories, a quantum impurity coupled to a free Dirac fermion, Wilson lines in Maxwell theory and $\mathcal N=4$ SYM, and
factorizing interfaces in two dimensions. They support the predicted
large-boost behavior when all the assumptions above apply and illustrate how
it is modified when they do not. These examples are discussed in
section~\ref{sec:examples}.

We also study cusps formed by quantum spin impurities in both the free and the interacting $O(3)$ symmetric CFTs
\cite{PhysRevB.61.4041,sachdev1999quantum,vojta2000quantum,Liu:2021nck,Cuomo:2022xgw}. Since the cusp anomalous dimension has not been studied previously in this setup, we discuss both the Euclidean and the Lorentzian regimes. In Euclidean signature in particular the cusp informs us about the Casimir force between defects and their fusion.\footnote{Our analysis of defect fusion in the interacting theory is restricted to tree level. We were informed about an upcoming work that will explore this problem at subleading orders \cite{Diatlyk:2026appear}.} In the Lorentzian large-boost limit, the degeneracy of the endpoint states makes the defect interaction
channel-dependent, providing a nontrivial test of the general framework
at both fixed and large spin. Our results are again consistent with the general
large-boost EFT picture; see section~\ref{sec_spin_imp}. 

Note that the endpoint assumption in \eqref{lorentzcusp} is essential. A Wilson line
charged under an exact one-form symmetry \cite{Gaiotto:2014kfa} cannot terminate on a local operator connecting it to the trivial line, and in the above AdS frame creates a flux tube that cannot break \cite{Alday:2007mf}.\footnote{Monodromy defects attached to topological surfaces similarly fall outside our argument.} The length of this flux tube grows proportionally to $\log y$. At strong coupling in gauge/string duality, this flux is
represented by a long string \cite{Drukker:1999zq,Kruczenski:2002fb}. By contrast, for Wilson lines that can be screened, the logarithmic growth \eqref{lightcuspgauge} may hold over a perturbative range of rapidities, while at parametrically larger boost $\eta \sim 1/g^2$ in weakly coupled examples, we expect a string-breaking crossover to the endpoint-dominated
behavior in \eqref{lorentzcusp}. It would be interesting to analyze further such crossover behavior, especially in the context of gauge theory scattering amplitudes.  

The rest of the paper is organized as follows. In section \ref{sec:generalities} we discuss the general theory of cusped defects in Lorentzian signature. We begin in subsec.~\ref{sec:continuations} by defining the Euclidean and Lorentzian cusp configurations and their different analytic continuations.  We discuss a positivity condition on the spacelike Lorentzian cusp anomalous dimension in subsec.~\ref{subsec:rindler-positivity}. We then explain in subsec.~\ref{subsec_HEFT} the role of conformal defects and the cusp anomalous dimension in the EFT description of real-time phenomena involving heavy particles, and use that to obtain another positivity bound for the Sudakov Lorentzian cusp anomalous dimension. In subsection~\ref{sec_Large_boost} we derive our main result~\eqref{eq_int_intro} and we discuss some generalizations in subsection~\ref{subsec_discussion}. In section~\ref{sec:examples}, we analyze several examples including free scalar and Maxwell theories, perturbative interacting CFTs, planar $\mathcal N=4$ SYM, and two-dimensional CFT interfaces. Section \ref{sec_spin_imp} computes and analyzes the cusp anomalous dimension for spin impurities. We conclude in section \ref{sec:outlook} with a discussion of open questions and future directions.
Appendix \ref{app_coordinates} contains a discussion of the various coordinate frames we use from the viewpoint of the conformal embedding space.  Appendix \ref{app_spectral_representation} analyzes the Lorentzian cusp and in particular the interaction \eqref{eq_int_intro} via spectral methods. Appendix \ref{app_Large_S} reviews the analysis of \cite{Alday:2007mf} of large spin operators. Appendix \ref{app_I12} and \ref{app_technical_spin} give technical details of the calculations.

\section{Cusped Defects and the Large-Boost Limit}\label{sec:generalities}
\subsection{Euclidean Cusps and Lorentzian Continuations}\label{sec:continuations}

\begin{figure}
\begin{subfigure}{0.43\textwidth}
    \centering
    \includegraphics[width=0.75\linewidth]{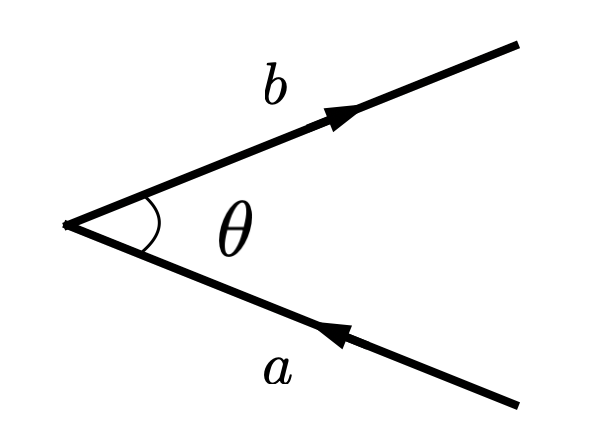}
        \caption{}
          \label{fig:EuclideanCusp}
    \end{subfigure}
    \hfill
   \begin{subfigure}{0.55\textwidth}
    \centering
    \includegraphics[width=\linewidth]{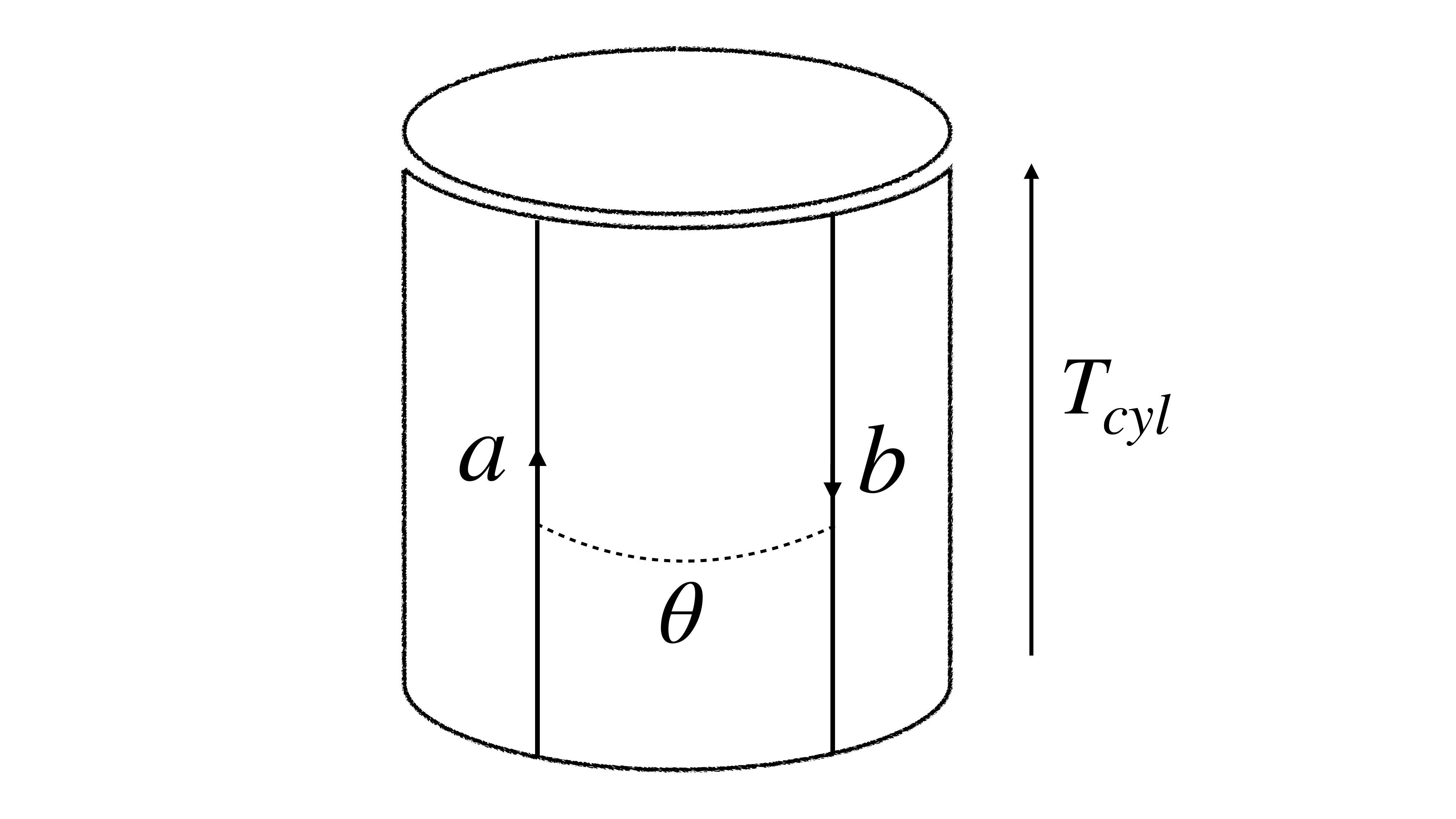}
        \caption{}
          \label{fig:EuclideanCusp_cyl}
    \end{subfigure}
    \caption{Cusped defects in Euclidean signature, both in flat space and on the cylinder.}
\end{figure}

The cusp anomalous dimension $\Gamma_{ab}(\theta)$ is defined as the coefficient appearing in the logarithmically divergent term of the partition function with line defects $a,b$ forming a cusped contour of angle $\theta$ as in fig.~\ref{fig:EuclideanCusp}:
\begin{equation}\label{cusp}
    \log Z_{ab}(\theta) = -\Gamma_{ab}(\theta)\log\left(\frac{L_{\textrm{IR}}}{\ell_{\textrm{UV}}}\right) + \dots~,
\end{equation}
where the terms appearing in the dots stand for non-logarithmically divergent terms. In the above equation, $L_{\textrm{IR}}$ is an infrared cutoff that we can interpret as the size of the defect, while $\ell_{\textrm{UV}}$ is an ultraviolet cutoff representing the length scale at which the cusp is smoothed out. 

For conformal field theories, the cusp configuration breaks the Euclidean conformal group $SO(d+1, 1)$ down to its subgroup of dilations $\mathbb{R}_D$ under which the cusp transforms linearly as with ordinary local operators. By a Weyl rescaling, the cusp anomalous dimension $\Gamma_{ab}(\theta)$ can also be thought of as the ground state energy on $\mathbb{R}\times S^{d-1}$ with two defects $a,b$ separated by an angle $\theta$, see fig.~\ref{fig:EuclideanCusp_cyl}:
\begin{equation}\label{cusp_cyl}
    \log\frac{Z_{ab}(\theta)}{\sqrt{Z_{aa}(\pi)Z_{bb}(\pi)}} = -\Gamma_{ab}(\theta)T_{\textrm{cyl}}~,
\end{equation}
where $T_{\textrm{cyl}}$ is the Euclidean time along the cylinder and the  normalization $\sqrt{Z_{aa}(\pi)Z_{bb}(\pi)}$ subtracts the scheme-dependent worldline mass counterterms by normalizing with the straight-line ($\theta=\pi$) defect partition functions.

\begin{figure}[t]
    \centering
    \begin{subfigure}{0.4\textwidth}
        \centering
        \includegraphics[width=0.8\linewidth]{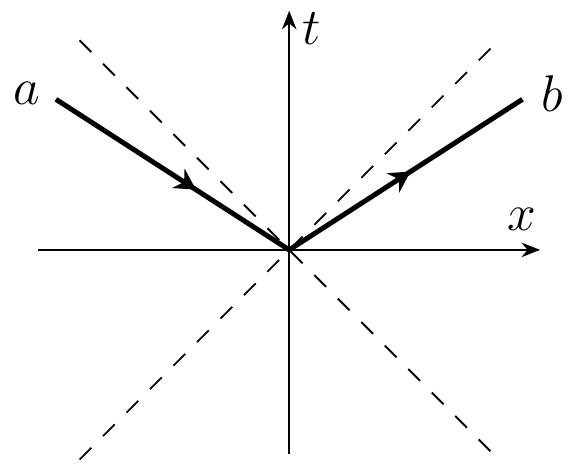}
        \caption{Spacelike}
        \label{fig:cusp-spacelike}
    \end{subfigure}
    \hspace{3em}
    \begin{subfigure}{0.4\textwidth}
        \centering
        \includegraphics[width=0.8\linewidth]{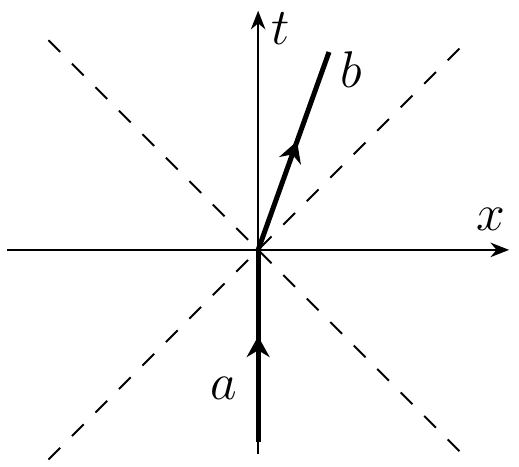}
        \caption{Quench}
        \label{fig:cusp-quench}
    \end{subfigure}\\
    \begin{subfigure}{0.4\textwidth}
        \centering
    \includegraphics[width=0.8\linewidth]{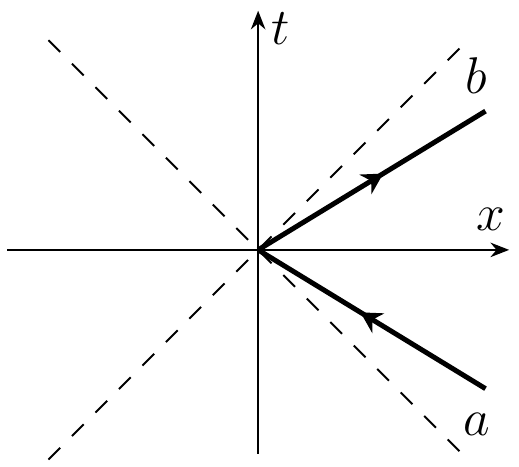}
       \caption{Spacelike$'$}
       \label{fig:cusp-spacelike-prime}
    \end{subfigure}   
     \hspace{3em}
     \begin{subfigure}{0.4\textwidth}
        \centering
        \includegraphics[width=0.8\linewidth]{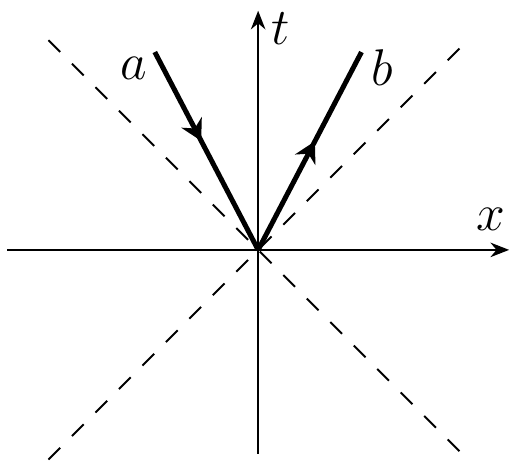}
        \caption{Sudakov}
        \label{fig:cusp-sudakov}
    \end{subfigure}
    \caption{Lorentzian cusp configurations considered in the text. The dashed lines denote the light cone \(t=\pm x\), while the thick lines denote the defects \(a\) and \(b\). Top left: the spacelike configuration defining \(\Gamma^L_{ab}(y)\). Top right: the quench configuration, where a static defect is kicked at the origin into the future light cone. Bottom left: the \(\text{spacelike}'\) configuration, where the defects run on spacelike contours causally connected with each other. Bottom right: the Sudakov configuration, with both defects inside the future light cone. }
    \label{fig:cusp-configurations}
\end{figure}

In this work, we will mainly be interested in the behavior of the cusp anomalous dimension under the analytic continuation
\begin{equation}\label{analyticont}
    \theta \to \pi - i \log y~.
\end{equation}
When $y>0$ this represents a boost in real time with rapidity
\begin{equation}
   \eta=\log y\,.
\end{equation}
The continuation~\eqref{analyticont} therefore allows us to define the Lorentzian cusp anomalous dimension as
\begin{equation}\label{def_cusp_analyticont}
    \Gamma_{ab}^L(y) = \Gamma_{ab}(\pi - i\log y)~.
\end{equation}
Here, the line defects $a$ and $b$ are taken to lie along trajectories that are everywhere spacelike separated, see  fig.~\ref{fig:cusp-spacelike}, so that the above quantity remains real. Without loss of generality, we can assume that both lines lie on the $(t,x)$ plane, and neglect the other coordinates. Then the two trajectories may be written explicitly as
\begin{equation}\label{eq_spacelike_traj}
    x_{1,2}^\mu=\frac{\lambda_{1,2}}{2}\left(1-\frac{1}{y},\,\pm\left(1+\frac{1}{y}\right)\right)\,,\qquad
    \lambda_{1,2}\in\mathds{R}^+\,,
\end{equation}
where $\lambda_{1,2}$ are the curves' affine parameters. From now on we take $y>1$, or equivalently positive relative rapidity, without loss of generality; the region $0<y<1$ is obtained by exchanging the two defect lines, $a\leftrightarrow b$, together with $y\leftrightarrow y^{-1}$.

When analytically continuing to configurations in which the two lines are in causal contact, we need to ensure the proper time-ordering. We suppress the defect labels $a,b$ below when no confusion can arise.

Let us consider first the \emph{Sudakov} cusp described in the introduction between two timelike lines (both inside the light cone of the origin), as in fig.~\ref{fig:cusp-sudakov}. Naively this is described by taking $y<-1$ in~\eqref{eq_spacelike_traj}. Since the Feynman prescription requires $t_{1,2}= \Re t_{1,2}-i\epsilon \Re t_{1,2}$ with $\epsilon>0$, we find that the analytic continuation describing the Sudakov configuration starting from $y>0$ is
\begin{equation}
    \Gamma^L_{\text{Sudakov}}(y)=\Gamma^{L}(e^{-i\pi+i\epsilon}y)=\Gamma^{L}(-y-i\epsilon)\,,\qquad
    \epsilon>0\,.
\end{equation}
Note that under this analytic continuation the cusp generically acquires a nontrivial complex phase. This describes the phase shift due to the exchange of soft quanta between the two heavy particles produced in the hard event.

The opposite continuation across the cut $y\rightarrow -y+i\epsilon$ also has a simple physical interpretation. To see this, note that we can boost the trajectories~\eqref{eq_spacelike_traj} to
\begin{equation}
    x^\mu_1=\lambda_1(0,1)\,,\qquad
    x^\mu_2=\lambda_2\left(\frac{y^2-1}{2 y},\,\frac{y^2+1}{2y}\right)\,.
\end{equation}
The continuation $y\rightarrow -y+i\epsilon$ then provides the correct Feynman prescription to describe a cusp configuration between spacelike but causally connected lines, such that the second line approaches the light cone on the right past of the origin---see fig.~\ref{fig:cusp-spacelike-prime}. 
Denoting this configuration as $\text{spacelike}'$, we conclude
\begin{equation}\label{eq_app_Gamma_spacelike2}
    \Gamma^L_{\text{spacelike}'}(y)=\Gamma^{L}(e^{i\pi-i\epsilon}y)=\Gamma^{L}(-y+i\epsilon)\,,\quad
    \epsilon>0\,.
\end{equation}

This causally connected spacelike configuration is related to a physical quench, where the heavy particle sits at rest, is suddenly \emph{kicked} at $t=0$ and moves with a nontrivial velocity for $t>0$. Indeed, the $\text{spacelike}'$ trajectories can be boosted to
\begin{equation}
    x_{1,2}^\mu=\frac{\lambda_{1,2}}{2}\left(\pm\left(1-\frac{1}{y}\right)\mp i\epsilon,\,1+\frac{1}{y}\right)\,,
\end{equation}
where we wrote the resulting $i\epsilon$ explicitly.
We now get timelike lines as in fig.~\ref{fig:cusp-quench} with the appropriate Feynman prescription sending $y\rightarrow -y-i\epsilon=e^{-i\pi+i\epsilon}y$:
\begin{equation}
\Gamma_{\text{quench}}^L(y)=\Gamma^L_{\text{spacelike}'}(e^{-i\pi+i\epsilon}y)=\Gamma^L_{\text{spacelike}'}(-y-i\epsilon)\,.
\end{equation}
Using~\eqref{eq_app_Gamma_spacelike2}, we conclude
\begin{equation}\label{eq_quench_spacelike}
    \Gamma_{\text{quench}}^L(y)=\Gamma^L(y)\,.
\end{equation}
This implies in particular that the cusp anomalous dimension describing the quench is real. All the above relations have been checked in the free theory examples discussed in sec.~\ref{subsec_free}.

Below we describe the positivity constraints obeyed by the Lorentzian cusp anomalous dimension and its physical significance using the language of heavy particle EFT. We then discuss the derivation of our main result~\eqref{lorentzcusp}.

\subsection{Positivity of the Lorentzian Cusp Anomalous Dimension}\label{subsec:rindler-positivity}

The spacelike configuration admits a simple positivity bound following from the Lorentzian analogue of reflection positivity: Rindler positivity \cite{Casini:2010bf,Hartman:2016lgu}. 

As a review, let
\begin{equation}
W_R=\left\{x^\mu:\ x>|t|\right\}
\end{equation}
be the right Rindler wedge, and denote by $\Theta_R$ the anti-linear Rindler reflection. Geometrically, it acts on the $(t,x)$ plane as
\begin{equation}
\mathcal R_R:(t,x)\longmapsto(-t,-x)
\end{equation}
and maps the right wedge to the left wedge. The operation $\Theta_R$ also includes Hermitian conjugation, as well as the appropriate conjugation of any internal quantum numbers. Rindler positivity states that, for any collection of operators $\mathcal O_i$ supported in $W_R$, the matrix
\begin{equation}\label{eq_rindler_gram_general}
M_{ij}=\left\langle \Theta_R(\mathcal O_i)\mathcal O_j\right\rangle
\end{equation}
is positive semidefinite, $M\succeq 0$. In particular, $M$ is Hermitian.

Rindler positivity has been proven for Wightman functions of local operators \cite{Casini:2010bf}. In CFTs, it also follows from positivity of local-operator two-point functions together with convergence of the OPE \cite{Hartman:2016lgu,Kravchuk:2018htv}. Below we assume that Rindler positivity extends to defects and apply it to defect half-lines contained in $W_R$. For defects admitting endpoint operators that connect them to the trivial line, this assumption can be justified by regulating the half-lines to finite segments. The resulting defect segment can then be expanded, through the OPE of its endpoint operators, as a convergent sum over local bulk operators, to which Rindler positivity applies directly.

A spacelike ray with boost parameter $\eta$ can be parametrized as
\begin{equation}\label{eq_rindler_ray}
X^\mu(\lambda;\eta)
=
\lambda\left(\sinh\eta,\cosh\eta\right),
\qquad \lambda>0\,.
\end{equation}
Let $\mathcal D_a(\eta)$ denote the regulated insertion of a defect of type $a$ along this ray. The Rindler reflection of a line with rapidity $\eta_i$, combined with a line at rapidity $\eta_j$, forms a cusp satisfying
\begin{equation}
\pd_{\lambda}[\mathcal R_R X](\lambda;\eta_i)\cdot \pd_{\lambda}X(\lambda;\eta_j)
=
-\cosh(\eta_j-\eta_i)
=
\cos\left[\pi-i(\eta_j-\eta_i)\right].
\end{equation}
Consequently,\footnote{\label{footnote_finite}As discussed above, for lines admitting endpoints connecting them to the trivial line, one may regulate each half-line by cutting it off at proper distances $\ell_{\rm UV}$ and $L_{\rm IR}$ from the origin, so that it becomes a finite defect segment and Rindler positivity follows from the endpoints OPE. The endpoint operators then contribute additional logarithmic terms, effectively replacing $\Gamma_{ab}^{L}\longrightarrow
\Gamma_{ab}^{L}-\Delta_{a0}-\Delta_{b0}$, where $\Delta_{a0}$ and $\Delta_{b0}$ are the corresponding endpoint-operator dimensions. These contributions cancel in the normalized ratio~\eqref{eq_rindler_cauchy_schwarz}, and therefore do not affect the resulting bound.}
\begin{align}\label{eq_rindler_defect_kernel}
\left\langle
\Theta_R\!\left(\mathcal D_a(\eta_i)\right)
\mathcal D_b(\eta_j)
\right\rangle
&=
Z_{ab}\left(\pi-i(\eta_j-\eta_i)\right)
\nonumber\\
&=
\exp\left[
-\Gamma_{ab}^{L}\!\left(e^{\eta_j-\eta_i}\right)
\log\left(\frac{L_{\textrm{IR}}}{\ell_{\textrm{UV}}}\right)
+O(1)
\right].
\end{align}
Here we used the definition $\Gamma_{ab}^{L}(y)=\Gamma_{ab}(\pi-i\log y)$, so that the relative rapidity $\eta_j-\eta_i$ corresponds to $y=e^{\eta_j-\eta_i}$. In our orientation conventions, illustrated in figs.~\ref{fig:EuclideanCusp} and~\ref{fig:cusp-spacelike}, the defect $a$ on the left is directed toward the cusp, while the defect $b$ on the right is directed away from it. Indeed, the geometric Rindler reflection maps the outward-directed right-wedge ray to the left wedge, while the adjoint operation contained in $\Theta_R$ reverses the defect ordering. Their combined action therefore produces an incoming left-wedge defect, with precisely the orientation used in the definition of the cusp anomalous dimension.

To obtain the desired inequality, we make the symmetric choice
\begin{equation}\label{eq_symmetric_rapidities}
\eta_1=-\frac{\eta}{2}\,,
\qquad
\eta_2=+\frac{\eta}{2}\,,
\qquad
\eta=\log y\,,
\end{equation}
where $y>0$ is arbitrary. Rindler positivity ensures that the following matrix is Hermitian and positive semidefinite
\begin{equation}\label{eq_full_rindler_matrix}
M=
\begin{pmatrix}
Z_{aa}(\pi)
&
Z_{ab}(\pi-i\eta)\\[2mm]
Z_{ba}(\pi+i\eta)
&
Z_{bb}(\pi)
\end{pmatrix}
\,.
\end{equation}
Hermiticity of the Rindler Gram matrix implies
\begin{equation}\label{eq_offdiagonal_conjugation}
Z_{ba}(\pi+i\eta)
=
Z_{ab}(\pi-i\eta)^*\quad\implies \quad
\Gamma_{ab}^{L}(y)=\Gamma_{ba}^{L}(y^{-1})\,,
\end{equation}
where we used that the cusp $\Gamma_{ab}^{L}(y)$ describing the configuration \ref{fig:cusp-spacelike} is real.

Positivity of the determinant of~\eqref{eq_full_rindler_matrix} gives the Cauchy--Schwarz inequality
\begin{equation}\label{eq_rindler_cauchy_schwarz}
\frac{\left|
Z_{ab}(\pi-i\eta)
\right|^2}{Z_{aa}(\pi)Z_{bb}(\pi)}
\leq 1\,.
\end{equation}
Importantly, the defect cosmological constant ambiguities cancel in the above ratio. Taking the coefficient of the large logarithm $\log(L_{\textrm{IR}}/\ell_{\textrm{UV}})$, we therefore conclude
\begin{equation}\label{eq_rindler_gamma_bound}
\Gamma^L_{ab}(y)\geq 0\,,
\qquad y>0\,,
\end{equation}
where we used the reflection symmetry \eqref{eq_offdiagonal_conjugation} to extend the result to arbitrary $y>0$. 

Note that \eqref{eq_rindler_gamma_bound} implies that the Lorentzian cusp anomalous dimension between two arbitrary defects is positive. This is to be contrasted with the Euclidean cusp anomalous dimension, which may take either sign for different lines, and is always negative (as well as decreasing and concave) for identical defects \cite{Cuomo:2024psk}.\footnote{These properties of the Euclidean cusp, as well as related bounds on the Casimir energy between defects \cite{Bachas:2006ti,Kravchuk:2024qoh}, also follow from a Cauchy-Schwarz argument analogous to \eqref{eq_rindler_cauchy_schwarz} \cite{Lanzetta:2026appear}.}

As a check, the bound~\eqref{eq_rindler_gamma_bound} can be readily verified in a neighborhood of $y=1$, corresponding to $\theta=\pi$. For two identical defects, the Lorentzian cusp anomalous dimension is determined by the normalization $C_D>0$ of the displacement-operator two-point function \cite{Correa:2012at,Cuomo:2024psk}:
\begin{equation}
\Gamma_{aa}^{L}(y)
=\Gamma_{aa}\left(\pi-i\log y\right)=\frac{C_D}{12}\log^2 y
+O\left(\log^4 y\right)\,,
\end{equation}
which is manifestly non-negative near $y=1$. For two different defects, the cusp anomalous dimension at $\theta=\pi$ coincides with the scaling dimension $\Delta_{ab}$ of the defect-changing operator. Therefore,
\begin{equation}
\Gamma_{ab}^{L}(y)=\Gamma_{ab}\left(\pi-i\log y\right)
=\Delta_{ab}
+O\left(\log y\right)\,,
\end{equation}
where $\Delta_{ab}\geq 0$ by the $SL(2,\mathbb{R})$ unitarity bounds, again in agreement with~\eqref{eq_rindler_gamma_bound} sufficiently close to $y=1$.

In view of \eqref{eq_quench_spacelike}, the result \eqref{eq_rindler_gamma_bound} also establishes a positivity bound for the quench configuration. The Sudakov and spacelike$'$ configurations are not directly constrained by the same Rindler-positivity argument. In the next section, we use different techniques to obtain a positivity bound on the real part of the Sudakov and spacelike$'$ cusp anomalous dimension in a broad class of examples.

\subsection{Defects, Born--Oppenheimer, and Heavy Particle Effective Theory}\label{subsec_HEFT}

Defects describe the interactions of heavy particles with the bath. The underlying physical picture is analogous to the Born--Oppenheimer approximation: on the time scales relevant for fast gapless modes, slow heavy particles behave as classical sources moving along fixed trajectories. As is well known from the physics of heavy quarks~\cite{Neubert:1993mb,Manohar:2000dt}, this statement can be made precise via heavy particle effective theory~\cite{Korchemsky:1991zp} (see \cite{DiRisi:2024nus,Berean-Dutcher:2025ohp,Tizzano:2026rgr} for recent applications). In this section we review this connection and show explicitly how the cusp anomalous dimension encodes the soft radiation emitted in the production of a heavy particle-antiparticle pair.

To be concrete, we consider a complex heavy scalar $\Phi$ immersed in a $d$-dimensional Ising CFT. We recall that the two lowest primary fields of the Ising CFT are the spin field $\sigma$, which is $\mathds{Z}_2$ odd, and the energy density field $\varepsilon$, which is even. In particular we have $\Delta_\sigma<1<\Delta_{\varepsilon}$ for $d<4$. Assuming that the interaction with the heavy particle breaks the $\mathds{Z}_2$ symmetry, the effective Lagrangian including the leading relevant coupling reads
\begin{equation}\label{eq_Ising_toy}
    \mL=\mL_{\text{Ising}}-|\pd\Phi|^2-m^2|\Phi|^2-g\,m^{2-\Delta_{\sigma}}\sigma|\Phi|^2\,.
\end{equation}
For simplicity, here we take $g$ to be a small dimensionless coupling, and we leave implicit the (scheme-dependent) counterterms which are needed to ensure that the bulk remains at criticality. Note that the $U(1)$ symmetry acting on $\Phi$ ensures that the heavy particle cannot decay. Below we would like to discuss how interactions between the Ising CFT and the scalar affect two basic quantities: the $\Phi-\Phi^*$ propagator near $p^2\approx -m^2$
\begin{equation}
    \langle\Phi(p)\Phi^*(p)\rangle\stackrel{g=0}{=}\frac{i}{-p^2-m^2+i 0}\,,
\end{equation}
and the \emph{generalized Sudakov form factor}~\cite{Collins:1989bt}
\begin{equation}
    \langle\Phi(p_1)\Phi^*(p_2)||\Phi|^2(0)|0\rangle\stackrel{g=0}{=}1\,.
\end{equation}
Diagrammatically, this last quantity is the amputated amplitude for producing a particle-antiparticle pair via a $|\Phi|^2$ vertex. This is therefore analogous to the current form factor in Deep Inelastic Scattering (suitably Wick-rotated to consider a production process, see e.g.~\cite{Karateev:2019ymz}).

The smallness of the dimensionless coupling implies that interactions between the heavy scalar field and Ising quanta of virtuality $|Q^2|\gtrsim m^2$ are negligible. The only important effects therefore arise from CFT quanta of virtuality $|Q^2|\ll m^2$ interacting with a nearly on-shell heavy field $\Phi$. Such interactions happen on time scales much larger than $1/m$. As in the Born--Oppenheimer approximation, where the slow heavy nucleus acts as a static source for the electronic wave-function, in this circumstance the $\Phi$ particle may be approximated by a rigid defect on its classical trajectory.

To make this intuition precise, we decompose the heavy field into two components as
\begin{equation}
    \Phi(x)=e^{ i m\,v\cdot x}\left[\Phi_{\text{near}}(x)+\Phi_{\text{off-shell}}(x)\right]\,,
\end{equation}
where $v$ is the particle's velocity, $\Phi_{\text{near}}(x)$ is the nearly on-shell component of the field, and $\Phi_{\text{off-shell}}(x)$ contains modes whose residual momenta are far from the heavy-particle mass shell. The exponential factor removes the large kinematic momentum $m v^\mu$, so that the remaining fields vary only on the scale of the residual momentum. In momentum space, this corresponds to writing
\begin{equation}
    p^\mu = m v^\mu + k^\mu\,,
\end{equation}
with $k^\mu \ll m$ for the near on-shell modes. The idea of heavy particle EFT is then simply to integrate out the \emph{fast} off-shell fluctuations $\Phi_{\text{off-shell}}$, and retain only the nearly on-shell component $\Phi_{\text{near}}$, that describes the slowly varying degrees of freedom.

Substituting the decomposition in the action~\eqref{eq_Ising_toy} and setting $\Phi_{\text{off-shell}}(x)=0$ to leading order at small $g$,\footnote{Integrating out $\Phi_{\text{off-shell}}(x)$ in general introduces perturbative corrections to the Wilson coefficient of the theory, as well as in the operator matching of, for instance, $|\Phi|^2$. Here we work at leading order in the coupling and neglect these.} we obtain
\begin{equation}\label{eq_HEFT}
    \mathcal L_{\rm near}
    \simeq
    2m\,\Phi_{\rm near}^*
    \left(
        i v\cdot \partial
        -\frac{g}{2}m^{1-\Delta_\sigma}\sigma
    \right)
    \Phi_{\rm near}\,.
\end{equation}
where we neglect terms with more than one derivative, which are suppressed by powers of $k/m$. The interaction with the CFT therefore appears as a one-dimensional
background potential along the worldline of the heavy particle. It can be removed from the leading heavy-particle Lagrangian by the field redefinition
\begin{equation}
    \Phi_{\rm near}(x)
    =
    \mathcal D_v(x,y)\,\widetilde \Phi_v(x)\,,
\end{equation}
with
\begin{equation}\label{eq_HEFT_pinning}
    \mathcal D_v(x,y)
    =
    \exp\left[
        -\frac{i g}{2}m^{1-\Delta_\sigma}
        \int_0^\lambda ds\,
        \sigma(y+s v)
    \right]\,,
    \qquad
    x^\mu=y^\mu+\lambda v^\mu \,,
\end{equation}
where $y$ is an arbitrary reference point. Indeed it is simple to check that
\begin{equation}
    \left(
        i v\cdot \partial
        -\frac{g}{2}m^{1-\Delta_\sigma}\sigma
    \right)
    \Phi_{\rm near}(x)
    =
    \mathcal D_v(x,y)\,i v\cdot \partial\,\widetilde\Phi_v(x).
\end{equation}
Thus the leading interaction with soft CFT modes has been traded for the insertion of a line defect supported on the heavy-particle trajectory. The line defect~\eqref{eq_HEFT_pinning} is just the pinning field defect studied in~\cite{Allais:2014fqa,2014arXiv1412.3449A,2017PhRvB..95a4401P,Cuomo:2021kfm,Hu:2023ghk,Zhou:2023fqu,Lanzetta:2025xfw}, which flows to an interacting conformal defect at large distances since $\Delta_{\sigma}<1$.

This immediately gives a defect representation of the heavy-particle propagator. One finds at leading
order in the heavy-mass expansion
\begin{equation}\label{eq_2pt_HEFT}
    \left\langle
        \Phi_{\rm near}(x)\Phi_{\rm near}^*(0)
    \right\rangle
    \simeq
    G^{(0)}_v(x)\,
    \left\langle
        \mathcal D_v(x,0)
    \right\rangle_{\rm CFT}\,,
\end{equation}
where
\begin{equation}
    G^{(0)}_v(x)
    =
    \frac{1}{2m}\,
    \theta(-v\cdot x)\,
    \delta^{(d-1)}(x_\perp)=\int d^dk\frac{-i\, e^{ikx}}{2m v\cdot k-i 0}\,,\qquad x_{\perp}^\mu=x^\mu+(v\cdot x)v^\mu\,,
\end{equation}
is the free propagator of the first-order heavy field. Fourier transforming with respect to the residual momentum \(k^\mu\), we obtain
\begin{equation}\label{eq_near_shell_prop}
    \langle \Phi(p)\Phi^*(p)\rangle
    \simeq
    \frac{1}{2m}
    \int_0^\infty d\tau\,
    e^{i (-v\cdot k+i0)\tau}\,
    \left\langle
        \mathcal D_v(\tau v,0)
    \right\rangle_{\rm CFT}\left[1+O\left(\frac{k}{m}\right)\right],
    \qquad
    p^\mu=m v^\mu+k^\mu \,.
\end{equation}

In the large distance limit, we expect the finite straight defect expectation value to decay according to the dimension $\Delta_{c0}$ of the lowest dimensional defect creation operator
\begin{equation}
    \left\langle \mathcal D_v(\tau v,0)\right\rangle_{\rm CFT}
    \propto
    \frac{1}{\tau^{2\Delta_{c0}}}.
\end{equation}
Using this result in~\eqref{eq_near_shell_prop}, and expressing $p^2+m^2\simeq 2m\,v\cdot k$, we obtain
\begin{equation}\label{eq_prop_NR}
    \langle \Phi(p)\Phi^*(p)\rangle
    \sim
    \frac{1}{m^{2\Delta_{c0}}
    \left(-p^2-m^2+i0\right)^{1-2\Delta_{c0}}
    } \,.
\end{equation}
Thus the coupling to the critical bath turns the heavy-particle pole into a branch point, with an exponent fixed by the scaling dimension of the creation operator of the defect.

We can treat the generalized Sudakov form factor in the same way. We define
\begin{equation}
    p_i^\mu=m v_i^\mu+k_i^\mu,\qquad
    \omega_i=-v_i\cdot k_i \,.
\end{equation}
The local operator $|\Phi|^2(0)$ creates a heavy particle and an antiparticle at the point
$x=0$. According to the analysis above, the subsequent soft evolution of the two heavy particles is described by two defect operators supported on the classical trajectories
\begin{equation}
    x_1^\mu(s)=s v_1^\mu,\qquad
    x_2^\mu(s)=s v_2^\mu,\qquad s>0 .
\end{equation}
Thus the unamputated production correlator in momentum space is, at leading order in the heavy-mass expansion,
\begin{equation}
  \langle \Phi^*(p_1)\Phi(p_2)|\Phi|^2(0)\rangle
    \simeq
    \frac{1}{(2m)^2}
    \int_0^\infty d\tau_1\,d\tau_2\,
    e^{i(\omega_1+i0)\tau_1+i(\omega_2+i0)\tau_2}
    \left\langle
        \mathcal D_{v_1}(\tau_1 v_1,0)
        \mathcal D_{v_2}(\tau_2 v_2,0)
    \right\rangle_{\rm CFT}.
\end{equation}
In the above, the two defect insertions meet at the production point and form a cusp whose opening angle is fixed by $v_1\cdot v_2$.  Using that the external two-point functions are given by~\eqref{eq_near_shell_prop} as before, we express the off-shell amputated soft factor as
\begin{equation}
\begin{split}
  \mathcal F(v_1,v_2;\omega_1,\omega_2) &\equiv\frac{\langle \Phi^*(p_1)\Phi(p_2)|\Phi|^2(0)\rangle}{\langle \Phi^*(p_1)\Phi(p_1)\rangle \langle \Phi^*(p_2)\Phi(p_2)\rangle } \\
  &\simeq
    \frac{
    \displaystyle
    \int_0^\infty d\tau_1\,d\tau_2\,
    e^{i(\omega_1+i0)\tau_1+i(\omega_2+i0)\tau_2}
    \left\langle
        \mathcal D_{v_1}(\tau_1 v_1,0)
        \mathcal D_{v_2}(\tau_2 v_2,0)
    \right\rangle_{\rm CFT}
    }{
    \displaystyle
    \prod_{i=1}^2
    \left[
    \int_0^\infty d\tau_i\,
    e^{i(\omega_i+i0)\tau_i}
    \left\langle
        \mathcal D_{v_i}(\tau_i v_i,0)
    \right\rangle_{\rm CFT}
    \right]
    } \,.
    \end{split}
\end{equation}
The generalized Sudakov form factor is the on-shell limit of the above
\begin{equation}\label{eq_Sudakov_final}
\langle
        \Phi(p_1)\Phi^*(p_2)
        |\,|\Phi|^2(0)\,|0\rangle
    =
   \lim_{\omega_1,\omega_2\rightarrow 0}
    \mathcal F(v_1,v_2;\omega_1,\omega_2)
    \,.
\end{equation}

Finally, we use again that in the infrared limit the defect is well described by a DCFT. In this case, scale invariance fixes
\begin{equation}
    \left\langle
        \mathcal D_{v_1}(\tau_1 v_1,0)
        \mathcal D_{v_2}(\tau_2 v_2,0)
    \right\rangle_{\rm CFT}=\frac{f(\tau_1/\tau_2)}{(\tau_1\tau_2)^{\Delta_{c0}+\Gamma^L_{\text{Sudakov}}(|v_1\cdot v_2|)/2}}\,,
\end{equation}
where $f(x)$ is an unknown function of the scale invariant ratio $\tau_1/\tau_2$ and $\Gamma^L_{\text{Sudakov}}(|v_1\cdot v_2|)=\Gamma^L(-|v_1\cdot v_2|-i\epsilon)$ is the Lorentzian cusp anomalous dimension analytically continued to describe two timelike defects, which is generically complex as explained in section~\ref{sec:continuations}.  Therefore, for small $\omega_i$, the off-shell amputated amplitude obeys
\begin{equation}
    \mathcal F(v_1,v_2;\lambda \omega_1,\lambda\omega_2) \simeq \lambda^{\Gamma^L_{\text{Sudakov}}(|v_1\cdot v_2|)-2\Delta_{c0}}\mathcal F(v_1,v_2; \omega_1,\omega_2)\,.
\end{equation}
We thus see that depending on the sign of $\Re \Gamma^L_{\text{Sudakov}}(|v_1\cdot v_2|)-2\Delta_{c0}$, the physical Sudakov form factor~\eqref{eq_Sudakov_final} may either diverge or vanish. In particular when $\Re\Gamma^L_{\rm Sudakov}(|v_1\cdot v_2|)> 2\Delta_{c0}$ the form factor vanishes
\begin{equation}
    \langle
        \Phi(p_1)\Phi^*(p_2)
        |\,|\Phi|^2(0)\,|0\rangle
    =
   \lim_{\omega_1,\omega_2\rightarrow 0}
    \mathcal F(v_1,v_2;\omega_1,\omega_2)=0\quad\text{if}\quad
    \Re\Gamma^L_{\rm Sudakov}(|v_1\cdot v_2|)> 2\Delta_{c0}\,.
\end{equation}
We shall see below however that unitarity is less strict and does not require the form factor to remain finite.

The striking changes in the behavior of correlators near the mass-shell are analogous to the vanishing of exclusive amplitudes in QED. The matrix element in \eqref{eq_Sudakov_final} is fully exclusive in the CFT sector: it asks for the production of the heavy particle-antiparticle pair with no additional radiation in the bath. However, the acceleration of the heavy sources at the production point necessarily excites arbitrarily soft CFT modes, which radically change its behavior. In condensed matter language, this phenomenon is an infrared effect analogous to the orthogonality catastrophe in Fermi surfaces~\cite{PhysRevLett.18.1049,Affleck:1996}.

In an actual physical measurement one should not take the strictly exclusive limit, since detectors have a finite energy resolution $E_{\rm det}$. The physical ``exclusive'' observable is really inclusive over all CFT states with energy below $E_{\rm det}$. We may estimate the probability for such an exclusive observable as
\begin{equation}\label{eq_prob_exclusive}
    P(E_{\rm det})\propto \int_{\omega_1+\omega_2<E_{\rm det}} d\omega_1 d\omega_2\rho_{v_1}(\omega_1)\rho_{v_2}(\omega_2)|\mathcal F(v_1,v_2;\omega_1,\omega_2)|^2\,,
\end{equation}
where we introduced the density of states associated with the \emph{dressed} heavy particle, which can be extracted by taking the discontinuity of~\eqref{eq_prop_NR}
\begin{equation}
    \rho_v(\omega)\sim \omega^{2\Delta_{c0}-1}\,.
\end{equation}
We then see that finiteness of the probability requires\footnote{Note that this is a necessary condition for the finiteness of the two-point function $\langle |\Phi|^2(x)  |\Phi|^2(0)\rangle$.}
\begin{equation}\label{eq_condition_Gamma}
     \Re\Gamma^L_{\rm Sudakov}(|v_1\cdot v_2|)\geq 0\,,
\end{equation}
which is weaker than the vanishing of the exclusive form factor, that would instead demand $\Re\Gamma^L_{\rm Sudakov}(|v_1\cdot v_2|)\geq 2\Delta_{c0}>0$. Under this condition the probability of measuring two heavy particles, in the state created by $|\Phi|^2$, with radiation up to the detector resolution scales as
\begin{equation}\label{eq_prob_Gamma}
    P(E_{\rm det})\propto (E_{\rm det})^{2\Re\Gamma^L_{\rm Sudakov}(|v_1\cdot v_2|)}\,.
\end{equation}
The imaginary part of $\Gamma^L_{\rm Sudakov}(|v_1\cdot v_2|)$ describes instead the phase shift accumulated by the heavy states due to the exchange of gapless modes.

Let us comment that our discussion focused on the leading scaling of the relevant observables with the off-shell regulator, or equivalently with the infrared cutoff. The heavy-particle EFT framework also provides a systematic way to compute corrections to this leading behavior. These corrections arise from subleading terms in the matching coefficients, from higher-derivative operators in the effective action~\eqref{eq_HEFT}, and from irrelevant operators localized on the defect.

Finally, we emphasize that analogous steps can be carried out in several other models including gauge theories. In that case, as is well known~\cite{Sudakov:1954sw,Polyakov:1980ca,Korchemsky:1987wg,Gubser:2002tv,Kruczenski:2002fb,Alday:2007mf}, the Sudakov exponent is identified with the Wilson-line cusp anomalous dimension. 
We can similarly generalize the positivity condition~\eqref{eq_condition_Gamma} to arbitrary defects that describe the propagation of heavy particles through the CFT:
\begin{equation}\label{eq_sudakov_gamma_bound}
    \Re\Gamma_{\mathrm{Sudakov},ab}^L(y)
    =
    \Re\Gamma_{ab}^L(-y-i\epsilon)
    \geq 0\,,
    \qquad y>0\,.
\end{equation}
We emphasize that the bound \eqref{eq_sudakov_gamma_bound} relies on heavy particle EFT techniques, which describe a broad class of examples, but may not apply to every conformal defect. We believe the result \eqref{eq_sudakov_gamma_bound} is general, but we did not prove it with CFT methods. Assuming the bound \eqref{eq_sudakov_gamma_bound} holds and using
\begin{equation}\label{eq_spacelikeprime_sudakov_conjugation}
    \Gamma_{\mathrm{spacelike}',ab}^L(y)
    =
    \left[
        \Gamma_{\mathrm{Sudakov},\bar b\bar a}^L(y)
    \right]^*\,,
\end{equation}
we also obtain
\begin{equation}
    \Re\Gamma_{\mathrm{spacelike}',ab}^L(y)=
    \Re\Gamma_{ab}^L(-y+i\epsilon)\geq 0\,.
\end{equation}

As a consistency check of the condition \eqref{eq_sudakov_gamma_bound}, note that, as $y\to1$, the Euclidean angle associated with the Sudakov continuation approaches $\theta=0$, and the cusp anomalous dimension is controlled by the fusion EFT of the defects $a$ and $\bar b$ \cite{Kravchuk:2024qoh,Cuomo:2024psk}. Denoting by $c$ the dominant line in this fusion channel, one finds
\begin{align}\nonumber
\Gamma_{\mathrm{Sudakov},ab}^L(y)&=
\Gamma_{ab}\left(\epsilon-i\log y\right)
\\ &=
\frac{C_{a\bar b c}}{\epsilon-i\log y}
+\Delta_{c0}
+\alpha_c\left(\epsilon-i\log y\right)^{\Delta_{\mathrm{irr}}-1}
+\ldots\,,
\quad \epsilon\to0^+ .
\end{align}
Here $C_{a\bar b c}$ is the Casimir-energy coefficient associated with the fusion $a\times\bar b\to c$, $\Delta_{c0}$ is the dimension of the lightest defect-creation operator for $c$, $\Delta_{\mathrm{irr}}>1$ is the dimension of the least irrelevant deformation of the fused line and $\alpha_c$ is an $O(1)$ coefficient. Taking $\epsilon\to0^+$ at fixed $y$, the leading Casimir term is purely imaginary, and does not contribute to the real part. Therefore we find
\begin{equation}
\lim_{y\to1}
\Re\Gamma_{\mathrm{Sudakov},ab}^L(y)=
\Delta_{c0}\geq0\,,
\end{equation}
in agreement with the bound~\eqref{eq_sudakov_gamma_bound}. When $\Delta_{c0}=0$ (e.g. when $c$ is the trivial line), the sign of the real part near $y=1$ is instead controlled by the first subleading term in the fusion EFT, and hence by the coefficient of the least irrelevant deformation of the fused line times its OPE coefficient with the defect endpoints.

\subsection{The Large Boost Limit}\label{sec_Large_boost}

In this section we derive the large boost behavior of the Lorentzian cusp anomalous dimension and explain the origin of the asymptotic form \eqref{lorentzcusp} in dimensions $d\geq 3$. The basic observation is that a cusp configuration between two null lines is invariant under both dilations and boosts in the plane spanned by the lines. For finite but large rapidity the boost symmetry is instead slightly broken. It is therefore convenient to find a Weyl frame in which these conformal symmetries are manifest isometries. To this aim we follow~\cite{Alday:2007mf} and perform a Weyl transformation to an AdS frame, where the boost parameter is mapped to the geodesic separation between the two defects in the coordinate conjugate to boosts. The large-boost limit $y\to \infty$ then becomes a large-separation limit, in which the interaction between the defects is governed by the minimal twist local operator that can be exchanged between the lines.

To make this structure explicit, let us place the two line defects along the trajectories    
\begin{equation}\label{eq_trj_2}
    x_{1,2}^\mu(\lambda_{1,2})=\lambda_{1,2}\left(
\frac{1}{2} \left(1+\frac{1}{\bar{y} }\right),\pm\frac{1}{2} \left(1-\frac{1}{\bar{y} }\right),\pm\frac{c}{\sqrt{\bar{y} }}\hat{n} \right)\,,\quad\lambda_{1,2}\in\mathds{R}^+\,,
\end{equation}
where $\hat{n}$ is a unit $d-2$-dimensional vector and $c>1$ is arbitrary for now.\footnote{In $d=3$, $\hat{n}=\pm 1$.}
For $c>1$ the lines are spacelike and become lightlike for $\bar{y}\to\infty$. As discussed in the previous section, this is precisely the relevant configuration we would like to consider to analyze the Lorentzian cusp $\Gamma^L(y)$. The corresponding cusp angle is given by:
\begin{equation}
    \frac{\dot{x}_1\cdot \dot{x}_2}{\sqrt{\dot{x}^2_1\dot{x}^2_2}}=-\frac{\bar{y}^2+2c^2\bar{y}+1}{2(c^2-1)\bar{y}}~.
\end{equation}
The relation between $\bar{y}$ and $y$ previously defined in~\eqref{analyticont} is thus given by
\begin{equation}
\begin{split}
    \bar{y} &=\frac{(c^2-1) y ^2-2 c^2 y +c^2-1+ (y -1)^2\sqrt{\left(c^2-1\right)\left[c^2 -\frac{(y +1)^2}{(y-1)^2}\right]}}{2 y }\\
    &=\left(c^2-1\right)y -2 c^2+\frac{c^2 \left(c^2-2\right)}{\left(c^2-1\right) y }+
O\left(\frac{1}{y^2}\right)\,.
    \end{split}
\end{equation}
We now perform a Weyl rescaling of the metric to AdS$_3\times S^{d-3}$:
\begin{equation}\label{eq_AdS3}
    ds^2_{\mathbb{R}^{1,d-1}}=r^2 ds^2_{\text{AdS}_3\times S^{d-3}}\,,\qquad
    ds^2_{\text{AdS}_3\times S^{d-3}}=\frac{-dt^2+dx^2+dr^2}{r^2}+d\hat{n}^2\,,
\end{equation}
where we work with coordinates $(t,x,r\,\hat{n})$.
Note that the trajectories defined in \eqref{eq_trj_2} lie at antipodal points on the sphere.\footnote{For $d=3$, the sphere reduces to two disconnected points, so the Weyl-rescaled geometry consists of two copies of AdS$_3$, with each line defect extending into one of them.}

At this point we may consider the following coordinate transformation \cite{Alday:2007mf}:
\begin{equation}\label{eq_alpha_coord}
    \frac{t\pm x}{r}=e^{\pm\gamma}\sin\alpha\,,\qquad
    \frac{1}{r}=e^{\tau}\cos\alpha\,.
\end{equation}
The coordinates $(\tau,\gamma,\alpha)$ cover the patch $0<t^2-x^2<r^2$, where the trajectories~\eqref{eq_trj_2} lie.
In terms of these new coordinates, the AdS$_3$ metric becomes:
\begin{equation}\label{eq_ds_alpha}
    ds^2_{\text{AdS}_3}=-d\alpha^2+d\gamma^2 \sin ^2\alpha +d\tau^2 \cos ^2\alpha \,.
\end{equation}
The above metric has two important isometries. The first one is the generator of conformal dilations which corresponds to translations in $\tau$. The second one is associated with boosts in the $(t,x)$-plane which corresponds to translations along $\gamma$.

The trajectories \eqref{eq_trj_2} correspond to
\begin{equation}\label{eq:traj-conditions}
\begin{aligned}
e^{\gamma_1+\tau_1}\tan\alpha_1 &= \lambda_1\,, & e^{2\gamma_1} &= \bar{y}, & \frac{1}{\sin\alpha_1} &= c\,, \\
e^{\tau_2-\gamma_2}\tan\alpha_2 &= \lambda_2\,, & e^{-2\gamma_2} &= \bar{y}, & \frac{1}{\sin\alpha_2} &= c\,,
\end{aligned}
\end{equation}
from which we see explicitly that $\gamma_{1,2}\rightarrow\pm \infty$ for $\bar{y}\rightarrow\infty$. Explicitly, the defects extend along the $\tau$-direction at a fixed value of the time coordinate $\alpha$ and at a separation given by:
\begin{equation}
    \Delta\gamma=\gamma_1-\gamma_2=\log \bar{y}\,.
\end{equation}

At this point we make our main assumption: the lines $a$ and $b$ admit endpoint operators connecting them to the trivial line. This means that the defects $a$ and $b$ are well-defined in isolation. Under this assumption the interaction between the defects decays in the $\Delta\gamma \to \infty$ limit, and hence the defects decouple in the large boost limit. To see this, we consider a suitable Wick rotation of the theory, that allows us to apply standard spectral methods. Note that a Wick rotation in $\tau$ or $\gamma$ alone is not globally well-defined. Let us consider instead performing the following analytic continuation on the metric \eqref{eq_ds_alpha}~\cite{Alday:2007mf}
\begin{equation}\label{eq_an_cont}
    \alpha=\frac{\pi}{4}+i\sigma\,,\qquad
    \tau=iu-\chi\,,\qquad
    \gamma=iu+\chi\,,
\end{equation}
giving rise to the following AdS$_3$ metric
\begin{equation}\label{eq_uc_metric}
    ds^2 = -du^2 +d\chi^2 -2\sinh(2\sigma)dud\chi +d\sigma^2~.
\end{equation}
Looking at the trajectories \eqref{eq:traj-conditions}, the analytic continuation~\eqref{eq_an_cont} is particularly natural if we take $c=\sqrt{2}$ so that $\alpha_1=\alpha_2=\pi/4$. In terms of the new coordinates the lines \eqref{eq_trj_2} are mapped to complex curves $u_{1,2}$ at distance $\Delta\chi$:
\begin{equation}
    u_{1,2} = -\frac{i}{2}\tau_{1,2}~\mp \frac{i}{4}\log \bar{y}~,\quad \Delta\chi\equiv\chi_1-\chi_2 = \frac{1}{2}\log \bar{y} ~-\frac{1}{2}\Delta\tau~,\quad
    \sigma_{1,2}=0\,,
\end{equation}
where $\Delta\tau \equiv \tau_1 -\tau_2$. The AdS$_3$ metric with local coordinates $(u,\chi,\sigma)$ appears prominently in the discussion of large spin operators in conformal field theory \cite{Alday:2007mf,Komargodski:2012ek,Fitzpatrick:2012yx}, as we review in app.~\ref{app_Large_S}. This is because it can be argued that the Hamiltonian $H_u$ associated with $u$-translations measures the twist spectrum of the theory, $\tau=\Delta-J$. We derive this fact in app.~\ref{app_coordinates}. Additionally, a double Wick-rotation $(u,\chi)\rightarrow (-i u,i\chi)$ results in an identical metric with $u$ and $\chi$ exchanged, hence also the Hamiltonian $H_{\chi}$ associated with $\chi$-translations is isomorphic to the twist generator. Note also that the warp factor $\sim dud\chi\sinh(2\sigma)$ localizes the excitations near the plane $\sigma=0$.

We may thus regard the cusp problem as a computation in a 2d gapped EFT with the $2d$ Lorentzian metric
\begin{equation}\label{2dmetric}
    ds^2|_{\sigma=0} = -du^2 + d\chi^2~.
\end{equation}
Correlations between the defects are mediated by the exchange of \emph{gapped} modes between them and are exponentially suppressed. Therefore the two defects decouple, and the corresponding contributions to the cusp anomalous dimension then factorize:
\begin{equation}\label{eq_Gamma_disconnected}
    Z_{ab}(\infty) = e^{-(f_a+f_b)\int d\tau}\qquad
    \Rightarrow\qquad
    \Gamma_{ab}^L(\infty) = f_a+f_b~,
\end{equation}
where we used that the logarithmic divergence is governed by the generator translating $\tau$.
The exponent is real because the defects extend along spacelike directions. The constants $f_a$ can be fixed by taking the second defect to be trivial:
\begin{equation}
    f_a = \Delta_{a0}~,
\end{equation}
where $\Delta_{a0}$ is the scaling dimension of the defect-creation operator for the line $a$, and similarly for $f_b$.

At large but finite $\Delta\gamma$ we turn on small interactions. Physically, the interaction between the defects is mediated by the exchange of (off-shell) massive modes. Since in the coordinates \eqref{eq_uc_metric} the gap of the lightest mode is measured by the twist $\tau$, we expect that the leading connected contribution at large boost parameter takes the form of a Yukawa-like potential $\sim e^{-\Delta\gamma \,\tau_{\rm exch}}$, where $\tau_{\rm exch}$ is the twist of the lightest operator that couples to the defects. 

In appendix \ref{app_spectral_representation} we discuss a spectral approach for estimating the interaction potential between the defects. Here we proceed more physically and model the interaction in terms of a two-body potential $\mathcal{L}_{\rm int}(\Delta\tau,\Delta\gamma)$:
\begin{equation}\label{eq_int_Gamma_L}
    \Gamma^L_{ab}(y)-\Gamma^L_{ab}(\infty)\simeq -\int d\Delta\tau
    \mathcal{L}_{\rm int}(\Delta\tau,\Delta\gamma)\,.
\end{equation}
In a large distance EFT valid at distances $\Delta\gamma\gg 1$, we can neglect the complicated local dynamics on the individual defects and we expect the interaction potential between two arbitrary far away sources to be described in terms of a bulk two-point function
\begin{equation}\label{eq_int_model}
    \mL_{\rm int}\simeq \lambda_{a}\lambda_b\langle\mO^{(a)}(x_1)\mO^{(b)}(x_2)\rangle\,,
\end{equation} 
where $\lambda_{a}$ and $\lambda_{b}$ are unknown Wilson coefficients,\footnote{Here we implicitly assumed that the defect endpoints are not degenerate. When they are, the coupling $\lambda^{(a)}$ and $\lambda^{(b)}$ should be promoted to matrices in the internal space---we will encounter an example in sec.~\ref{sec_spin_imp}.} and $\mO^{(a)}$ and $\mO^{(b)}$ are linear combinations of all bulk operators that create the lowest twist state when acting on the vacuum. Note that an analogous ansatz is also used to derive that the leading force between two far-away static (gauge neutral) particles in a flat space gapped theory is Yukawa like.\footnote{To see this, one expresses the correlator via the K\"all\'en-Lehmann representation, with the leading contribution arising from the lowest massive particle. It might be possible to justify more rigorously such ansatzes for long distance potentials in terms of an analytically continued non-relativistic EFT as suggested in~\cite{Lopes:2026erz}.} 
On general grounds therefore $\mO^{(a)}$ can be written in terms of the lowest twist primary state $\mO_{\rm exch}$ and its derivatives as
\begin{equation}
    \mO^{(a)}(\tau,\gamma,\alpha,\hat{n})=\sum_{n,m,k}c_{n,m,k}^{(a)}\pd_{\gamma}^n\pd^m_{\alpha}(\nabla_{\hat{n}}^2)^k\mO_{\rm exch}(\tau,\gamma,\alpha,\hat{n})\,,
\end{equation}
where we suppressed spin indices since they are inessential for our argument, and we neglected $\tau$-derivatives since these are total derivatives and thus do not contribute to the potential. 

Using the form of the primary two-point function to leading exponential accuracy
\begin{multline}\label{eq_2pointexch}
    \langle\mO_{\rm exch}(x_1)\mO_{\rm exch}(x_2)\rangle
    \\
    \propto\frac{\left[\cosh(\Delta\gamma) \cosh(\Delta\tau)\right]^J+\ldots}{[2\sin(\alpha_1)\sin(\alpha_2)\cosh( \Delta \gamma)+2\cos(\alpha_1)\cos(\alpha_2)\cosh(\Delta\tau)-2\hat{n}_1\cdot\hat{n}_2]^{\Delta+J}}\,,
\end{multline}
we can give more details on the integral appearing in \eqref{eq_int_Gamma_L}. For the configuration relevant here, we can neglect derivatives in $\alpha$ and $\hat{n}$ and set $\alpha_1=\alpha_2=\pi/4$ and $\hat{n}_1\cdot \hat{n}_2=-1$ to the order of interest. Then the contribution of \eqref{eq_2pointexch} is proportional to: 
\begin{equation}
-\sum_{n,m}c_{n,0,0}^{(a)}c_{m,0,0}^{(b)}(-)^{m}\pd_{\Delta\gamma}^{n+m}\int_{-\infty}^{\infty} d\Delta\tau \frac{\left[\cosh(\Delta\gamma) \cosh(\Delta\tau)\right]^J}{[\cosh( \Delta \gamma)+\cosh(\Delta\tau)+2]^{\Delta+J}}~.
\end{equation}
The large-$\Delta\gamma$ behavior of the above integral depends qualitatively on whether the exchanged operator is a scalar. For $J=0$, one obtains
\begin{equation}
    \int_{-\infty}^{\infty} d\Delta\tau \frac{1}{[\cosh( \Delta \gamma)+\cosh(\Delta\tau)+2]^{\Delta}} =2^{\Delta+1} \left[\Delta\gamma
    -\psi^{(0)}(\Delta )-\gamma_E
    \right] e^{-\Delta\Delta\gamma}[1 + O\left(e^{-\Delta\gamma}\right)]~,
\end{equation} 
where $\psi^{(0)}(x)\equiv \frac{d}{dx}\log \Gamma(x)$ is the digamma function, and $\gamma_E$ is the Euler--Mascheroni constant. In words, the local interaction is roughly constant for $|\Delta\gamma|\gg |\Delta \tau| $. This region gives rise to a logarithmic term in the cusp anomalous dimension. For $J>0$, the numerator suppresses the $\Delta\gamma$ enhancement and we have instead:
\begin{equation}
\int_{-\infty}^{\infty} d\Delta\tau \frac{\left[\cosh(\Delta\gamma) \cosh(\Delta\tau)\right]^J}{[\cosh( \Delta \gamma)+\cosh(\Delta\tau)+2]^{\Delta+J}}=2^{\Delta -J+1}\frac{\Gamma (\Delta )  \Gamma (J)}{\Gamma (J+\Delta )} e^{-(\Delta-J)\Delta\gamma}\left[1+O\left(e^{-\Delta\gamma}\right)\right]~.
\end{equation}
Finally, using that $\tau_{\rm exch} =\Delta-J$ and $\Delta\gamma = 
\log{y} + O(y^{-1})$ we find
\begin{equation}\label{eq_Gamma_L_int_2}
\begin{split}
    \Gamma^L_{ab}(y)-\Gamma^L_{ab}(\infty)
    &\simeq-\frac{\lambda_a\lambda_b}{y^{\tau_{\rm exch}}}\times\begin{cases}
        \#\log y+\#  &J=0\\[0.4em]
       \# &J>0\,,
    \end{cases}
    \end{split}
\end{equation}
where we denote by $\#$ the constants that depend on the coefficients $c_{n,0,0}^{(a)}$ and $c_{m,0,0}^{(b)}$ and the spin-index contractions. These are not determined by our arguments. We relate the coefficients in \eqref{eq_Gamma_L_int_2} to certain singularities of the analytic continuation of a suitably defined spectral density in app.~\ref{app_spectral_representation}. This concludes the derivation of our main result~\eqref{lorentzcusp}.

\subsection{Discussion}\label{subsec_discussion}

Several comments are in order. First, we note that arguments analogous to the above can be used to estimate the boost suppression of bulk operator one-point functions in the presence of a boosted line ending at the origin. Consider a bulk operator of twist $\tau_{\mO}$ in the presence of a boosted line ending at the origin. This is the three-point function of the bulk operator with the two defect endpoints, and it is not fixed by symmetry.  Let $\log y$ be the rapidity angle between the boosted line and the unique geodesic connecting the origin with the bulk operator insertion point. Since in the Weyl frame~\eqref{eq_ds_alpha}, the correlation is mediated by the propagation of a mass $\tau_{\mO}$ mode, we can estimate the $y$-dependence of the correlator (up to logarithms)
\begin{equation}\label{eq_1pt}
    \langle\mO(x)\mathcal{D}\rangle\stackrel{y\rightarrow\infty}\propto y^{-\tau_{\mO}}\,,
\end{equation}
where $\mathcal{D}$ refers to the defect connecting the origin and spatial infinity. In~\eqref{eq_1pt} we have neglected the dependence on the distance $x^2$ between the insertion point and the defect endpoint at the origin, that is kept fixed as $y\rightarrow \infty$. Note that although we have implicitly assumed the bulk operator and the line to be everywhere spacelike separated, we expect a similar suppression for other configurations related by analytic continuation.  

Analogously, we can consider a cusp configuration with a defect operator insertion on one of the two segments. In the large boost limit then we expect that the correlator will reduce to the single segment three point function of the defect operator with the line endpoints.

It should also be clear that our arguments similarly predict the cusp between several quasi-lightlike lines $a_1,a_2,\ldots$ meeting at the origin, a setup which arises naturally in gauge theories \cite{Korchemsky:1993hr,Korchemskaya:1994qp,Dixon:2016epj,Giombi:2020pdd}. Denoting with $y_{ij}$ their relative boost angles, we claim that when all of these are large we have
\begin{equation}
    \Gamma^L_{a_1,a_2,\ldots}(\{y_{ij}\})=\sum_{i}\Delta_{a_i0}+O\left(\left(\text{max}_{i,j}|y_{ij}|^{-\tau_{ij}}\right)\right)\,,
\end{equation}
where $\tau_{ij}$ is the minimal twist operator exchanged between the two lines. 

Similarly, we can consider a bulk operator in the presence of two or multiple highly boosted defect lines meeting at a cusp. Let us denote with $\log y_{i\mO}$ the rapidity angle between the defect $a_i$ and the geodesic connecting the bulk operator to the cusp. To leading order at large $|y_{i\mO}|$ the bulk operator expectation value is then the sum of the interactions with the individual defects and decays as $\text{max}(|y|_{i\mO})^{-\tau_{\mO}}$.

Another comment is that, in many perturbative examples below, we will observe that the interaction term is captured by replacing the potential~in~\eqref{eq_int_model} \emph{exactly} with a scalar primary two-point function, i.e. taking $\mO^{(a)}=\mO^{(b)}=\mO_{\rm exch}$. In this case the ratio between the logarithmic and the constant term can be calculated: 
\begin{equation}\label{eq_int_scalar}
    \Gamma^L_{ab}(y)-\Gamma^L_{ab}(\infty)
    =-\frac{2^{\tau_{\rm exch} +1}\lambda_{a}\lambda_b}{y^{\tau_{\rm exch}}}\left[\log y -\psi ^{(0)}(\tau_{\rm exch} )-\gamma_E \right]+\ldots\,,
\end{equation}
where the dots stand for terms which decay faster than  $1/y^{\tau_{\rm exch}}$. We do not know if the prediction~\eqref{eq_int_scalar} is a perturbative accident or holds more generally.

Finally we remark that our conclusions do not apply to lines that do not admit endpoint operators connecting them to the trivial line.  Relevant examples include lines charged under one-form symmetries, such as the Wilson line in the fundamental representation of $\mathcal{N}=4$ SYM, and monodromy defects, which are attached to topological surfaces and hence are not genuine line defects. 

The behavior of the cusp anomalous dimension for Wilson lines in gauge theories is well known~\cite{Sudakov:1954sw,Polyakov:1980ca,Korchemsky:1987wg,Gubser:2002tv,Kruczenski:2002fb,Alday:2007mf}. Wilson lines couple to the gauge field, whose twist is zero, leading to the formation of a flux tube in AdS$_3$. The cusp anomalous dimension is therefore extensive in the separation $\Delta\gamma\simeq \log y$ between the lines,
\begin{equation}
    \Gamma^L(y)=T\log y+\ldots\,,
\end{equation}
where $T$ is the string tension. We expect the subleading terms at large $y$ to be related to the spectrum of excitations of the flux tube. We will discuss one such example in sec.~\ref{subsec_N4}.

The behavior $\Gamma^L(y)\sim T\log y+\ldots$ also holds perturbatively for Wilson lines that are not protected by an exact one-form symmetry, with $T\sim g^2$ in terms of the gauge coupling. In this case, however, we expect a transition in the nature of the cusp ground state at nonperturbatively large values of $\log y\sim 1/g^2$. Indeed, when $T\log y\gtrsim 2\Delta_{c0}$, it becomes energetically favorable for the AdS$_3$ flux tube to break by pair creation, and the lowest-energy state crosses over to two isolated neutral partons, weakly interacting with one another as in our description. This is a standard string-breaking transition, analogous for instance to the screening of flux tubes over lengths of order $L\sim m/e^2$ in the two-dimensional massive Schwinger model.\footnote{As usual, in the planar limit string breaking effects are $1/N$ suppressed.} It would be interesting to understand the implications of this phenomenon for the many factorization theorems involving the cusp anomalous dimension of Wilson lines~\cite{Collins:1989gx,Alday:2010zy,Chen:2025ffl,Korchemsky:2019nzm,Moult:2025nhu}. 

Much less is known about cusp anomalous dimensions for monodromy defects or other lines charged under one-form symmetries. In some examples, such as the monopole $U(1)_T$ symmetry in QED$_3$ with matter, monodromy defects can also be viewed as Wilson lines of fractional charge. We therefore expect their cusp anomalous dimension to grow as $\Gamma^L(y)\propto \log y+\ldots$ for $y\gg 1$. In general, the behavior of the Lorentzian cusp anomalous dimension of these lines depends on whether the reduced theory on the $2d$ space \eqref{2dmetric} confines or not.\footnote{A related question arises in the study of CFTs on the pp-wave geometry \cite{Komargodski:2026ain}. We thank Z. Komargodski for useful discussions on this point.}
We leave a more detailed investigation of cusp anomalous dimensions for monodromy defects to future work.

\section{Examples}\label{sec:examples}

\subsection{Free Theories}\label{subsec_free}

In this section we study the Lorentzian cusp in free theories. In subsec.~\ref{SubSec_One_pt} we will study the expectation value of simple bulk operators in the same cusped defect backgrounds. 

As a first example, we consider a theory of $N$ free scalars $\vec{\phi}=(\phi_1,\ldots,\phi_N)^T$ in $d=4$. The pinning field defect, written as
\begin{equation}\label{eq_pinning_field}
\mathcal{D}_{\vec{h}}=e^{-\int d\tau \vec{h}\cdot\vec{\phi}}\,,
\end{equation}
defines a conformal defect for every value of $\vec{h}$. The Euclidean cusp anomalous dimension is given by~\cite{Cuomo:2024psk}
\begin{equation}\label{eq_tree_level}
\Gamma_{\vec h_1\vec h_2}(\theta)=-\frac{\vec{h}_1\cdot\vec{h}_2}{4\pi^2}\frac{\pi-\theta}{\sin\theta}+\frac{\vec{h}_1^2+\vec{h}_2^2}{8\pi^2}\,.
\end{equation}
The Lorentzian cusp is obtained by performing the analytic continuation~\eqref{analyticont}:
\begin{equation}\label{eq_free_pinning}
\begin{split}
\Gamma^L_{\vec h_1\vec h_2}(y) &=
\frac{\vec{h}_1^2+\vec{h}_2^2}{8\pi^2}-\frac{\vec{h}_1\cdot\vec{h}_2}{2\pi^2}
\frac{ y  \log y }{y^2-1} \\
&=
\frac{\vec{h}_1^2+\vec{h}_2^2}{8\pi^2}-\frac{\vec{h}_1\cdot\vec{h}_2}{2\pi^2}\left[
\frac{ \log y }{y }+\frac{ \log y }{y^3}+\ldots\right]\,.
\end{split}
\end{equation}
This result is in perfect agreement with the expectation~\eqref{lorentzcusp}. Indeed the defect creation operator dimension is
\begin{equation}
    \Delta_{\vec{h},0}=\frac{\vec{h}^2}{8\pi^2}\,.
\end{equation}
Additionally, the lowest twist operator is $\phi$ itself, which is a scalar and has $\tau=\Delta=1$. In fact using $\psi^{(0)}(1)=-\gamma_E$, the result~\eqref{eq_free_pinning} can be seen to be in agreement with the simplified form~\eqref{eq_int_scalar}, since to this order the defect interaction potential is proportional to the scalar primary two-point function. Similarly, the terms $\sim y^{-2n-1}$ are due to the exchange of the descendants of the free scalar with twist $\tau=2n+1$. Technically, the suppression of the exchange diagram can easily be seen using the following representative trajectories
\begin{equation}\label{contours}
x^\mu_{1,2}=\frac12\lambda\left(1-\frac{1}{y},\pm\left(1+\frac{1}{y}\right),\vec{0}\right)\,,
\end{equation}
in which case the measure reads $\sqrt{\dot{x}^2}=1/\sqrt{y}$.

Note that~\eqref{eq_free_pinning} satisfies the expected bounds
\begin{equation} 
\Gamma^L_{\vec{h}_1\vec{h}_2}(y)\geq 0\,,\quad
\Re\Gamma^L_{\vec{h}_1\vec{h}_2}(-y-i\epsilon)\geq 0\quad\text{for}\quad y> 0\,.
\end{equation}
As commented before, the bound on $\Re\Gamma^L_{\vec{h}_1\vec{h}_2}(-y-i\epsilon)$ ensures that the probability~\eqref{eq_prob_exclusive} for exclusive production of a heavy pair remains finite as we remove the infrared regulator. It is however easy to see that for $\vec{h}_1\cdot\vec{h}_2<0$ we can have $\Re\Gamma^L_{\vec{h}_1\vec{h}_2}(-y-i\epsilon)-\Delta_{{\vec{h}_1,0}}-\Delta_{{\vec{h}_2,0}}<0$, implying that there exist heavy field form factors similar to~\eqref{eq_Sudakov_final} which diverge for $\omega_{1,2}\rightarrow 0$. Note also that the spacelike cusp is real but the analytic continuation produces a nonzero imaginary part $\Im\Gamma^L_{\vec{h}_1\vec{h}_2}(-y
\pm i\epsilon)=\pm\frac{\vec{h}_1\cdot\vec{h}_2}{2\pi}\frac{y}{y^2-1}$.

Let us now consider a charge $q$ Wilson line in free Maxwell theory
\begin{equation}
W_q=e^{i q \int d x^\mu A_\mu(x(\tau))}\,.
\end{equation}
The Euclidean cusp anomalous dimension is given by~\cite{Polyakov:1980ca}
\begin{equation}\label{eq_tree_level_gauge}
\Gamma_{qq}(\theta)=-\frac{e^2q^2}{4\pi^2}\left(\frac{ \pi-\theta}{\tan\theta}+1\right)\,.
\end{equation}
The analytic continuation to the Lorentzian signature then yields a qualitatively different result compared to the free scalar case:
\begin{equation}\label{eq_Maxwell_Gamma_L}
\begin{split}
\Gamma^L_{qq}(y)&=\frac{e^2 q^2 }{4 \pi ^2}\left(\frac{y^2+1 }{y^2-1} \log y-1\right) \\
&=\frac{e^2 q^2 }{4 \pi ^2}\left(\log y-1+2\frac{\log y}{y^2}+\ldots\right)\,.
\end{split}
\end{equation}
As is well known, the cusp anomalous dimension displays a logarithmic enhancement in the large boost limit. As reviewed before, this can be interpreted as a string in AdS$_3$ with tension $\frac{e^2 q^2}{4\pi^2}$ stretching between the charges. The string is due to the flux of the twist-$0$ gauge field, and cannot break due to the electric one-form symmetry. The subleading $\sim y^{-2n}$ terms are associated with excited states on the string, such as the field strength and derivatives thereof.\footnote{The constant term can be understood as the contribution due to effective mass terms on the string endpoints.} Physically, the emission of photons leads to the well known doubly-logarithmic Sudakov suppression of the exclusive amplitude. Note also that \eqref{eq_Maxwell_Gamma_L} satisfies the conditions \eqref{eq_rindler_gamma_bound} and \eqref{eq_sudakov_gamma_bound}.

\subsection{Pinning Field in the \texorpdfstring{$O(N)$}{O(N)} model}

As a first interacting example, we consider the pinning field defect~\eqref{eq_pinning_field} at the $O(N)$ Wilson--Fisher fixed point in the $\varepsilon$-expansion. As a reminder, the Euclidean bulk action is given by
\begin{equation}\label{eq_BulkAction_epsilon}
S=\int d^dx\left[\frac{1}{2}\pd_\mu\vec{\phi}\cdot\pd^\mu\vec{\phi}+
\mu^{\varepsilon}
\frac{\lambda}{4!}\left(\vec{\phi}\cdot\vec{\phi}\right)^2\right]\,,
\end{equation}
where $\mu$ is the sliding scale. We work in dimensional regularization within the minimal subtraction scheme, where the beta function of the coupling reads \cite{Kleinert:2001ax}
\begin{equation}\label{eq_beta_BULK}
\beta_{\lambda}=
-\varepsilon\lambda+\frac{N+8}{3}\frac{\lambda^2}{(4\pi)^2}
-\frac{3N+14}{3}\frac{\lambda^3}{(4\pi)^4}
+O\left(\frac{\lambda^4}{(4\pi)^6}\right)\,.
\end{equation}
The beta function \eqref{eq_beta_BULK} admits a zero $\beta(\lambda^*)=0$ at the Wilson--Fisher fixed point, for which:
\begin{equation}\label{eq_lambda_fix}
\frac{\lambda^*}{(4\pi)^2}=\frac{3 \varepsilon }{N+8}+
\frac{9 (3 N+14) \varepsilon ^2}{(N+8)^3}
+O\left(\varepsilon^3\right)\,.
\end{equation}
This fixed point describes the long distance behavior of correlation functions and is $O(N)$ invariant and weakly coupled for $\varepsilon \ll 1$.

The magnetic field $\vec{h}$ in the pinning field defect~\eqref{eq_pinning_field} explicitly breaks the $O(N)$ symmetry to $O(N-1)$ for $N>1$ and fully breaks the $\mathbb{Z}_2$ symmetry for $N=1$. The beta-function of the defect coupling $h=|\vec{h}|$ to order $O\left(\lambda^2\right)$ is given by \cite{Allais:2014fqa,Cuomo:2021kfm}:
\begin{equation}\label{eq_beta_h}
\beta_{h^2} =-\varepsilon h^2+\frac{\lambda }{(4\pi)^2}\frac{h^4}{3}+
\frac{\lambda^2}{(4\pi)^4}\left(
\frac{N+2}{18} h^2-\frac{N+8}{18}h^4-\frac{h^6}{6}
\right)
+O\left(\frac{\lambda^3}{(4\pi)^6}\right)\,.
\end{equation}
This beta function admits an infrared fixed point that describes the pinning field DCFT at
\begin{equation}\label{eq_h_fix}
(h^*)^2=(N+8)+
\varepsilon\frac{4 N^2+45 N+170}{2 N+16}+
O\left(\varepsilon^2\right)\,,
\end{equation}
where we used eq.~\eqref{eq_lambda_fix}. Although the defect coupling at the fixed point is not small, it is nonetheless possible to study diagrammatically the defect fixed point for small bulk coupling.

In~\cite{Cuomo:2024psk} the cusp anomalous dimension for the pinning field in the $4-\varepsilon$ dimensional $O(N)$ model was computed to be
\begin{equation}\label{eq_Gamma12_res}
\begin{split}
&\Gamma_{\vec{h}_1^*\vec{h}_2^*}(\theta)
=-\frac{N+8}{4\pi^2}\left(\hat{m}_1\cdot\hat{m}_2\frac{\pi-\theta}{\sin\theta}-1\right)\\
&+\varepsilon\frac{N+8}{4\pi^2}\left[
\left(\hat{m}_1\cdot\hat{m}_2\frac{\pi-\theta}{\sin\theta}-1\right)
\frac{N^2 (\log 64-5)+N (96 \log 2-61)+384 \log 2-234}{2 (N+8)^2}\right.
\\
&\left.\hspace*{4.5em}-
\left.\hat{m}_1\cdot\hat{m}_2\frac{d f_{4-\varepsilon}(\cos\theta)}{d
\varepsilon}\right\vert_{\varepsilon=0}
+\frac{\hat{m}_1\cdot\hat{m}_2}{4\pi^4}I_{11}(\cos\theta)
+\frac{1+2\left(\hat{m}_1\cdot\hat{m}_2\right)^2}{16\pi^4}I_{12}(\cos\theta)\right.\\
&
\left.\hspace*{4.5em}+
\frac{3 \zeta (3)}{2 \pi ^2}+\log 8-1
\right]+O\left(\varepsilon^2\right)\,.
\end{split}
\end{equation}
In the above, we introduced the following quantities
\begin{align}
\label{eq_fd_def}
f_d(\cos\theta)&=\int d\tau 
\frac{e^{-\frac{d-2}{2}|\tau|}}{(1-2\cos\theta e^{-|\tau|}+e^{-2|\tau|})^{\frac{d-2}{2}}}
\stackrel{d=4}{=}\frac{\pi -\theta }{ \sin (\theta )}\,,\\
I_{11}(\hat{n}_1\cdot\hat{n}_2)&=\int d^3\hat{n}\left[
f_4^3(\hat{n}\cdot\hat{n}_1)f_4(\hat{n}\cdot\hat{n}_2)-
\frac{\pi^3 f_4(\hat{n}_1\cdot\hat{n}_2)}{\left(2-2\hat{n}\cdot\hat{n}_1\right)^{\frac{3}{2}}}
\right]\,,\\
\label{eq_I12}
I_{12}(\hat{n}_1\cdot\hat{n}_2)&=\int d^3\hat{n}f_4^2(\hat{n}\cdot\hat{n}_1)f_4^2(\hat{n}\cdot\hat{n}_2)\,.
\end{align}
Both $\left.\frac{d f_{4-\varepsilon}(\cos\theta)}{d\varepsilon}\right\vert_{\varepsilon=0}$ and $I_{11}(\cos\theta)$ can be expressed in closed form as lengthy sums of polylogarithms, but we will not report the explicit expressions. The integral $I_{12}(\cos\theta)$ can be evaluated numerically straightforwardly for arbitrary values of $\theta$.

We are interested in the large boost limit of the Lorentzian cusp. Under the analytic continuation~\eqref{analyticont}, we find
\begin{align}\label{eq_cont_ex1}
  \left.  \frac{d f_{4-\varepsilon}(\cos\theta)}{d
\varepsilon}\right\vert_{\varepsilon=0}&\rightarrow\frac{\log^2 y+\pi^2/6}{ y}+O\left(\frac{\log^2 y}{ y^3}\right) \,,\\
\label{eq_cont_ex2}
I_{11}(\cos\theta)&\rightarrow
-\frac{4 \log  y \left[12 \pi ^2 \zeta (3)+\pi ^4 (\log 64-4)\right]}{ y}+
O\left(\frac{\log y}{ y^3}\right)\,,\\ \label{eq_cont_ex3}
I_{12}(\cos\theta)&\rightarrow \frac{4\pi^4}{3}\frac{\log^2y+O\left(\log y\right)}{y^2}\,.
\end{align}
The analytic continuations of all of the above are straightforward except for $I_{12}$, that we discuss in detail in appendix \ref{app_I12}. Using these, we see that the result for the cusp anomalous dimension is in perfect agreement with our prediction, including the form~\eqref{eq_int_scalar} of the scalar potential:
\begin{equation}\label{eq_pinning_one_loop}
\begin{split}
    \Gamma^L_{\vec{h}_1^*\vec{h}_2^*}(y) &\simeq 2\Delta_{\vec{h}^*0}-\hat{m}_1\cdot\hat{m}_2\frac{\lambda^2_{\rm int}}{y^{\Delta_{\phi}}}\left[\log y-\psi^{(0)}(\Delta_\phi)-\gamma_E\right]\,.
    \end{split}
\end{equation}
The leading term for large $y$ is determined as expected by the dimension of the defect creation operator~\cite{Cuomo:2024psk}:
\begin{equation}\label{eq_Delta_defect_creation}
\Delta_{\vec{h}^*\,0}=
\frac{N+8}{8 \pi ^2}+\varepsilon
\left[\frac{3 (N+8)^2 \zeta (3)+\pi ^2 (3 N^2+29N+106)}{16 \pi ^4 (N+8)}\right]+O\left(\varepsilon^2\right)\,.
\end{equation}
The subleading term in~\eqref{eq_pinning_one_loop} is due to the exchange of the fundamental scalar,
\begin{align}
    \Delta_{\phi}&=1-\frac{\varepsilon}{2}
    +O\left(\varepsilon^2\right)
    \quad\implies\quad 
    \psi^{(0)}(\Delta_\phi)= -\gamma_E-\frac{\pi ^2 \varepsilon }{12}+O\left(\varepsilon^2\right)\,,
\end{align}
and we determined the interaction coefficient:
\begin{equation}
    \lambda_{\rm int}^2=\frac{N+8}{2 \pi ^2}+
    \varepsilon  \left[\frac{3 (N+8) \zeta (3)}{\pi ^4}+\frac{(N-3) N-22}{4 \pi ^2 (N+8)}\right]+O\left(\varepsilon^2\right)\,.
\end{equation}
The corrections from the expansion of $I_{12}(\cos\theta)$ introduce $1/y^2$ corrections, associated with the exchange of double-trace operators of the form $\sim\pd^{n-k}\phi\pd^k\phi$. 

\subsection{Pinning Field in Yukawa CFTs}

A similar example is given by the line defect in the scalar-fermion theories with action \cite{Giombi:2022vnz, Pannell:2023pwz, Barrat:2023ivo} 
\begin{equation}
\label{YukawaCFT}
S=\int d^dx\left[\frac{1}{2}(\partial\vec{\phi})^2+\bar{\Psi}\slashed{\partial}\Psi  +g_1\sum_a \phi_a\bar{\Psi}\Sigma_a\Psi+ \frac{g_2}{24}(\vec{\phi}\cdot\vec{\phi})^2\right] + \int_{\gamma}d\tau \vec{h}\cdot \vec{\phi}
\end{equation}
where $\vec{\phi}$ are $N_s$ real scalars, $\Psi$ is a collection of $N_f$ Dirac fermions, and $\vec{h}$ is an $N_s$-dimensional vector of defect couplings. The cases $N_s=1,2,3$ (with a specific choice of the matrices $\Sigma_a$, $a=1,\ldots,N_s$ in the Yukawa couplings) correspond to the Gross--Neveu--Yukawa (GNY), Nambu--Jona-Lasinio--Yukawa (NJLY) and chiral-Heisenberg models, respectively. These models 
have IR fixed points in $d=4-\varepsilon$ with the fixed point value of the bulk couplings given by
\begin{equation}
\begin{aligned}
&g_{1*}^2=\frac{16\pi^2 \varepsilon}{N+8-2N_s}+O(\varepsilon^2)\\
&g_{2*} = \frac{24 \pi^2 \varepsilon
\Big( 8 - N - 2 N_s + \sqrt{ N^2 + 4 (-4 + N_s)^2 + 4 N (28 + 5 N_s) } \Big)
}
{ \Big( 8 + N - 2 N_s \Big) \Big( 8 + N_s \Big) }+O(\varepsilon^2)
\end{aligned}
\end{equation}
where $N\equiv 4 N_f$. The beta function of the defect coupling is known to 2-loop order \cite{Pannell:2023pwz, Giombi:2025evu}.  Writing $\vec{h}=h \hat{m}$, the corresponding fixed point is given by
\begin{equation}
h_{*}^2=h_{*,0}^2+h_{*,1}^2\varepsilon+O(\varepsilon^2)\,,
\end{equation}
where
\begin{equation}
    h_{*,0}^2=\frac{4(4-N_s)(N_s+8)}{\Big( 8 - N - 2 N_s + \sqrt{ N^2 + 4 (-4 + N_s)^2 + 4 N (28 + 5 N_s) } \Big)}
\end{equation}
and the (rather long) explicit expression for $h_{*,1}^2$ can be found in \cite{Giombi:2025evu}.

The anomalous dimension of the Euclidean cusp in these theories was computed to one-loop order in \cite{Giombi:2025evu}. Using the same notation as in the previous section, it is given by
\begin{equation}\label{gengammaexp}
    \begin{split}
        \Gamma_{\vec{h}_{1*}\vec{h}_{2*}}(\theta) =& -\frac{h_*^2}{4\pi^2}\left(\hat{m}_1\cdot\hat{m}_2\frac{\pi-\theta}{\sin\theta}-1\right) \\& +\frac{h_*^2}{64\pi^4}\left(\hat{m}_1\cdot\hat{m}_2\frac{\pi-\theta}{\sin\theta}-1\right)\left[g_{1*}^2N+\frac{g_{2*}h_*^2}{6}\log(64e^{-1})\right]\\&
-\left(\varepsilon\frac{h_*^2}{4\pi^2}-\frac{g_{1*}^2Nh_*^2}{64\pi^4}\right)\left[\left.\hat{m}_1\cdot\hat{m}_2\frac{d f_{4-\varepsilon}(\cos\theta)}{d
\varepsilon}\right\vert_{\varepsilon=0}-1\right]\\&        +\frac{g_{2*}h_*^4}{768\pi^8}\left[\hat{m}_1\cdot\hat{m}_2\,I_{11}(\cos\theta)-8\pi^4+4\pi^4\log8+24\pi^2\zeta(3)\right]\\&
+\frac{g_{2*}h_*^4}{1024\pi^8}\left[\left(\frac{1+2(\hat{m}_1\cdot\hat{m}_2)^2}{3}\right)I_{12}(\cos\theta)-24\pi^2\zeta(3)\right]+O\left(\varepsilon^2\right)
    \end{split}
\end{equation}

Setting $\theta=\pi-i\log y$ and using the analytic continuations~\eqref{eq_cont_ex1},~\eqref{eq_cont_ex2} and~\eqref{eq_cont_ex3}, we find that the structure of the Lorentzian cusp in the large boost limit is again in agreement with the expectations
\begin{equation}\label{eq_pinning_one_loop_Yukawa}
\begin{split}
    \Gamma^L_{\vec{h}_1^*\vec{h}_2^*}(y) &\simeq 2\Delta_{\vec{h}^*0}-\hat{m}_1\cdot\hat{m}_2\frac{\lambda^2_{\rm int}}{y^{\Delta_{\phi}}}\left[\log y-\psi^{(0)}(\Delta_\phi)-\gamma_E\right]\,,
    \end{split}
\end{equation}
where
\begin{align}
    \Delta_{\phi}&=1-\frac{\varepsilon}{2}+\frac{g_{1*}^2N}{32\pi^2}
    +O\left(\varepsilon^2\right)
    \quad\implies\quad 
    \psi^{(0)}(\Delta_\phi)= -\gamma_E-\frac{\pi ^2 \varepsilon }{12}+\frac{g_{1*}^2N}{192}+O\left(\varepsilon^2\right)\,,
\end{align}
the dimension of the defect creation operator computed in \cite{Giombi:2025evu} is given by
\begin{equation}
\Delta_{\vec{h}_*0}=\frac{h_{*}^2}{8\pi^2}(1+\varepsilon)-\frac{g_{1*}^2Nh_{*,0}^2}{64\pi^4} -\frac{g_{2*}h_{*,0}^4}{256\pi^6}\left[\pi^2-\zeta(3)\right]+O(\varepsilon^2)\,,
\end{equation}
and we find the ``effective coupling constant'' $\lambda^2_{\rm int}$ to be
\begin{equation}
\lambda^2_{\rm int}= \frac{h_{*}^2}{2\pi^2}
-\frac{g_{1*}^2Nh_{*,0}^2}{32\pi^4}-\frac{g_{2*}h_{*,0}^4}{64\pi^6}\left[\pi^2-4\zeta(3)\right]+O(\varepsilon^2)\,.
\end{equation}
Note that the fundamental field admits a nontrivial anomalous dimension, unlike at the Wilson--Fisher fixed point. This contributes nontrivially to the final result~\eqref{eq_pinning_one_loop_Yukawa}.

\subsection{Examples with \texorpdfstring{$J>0$}{J>0}}

In this section, we discuss examples in which the leading operator exchange has nonzero spin $J>0$, so that the large-boost interaction is not accompanied by the logarithmic enhancement characteristic of scalar exchange.

\paragraph{Critical $O(2)$ Wilson--Fisher Model}
Let us consider two orthogonal pinning-field defects \eqref{eq_pinning_field} in the three-dimensional $O(2)$ Wilson--Fisher CFT. At the infrared fixed point, we take their orientations to be:
\begin{equation}
    \vec{h}^*_1 = h_{*}(1,0)~,\quad  \vec{h}^*_2 = h_{*}(0,1)~, \quad \hat{m}_1\cdot \hat{m}_2 =0~.
\end{equation}
For a straight pinning field defect, $O(2)$ covariance fixes the one-point function of the fundamental field to have the form
\begin{equation}
    \langle\phi_i(x)\rangle_{\mathcal{D}} \propto \frac{\hat{m}_i}{|x_\perp|^{\Delta_\phi}}~.
\end{equation}
The coefficient for exchanging the fundamental scalar between the two pinning field defects is therefore proportional to\footnote{More precisely, for a field to be exchanged between the defects its one-point functions with an open defect segment should be nonzero. This distinction is irrelevant in the present example.} 
\begin{equation}
    \sum_i \hat{m}_{1i}\hat{m}_{2i} = \hat{m}_1\cdot\hat{m}_2 =0~.
\end{equation}

Orthogonality does not, however, project out all the scalar operators. For example the lowest scalar in the symmetric-traceless representation is allowed as well as the lowest $O(2)$-singlet scalar. They can, however, be ruled out as the leading exchange using numerical bootstrap estimates for the relevant low-lying scalar dimensions \cite{Chester:2019ifh}
\begin{equation}
\Delta_\phi=0.519088(22)\,,
\qquad
\Delta_{\rm charge\text{-}2}=1.23629(11)\,,
\qquad
\Delta_{\rm singlet}=1.51136(22)\,.
\end{equation}
By contrast, the stress tensor $T_{\mu\nu}$ has the exact quantum numbers $\Delta_T =3, J_{T}=2$ and twist $\tau_T = 1$. The fundamental scalar has smaller twist but has been removed by the orthogonality condition, while the surviving scalar primaries displayed above have twist strictly larger than one. Similarly, all multi-trace operators, including the charge $2$ ones in the $\phi-\phi$ OPE and the neutral ones in the stress-tensor family, have $\tau>1$ by the unitarity bounds---see \cite{Liu:2020tpf} for explicit results.

The $O(2)$ current $J_{\mu}$ is another operator of twist one, but it does not couple to the defect identity channel. For example, the defect oriented along $(1,0)$ preserves the reflection $\phi_2\to-\phi_2$, under which $J_\mu \longrightarrow -J_\mu$. Its defect one-point function therefore vanishes.

To conclude, the stress tensor is an $O(2)$ singlet and is not removed by orthogonality. Since its one-point function in the presence of the defect is perturbatively nonzero near four dimensions \cite{Bianchi:2022sbz}, we expect its coupling to the pinning defect to be nonzero; assuming that this coefficient does not vanish accidentally in three dimensions, the stress tensor is therefore the leading operator exchanged. Applying the large-distance exchange ansatz of sec.~\ref{sec_Large_boost}, we predict
\begin{equation}\label{eq_pinning_m1m20}
    \Gamma^L_{\vec h_1^*\vec h_2^*}(y)
=
2\Delta_{\vec h^*0}
+
\frac{\#}{y}
+
\ldots\,,
\qquad
\hat m_1\cdot\hat m_2=0~.
\end{equation}
By similar arguments, we also expect~\eqref{eq_pinning_m1m20} to hold in the $O(3)$ model (see~\cite{Chester:2020iyt} for high-precision bootstrap results on its conformal spectrum).

It is important to emphasize that this prediction cannot be checked directly using the available \(\varepsilon\)-expansion result. For orthogonal orientations, the remaining angle dependence in eq.~\eqref{eq_Gamma12_res} is contained in \(I_{12}(\cos \theta)\). In perturbation theory around \(d=4\), this term receives contributions from an infinite tower of bilinear operators that are degenerate at twist two in the free-theory limit, including the stress tensor. At the perturbative order currently available, their individual conformal-family contributions cannot be disentangled, and our prediction based on a single or finite number of exchanged states does not apply.

\paragraph{Two-state Impurity Coupled to a Free Dirac Fermion}

As another example, consider a free Dirac fermion in $d=3$ Euclidean dimensions
\begin{equation}
    S_{\mathrm{bulk}}
    =
    \int \dd^3x\,\bar\psi\slashed{\partial}\psi~.
    \end{equation} 
We can define a classically conformal defect as follows. Let \(x^\mu(s)\) be an oriented curve, let
\(\hat v^\mu=\dot x^\mu/|\dot x|\) be its unit tangent, and choose a normalized spinor \(u(\hat v)\) satisfying
\begin{equation}
    \slashed{\hat v}\,u(\hat v)=u(\hat v)\,,
    \qquad
    \bar u(\hat v)\slashed{\hat v}=\bar u(\hat v)\,,
    \qquad
    \bar u(\hat v)u(\hat v)=1\,.
    \label{eq:polarization}
\end{equation}
We then define the following matrix-valued connection
\begin{equation}
    \mathbb L(s)
    =
    g|\dot x(s)|
    \begin{pmatrix}
        0 & \bar u(\hat v)\psi(x(s))\\
        \bar\psi(x(s))u(\hat v) & 0
    \end{pmatrix}\,,
    \label{eq:matrix-connection}
\end{equation}
from which we obtain the following line defect
\begin{equation}
   \mathcal{D}=\text{Tr}\left[
    P\exp\left(\int_{\mathcal{C}}\dd s\,\mathbb L(s)\right)\right]\,.
    \label{eq:defect}
\end{equation}
Physically,~\eqref{eq:defect} describes a localized two-state impurity interacting with a free fermion. This is equivalent to the pseudogap resonant-level model studied in~\cite{FritzVojta:2004}.

Note that when expanding~\eqref{eq:defect} only quadratic terms in $g$ survive the trace. For a straight or circular line, these can all be simplified using $u\bar{u}=(\mathbf{1}_2+\slashed{v})/2$. Therefore, the defect classically preserves
\begin{equation}
    SO^+(2,1)_{\parallel}
    \times
    \left[
        \frac{
            \operatorname{Spin}(2)_{\perp}\times U(1)
        }{
            \mathbb Z_2^{\,F}
        }
        \rtimes
        \mathbb Z_2^{\,C\mathcal R_\perp}
    \right],
\end{equation}
where the $U(1)$ acts as $\psi\rightarrow e^{i\alpha}\psi$ and the $\mathbb Z_2^{\,F}$ quotient identifies the central
elements of the internal and the transverse $\operatorname{Spin}(2)_{\perp}$ symmetries. Note also that the defect breaks transverse reflections $\mathcal R_\perp$ and time reversal $\mathcal T$ since we specified a polarization in~\eqref{eq:polarization}. It does, however, preserve the discrete $C\mathcal R_\perp$ symmetry, where $C$ stands for charge conjugation; importantly, the $U(1)\rtimes \mathbb{Z}_2^{\,C\mathcal R_\perp}$ symmetry prevents a non-diagonal mass renormalization $\propto m_i \sigma^i$, that would break the degeneracy between the two qubit states. At the quantum level the coupling becomes marginally irrelevant, and it admits a fixed point in $d=3-\eps$ dimensions \cite{FritzVojta:2004}:
\begin{equation}\label{eq:beta_fer}
    \beta_g=-\frac{\eps}{2}g+\frac{g^3}{4\pi}+O\left(g^5\right)\,.
 \end{equation}

Let us now consider the cusp anomalous dimension. We work at order $O(g^2)$, for which the quantum RG is negligible and we can treat the defect as conformal already in $d=3$. The tangents of the lines meeting at the origin are \(v_1=-\hat{n}_1\) and \(v_2=\hat{n}_2\) such that the cusp angle is $\hat{n}_1\cdot \hat{n}_2=\cos\theta$. Given polarization vectors satisfying 
\begin{equation}
    \slashed v_i u_i=u_i,
    \qquad
    \bar u_i\slashed v_i=\bar u_i\,,
\end{equation}
we can choose their relative phase so that
\begin{equation}
    \bar u_2u_1=\sqrt{|\bar u_2u_1|^2}=\sqrt{\text{Tr}\left[\left(\frac{\mathbf{1}_2+\slashed{v}_2}{2}\right)\left(\frac{\mathbf{1}_2+\slashed{v}_1}{2}\right)\right]}
    % =
    % \cos\left(\frac{\pi-\theta}{2}\right)
    =
    \sin\frac{\theta}{2}\,.
    \label{eq:spinor-overlap}
\end{equation}

The only angle-dependent contribution to the cusp is due to a fermion propagator connecting the two lines. Parametrizing the contours as
\begin{equation}
    x_1=-s v_1\,
    \qquad
    x_2=t v_2\,,
    \qquad
    s,t>0\,,
\end{equation}
and using the free fermion propagator
\begin{equation}
    \langle\psi(x)\bar\psi(0)\rangle
    =
   \frac{\slashed{x}}{4\pi|x|^3}\,,
    \end{equation}
the relevant correlator is
\begin{align}
    \left\langle\mathbb L_2(t)\mathbb L_1(-s)\right\rangle
    &=
    \frac{g^2}
    {4\pi\left(s^2+t^2-2st\cos\theta\right)^{3/2}}
    \begin{pmatrix}
        \bar u_2\left(t\slashed v_2+s\slashed v_1\right)u_1 & 0\\
        0 & \bar u_1\left(t\slashed v_2+s\slashed v_1\right)u_2
    \end{pmatrix}
    \nonumber\\
    &=
    \frac{g^2}{4\pi}\sin\frac{\theta}{2}
    \frac{s+t}{
        \left(s^2+t^2-2st\cos\theta\right)^{3/2}
    }
    \,\mathbf 1_2 \,.
    \label{eq:cross-contraction}
\end{align}
The two diagonal entries agree after choosing the overlap
\eqref{eq:spinor-overlap} to be real.\footnote{For a different choice of the phase, the cusp ground state is obtained inserting a diagonal matrix that cancels the phase difference.} Setting $t=s u$ and isolating the logarithmic divergence $\int ds/s$, the relevant integral is
\begin{equation}
    \int_0^\infty\dd u\,
    \frac{1+u}{
        \left(1+u^2-2u\cos\theta\right)^{3/2} }
    =
    \csc^2\frac{\theta}{2}\,.
\end{equation}
Demanding that the cusp anomalous dimension vanishes at $\theta=\pi$ to fix the angle-independent contribution, we arrive at
\begin{equation}
    \Gamma(\theta)
    =
    \frac{g^2}{4\pi}
    \left(1-
        \csc\frac{\theta}{2}
    \right)
    +O(g^4)\,.
    \label{eq:cusp-result}
\end{equation}
Since the angle-independent term is twice the self-energy diagram for a defect with an endpoint, we also obtain the defect creation operator dimension:
\begin{equation}
 \Delta_{\mathcal D\mathbf 1}= \frac{g^2}{8\pi} +O(g^4)\,.
\end{equation}

The result~\eqref{eq:cusp-result} is negative and concave for $\theta\in (0,\pi]$, in agreement with the general expectation~\cite{Cuomo:2024psk}. Setting $\theta=\pi-i\log y$ we obtain the Lorentzian cusp
\begin{equation}
\begin{split}\label{eq_GammaL_fermion}
    \Gamma^L(y) &=\frac{g^2 \left(\sqrt{y}-1\right)^2}{4 \pi  (y+1)} \\
    &=\frac{g^2}{4 \pi }-\frac{g^2 }{2 \pi \sqrt{y}}+\frac{g^2 }{2 \pi \,y^{3/2}}+O\left(\frac{1}{y^{5/2}}\right)\,.
    \end{split}
\end{equation}
The large boost expansion agrees with our expectation. In particular, the leading interaction term decays as $1/\sqrt{y}$ due to the propagation of the bulk fermion which has twist $1/2$.\footnote{Note that the bulk fermion has a non-zero matrix element with an open defect segment with oppositely charged endpoints.} Note also that~\eqref{eq_GammaL_fermion} satisfies both $\Gamma^L(y)\geq 0 $ and $\text{Re}\,\Gamma^L(-y-i\epsilon)\geq 0$ for $y>0$.

\subsection{\texorpdfstring{$\mathcal{N}=4$}{N=4} SYM}\label{subsec_N4}

In this section, we focus on cusp configurations formed by 1/2 BPS Wilson lines in the fundamental representation of $SU(N)$ in $\mathcal{N}=4$ super Yang-Mills theory (SYM). The cusp anomalous dimension associated with these Wilson lines is one of the most extensively studied observables in the literature (see e.g. \cite{Drukker:1999zq,Makeenko:2006ds,Drukker:2007qr,Drukker:2011za,Correa:2012at,Correa:2012hh,Gromov:2015dfa,Grozin:2015kna,Henn:2019swt}). In the large boost limit, it is well known that this cusp anomalous dimension grows logarithmically $\sim \log y$. Here we review these results in the planar limit, both at weak and at strong coupling, with particular focus on the subleading corrections at large $y$. 

Let us first recall the definition of 1/2 BPS Wilson lines which is given by
\begin{equation}
    W_F = \mathrm{Tr}_F\left[\mathrm{P}\,\mathrm{exp}\int d\tau \left(i\dot{x}^\mu A_\mu + \sqrt{-\dot{x}^2}\, \zeta^I\Phi_I \right)\right]~,
\end{equation}
where $\zeta^I$ is a constant unit $SO(6)_R$ vector such that $\zeta_I\zeta^I = 1$ and $\Phi_I$ is the bottom component of the $\mathcal{N}=4$ vector multiplet. The choice of a fixed vector $\zeta^I$ singles out one direction in the $SO(6)_R$ space; therefore the insertion of the Wilson line breaks $SO(6)_R$ down to the subgroup $SO(5)$ that preserves $\zeta^I$. We are interested in a cusp formed by two $1/2$-BPS Wilson lines meeting at a geometric angle $\theta$. More generally, one can consider the generalized cusp anomalous dimension $\Gamma(\theta,\phi)$, where $\phi$ is the angle between the two scalar couplings $\zeta_1^I$ and $\zeta_2^I$.\footnote{In much of the $\mathcal{N}=4$ SYM literature, the standard convention is to denote by $\pi - \phi$ the geometric angle between the two lines and by $\theta$ the angle between the scalar couplings. Here we adopt the opposite convention and hope that this choice of notation will not lead to confusion.} In this work, for simplicity, we restrict our attention to a cusp between two identical Wilson lines, so we will not discuss the dependence on the internal angle $\phi$ any further.

The leading-order result for the Euclidean cusp anomalous dimension $\Gamma(\theta)$ at weak 't Hooft coupling $\lambda$ is given by
\begin{equation}
    \Gamma(\theta) = -\frac{\lambda}{8\pi^2}\frac{\pi-\theta}{\sin \theta}(1+\cos \theta ) + O\left(\frac{\lambda^2}{(4\pi)^4}\right)~.
\end{equation}
As in our previous examples, setting $\theta = \pi -i\log y$ and sending $y \to \infty$ we find:
\begin{equation}
    \Gamma^L(y) = \frac{\lambda}{8\pi^2}\left[\log y  - 2\frac{\log y}{y} +\frac{2 \log y}{y^2} + O\left(\frac{\log y}{y^3}\right)\right]~. 
\end{equation}
Similarly to Maxwell theory, the cusp is extensive in $\log y$. The main difference compared to pure gauge theory is that exchanges of the scalar state $\Phi^I$, which can be thought of as an excitation of the AdS flux tube created by the line, lead to corrections proportional to odd powers of $1/y$.

Let us now consider the calculation at strong coupling. For this we follow the analysis in \cite{Kruczenski:2002fb}. The Lorentzian cusp at strong 't Hooft coupling is obtained by performing the following integral:
\begin{equation}\label{cuspstrongeuclidean}
    \Gamma^L(y) = \frac{\sqrt{\lambda}}{2\pi} \int_{-\infty}^{\infty}du\left[\sqrt{\frac{1+u^2-f^2_0}{1+u^2-2f_0^2}} -1 \right]~,
\end{equation}
where the cusp angle is related to $f_0$ via
\begin{equation}\label{y_strongcoupling}
   -\log y= \mathrm{P.V.} \int_{-\infty}^{\infty} du \frac{f_0\sqrt{1-f^2_0}}{(u^2-f^2_0)\sqrt{(1+u^2-f^2_0)(1+u^2-2f^2_0)}}\,.
\end{equation}
To perform the integral~\eqref{y_strongcoupling} it is convenient to use invariance under $u$ reflections of the integrand to reduce the contour to $u>0$ and isolate the non-integrable divergence at $u=f_0$ writing
\begin{equation}\label{eq_y_strong2}
    -\log y=\mathrm{P.V.} \int_{0}^{\infty} du \frac{2 f_0^2}{(u-f_0) \left(f_0^2+u^2\right)}
    +\int_0^{\infty}du g(f_0,u)\,,
\end{equation}
where the second integral does not require regularization and its integrand reads
\begin{equation}
    g(f_0,u)=\frac{2f_0\sqrt{1-f^2_0}}{(u^2-f^2_0)\sqrt{(1+u^2-f^2_0)(1+u^2-2f^2_0)}}-\frac{2 f_0^2}{(u-f_0) \left(f_0^2+u^2\right)}\,.
\end{equation}
The first integral in~\eqref{eq_y_strong2} can be computed explicitly and is independent of $f_0$
\begin{equation}
    \mathrm{P.V.} \int_{0}^{\infty} du \frac{2 f_0^2}{(u-f_0) \left(f_0^2+u^2\right)}=-\frac{\pi}{2}\,.
\end{equation}
To evaluate the second integral in~\eqref{eq_y_strong2} we set
\begin{equation}
    f_0=\sqrt{\frac{1-\delta^2}{2}}\,,\quad
    0<\delta\ll 1\,.
\end{equation}
The integral may then be evaluated using the method of matched asymptotic expansions.\footnote{The same method is adopted in appendix \ref{app_I12} to analyze the integral $I_{12}$.} Namely we split the integral in two contours as $(0,\infty)=(0,c]\cup(c,\infty)$, with $\delta\ll c\ll 1$ and set $u=\delta\bar{u}$ in the first interval $(0,c]$. The two integrals are then evaluated by expanding the integrands at small $\delta$. The dependence on the arbitrary regulator scale $c$ cancels when taking the sum and we arrive at
\begin{equation}
    \log y=-\left[2 \sqrt{2} \log (\delta )+2\, \text{arccoth}\left(\sqrt{2}\right)-3 \sqrt{2} \log (2)\right]+\frac{\delta ^4 \left[4 \log (\delta )-6-\log (64)\right]}{16 \sqrt{2}}+O\left(\delta^8\right)\,.
\end{equation}
This equation can be inverted as
\begin{equation}
    \delta=
    \frac{2 \sqrt{2}}{x}-\frac{4 \sqrt{2} (2 \log x+3)}{x^5}+O\left(\frac{1}{x^9}\right)\,,\qquad
    x=\left(2 \sqrt{2}+3\right)^{\frac{1}{2 \sqrt{2}}} y^{\frac{1}{2 \sqrt{2}}}\,.
\end{equation}
The integral giving the cusp is performed similarly and yields 
\begin{equation}
\begin{split}
    \Gamma^L(y) &= \sqrt{\lambda}\left[\frac{\log x-1}{\sqrt{2} \pi }+
    \frac{4 \sqrt{2}}{\pi  x^4}+O\left(\frac{1}{x^8}\right)
    \right]\\
    &=\frac{\sqrt{\lambda}}{2\pi}\left[\frac{1}{2}\log y+\frac{1}{2} \log \left(2 \sqrt{2}+3\right)-\sqrt{2}
    +8 \sqrt{2} \left(2 \sqrt{2}+3\right)^{-\sqrt{2}} y^{-\sqrt{2}}
    +O\left(y^{-2\sqrt{2}}\right)
    \right]\,.
    \end{split}
\end{equation}
It is interesting to interpret the power corrections to the leading $\log y$ contribution in terms of the spectrum of excitations above the long GKP string~\cite{Gubser:2002tv}. At strong coupling this spectrum contains a bosonic mode of mass $m=\sqrt{2}$ \cite{Basso:2010in}. Since the large $y$ limit plays the role of a long propagation time on the GKP worldsheet, the exchange of an excitation of energy $E$ is expected to produce a correction
of order $y^{-E}$. The first correction in our result is therefore naturally identified with the propagation of the lightest massive mode, $y^{-\sqrt{2}}$.

\subsection{\texorpdfstring{$2d$}{2d} CFTs}

It is instructive to contrast the $d>2$ large boost expansion with the special case of two-dimensional CFTs. Consider two interfaces $a$ and $b$, which factorize into a product of boundary conditions. The cusp anomalous dimension between these interfaces is given by~\cite{Cardy:1988tk,Cuomo:2024psk}
\begin{equation}\label{eq_cusp_2d}
\Gamma_{ab}(\theta)=\frac{2 \pi ^2 \Delta_{|a\rangle |b\rangle }}{(2 \pi -\theta ) \theta }-\frac{c (\pi -\theta )^2}{12 (2 \pi -\theta ) \theta }\,,
\end{equation}
where $\Delta_{|a\rangle |b\rangle }$ is the dimension of the interface changing operator. Using the analytic continuation adopted throughout this paper, we therefore obtain
\begin{equation}\label{eq_cusp_2d_Lor}
\begin{split}
    \Gamma_{ab}^L(y) &=\frac{c \log ^2y +24 \pi ^2 \Delta_{|a\rangle |b\rangle } }{12 \left(\log ^2y +\pi ^2\right)} \\
    &=\frac{c}{12}+\frac{24 \pi ^2 \Delta_{|a\rangle |b\rangle } -\pi ^2 c}{12 \log^2y}+O\left(\frac{1}{\log^4y}\right)\,.
    \end{split}
\end{equation}
This behavior is qualitatively different from the higher-dimensional examples. In $d>2$, a finite twist gap leads to power corrections controlled by the lowest-twist exchanged operators. In two dimensions, however, the vacuum module contains infinitely many twist-zero Virasoro descendants. These states are not separated by a twist gap, and their contributions must be resummed. The
result of this resummation is not a power series in inverse powers of $y$, but rather an expansion in inverse powers of $\log y$. The leading term is universal and depends only on the central charge $c$. The dependence on the interface-changing operator dimension $\Delta_{|a\rangle|b\rangle}$ first appears at order \(1/\log^2 y\). Thus the large boost limit washes out the boundary-condition data at leading order, leaving only the universal Virasoro vacuum contribution. In the future, it would be interesting to understand the expansion in $1/\log y$ systematically.

Note finally that \eqref{eq_cusp_2d_Lor} satisfies the two positivity conditions \eqref{eq_rindler_gamma_bound} and \eqref{eq_sudakov_gamma_bound}. The latter in particular follows from
\begin{equation}
\operatorname{Re}\Gamma_{ab}^L(-y-i\epsilon)
=
\frac{
c\left(\log^2 y+3\pi^2\right)
+24\pi^2\Delta_{|a\rangle |b\rangle }
}{
12\left(\log^2 y+4\pi^2\right)
}\geq 0\quad \text{for}\quad y>0\,.
\end{equation}

\subsection{Bulk Operator Expectation Values in Free Theories}\label{SubSec_One_pt}

In this section we study the expectation value of bulk operators in the free theories studied in sec.~\ref{subsec_free}. For concreteness, we focus on the Sudakov configuration that is obtained by sending $y\rightarrow -y$ in~\eqref{contours} and on time ordered expectation values in the forward light cone of the cusp:
\begin{equation}
    t>\sqrt{x^2+\vec{x}_{\perp}^2}\,,
\end{equation}
where $\vec{x}_{\perp}$ denotes the coordinates transverse to the plane spanned by the defects.

In the free scalar theory, the time-ordered one-point function follows from
\begin{equation}
  \vec{\phi}_F(t,\vec{x})
\equiv
\frac{
\langle0|T\{\vec{\phi}(t,\vec{x})\mathcal D\}|0\rangle
}{
\langle0|\mathcal D|0\rangle
}
=
\int d^4x'\,G_F(x-x')
\sum_{i=1,2}\vec{h}_{i}\int d\lambda\sqrt{-\dot{x}_i^2}\,\delta^{(4)}\left(x'-x_i(\lambda)\right)\,,
\end{equation}
where $\mathcal{D}$ denotes the cusped defect and the Feynman propagator is
\begin{equation}
    G_F(x)=\frac{i}{4\pi^2[t^2-\vec x^{\,2}-i\epsilon]}\,.
\end{equation}
In the region of interest, this gives
\begin{equation}
\vec\phi_F(t,\vec{x})=
-\sum_{i=1,2}
\frac{\vec h_i}{4\pi R_i}
\left[
1+\frac{i}{\pi}
\log\left(
\frac{V_i+R_i}{\sqrt{t^2-\vec{x}^{2}}}
\right)
\right]\,
\end{equation}
where in terms of $\eta=\log y$ we defined
\begin{equation}
R_{1,2}^2
=
\left(
x\cosh\frac{\eta}{2}
\mp \,t\sinh\frac{\eta}{2}
\right)^2
+
\vec x_\perp^{\,2}\,,\quad
V_{1,2}
=
t\cosh\frac{\eta}{2}
\mp x\sinh\frac{\eta}{2}\,;
\end{equation}

We now focus for simplicity on the plane $\vec{x}_\perp=0$ and parametrize
\begin{equation}\label{eq_1pt_coords}
    t=L\cosh\bar\eta\,,\quad
    x=L\sinh\bar\eta\,,\quad
    -\frac{\eta}{2}<\bar\eta<\frac{\eta}{2}\,.
\end{equation}
We then obtain
\begin{equation}
    \vec{\phi}_F=-\frac{\vec{h}_1}{4\pi L\sinh\left(\frac{\eta}{2}-\bar{\eta}\right)}\left[1+
    \frac{i(\eta-2\bar\eta)}{2\pi}\right]
    -\frac{\vec{h}_2}{4\pi L\sinh\left(\frac{\eta}{2}+\bar{\eta}\right)}\left[1+
    \frac{i(\eta+2\bar\eta)}{2\pi}\right]\,,
\end{equation}
which decays exponentially with the rapidity difference in agreement with~\eqref{eq_1pt}. Similarly, choosing for instance $1\ll\eta-2\bar{\eta}\ll \eta+2\bar{\eta}$, the expectation value of the energy density measured by the improved stress tensor,
\begin{equation}
T_{00}
=
\frac12\dot{\vec\phi}^{\,2}
+
\frac16\left(\nabla\vec\phi\right)^2
-
\frac13\vec\phi\cdot\ddot{\vec\phi}\,,
\end{equation}
receives a contribution from a single defect and simplifies to
\begin{equation}
\frac{\left\langle T\{T_{00}(t,x,\vec{0}_\perp)\mathcal D\}\right\rangle}{\left\langle\mathcal{D}\right\rangle}
\simeq
\frac{\vec{h}_1^2e^{-2\left(\frac{\eta}{2}-\bar{\eta}\right)}}{24\pi^4L^4}\left[
\cosh(2\bar{\eta})\left(\eta-2\bar{\eta}-2i\pi-2
\right)+1\right]
\, ,
\end{equation}
which decays exponentially with the rapidity difference $\frac{\eta}{2}-\bar{\eta}$ as expected. The factor $\cosh(2\bar{\eta})\sim e^{2\bar{\eta}}$ is also explained by the change of coordinates to the AdS$_3\times S^1$ frame, which schematically gives $T_{\text{flat}}\sim e^{2\gamma}T_{\text{AdS}_3\times S^1}$.
Note that at $x=0\iff \bar{\eta}=0$ the stress tensor is instead equidistant in rapidity from the lines, and thus receives contributions from both defects. Explicitly
\begin{equation}
\begin{split}
\frac{\left\langle T\{T_{00}(t,\vec{0})\mathcal D\}\right\rangle}{\left\langle\mathcal{D}\right\rangle}
&\simeq
\frac{\vec{h}_1^2\,e^{-\eta}}{24\pi^4 t^4}\left(\eta-1-2i\pi\right)
+\frac{\vec{h}_2^2\,e^{-\eta}}{24\pi^4 t^4}\left(\eta-1-2i\pi\right)\\
&+
\frac{\vec{h}_1\cdot\vec{h}_2\,e^{-\eta}}{24\pi^4 t^4}\left[\eta ^2-(2+4 i \pi ) \eta -4 \pi ^2+4 i \pi +2\right]
\, .
\end{split}
\end{equation}

Let us contrast the above expressions with Maxwell theory. Working in Feynman gauge, the field sourced by Sudakov Wilson lines is 
\begin{equation}
A_F^\mu(x)
=
e^2 q\sum_{i=1,2} \frac{\epsilon_i u_i^\mu}{4\pi R_i}
\left[
1+\frac{i}{\pi}
\log\left(\frac{V_i+R_i}{\sqrt{t^2-\vec{x}^{\,2}}}\right)
\right]\,,
\end{equation}
where 
\begin{equation}
    \epsilon_{1,2}=\pm 1\,,\qquad
    u_{1,2}^\mu=\left(\cosh\frac\eta 2,\pm\sinh\frac\eta 2,\vec{0}\right)\,.
\end{equation}
In this case however the expectation value does not decay exponentially, as can be seen by considering the electric field
\begin{equation}
\frac{\left\langle T\{F_{0x}(t,\vec{x})\mathcal D\}\right\rangle}{\left\langle\mathcal{D}\right\rangle}\simeq\frac{i e^2 q}{2 \pi ^2 \left(t^2-\vec{x}^2\right)}\,,
\end{equation}
where we neglected terms that decay exponentially for $\eta=\log y\rightarrow\infty$. Note that the electric field is complex and, correspondingly, the energy density one-point function is negative.

Let us finally comment that standard physical measurements are related to in-in correlators, namely
\begin{equation}
\langle \mO(x)\rangle_{\mathcal D}=
\frac{\langle 0|\mathcal D^\dagger \mO(x)\mathcal D|0\rangle}
{\langle 0|\mathcal D^\dagger\mathcal D|0\rangle}\,,
\end{equation}
that cannot be obtained as analytic continuation of time-ordered correlators in general. On physical grounds, one might also expect these to behave similarly to the above correlators. In practice, we show below that these sometimes decay faster with the boost parameter than the Feynman correlators. 

In the present free examples, the in-in classical one-point functions are obtained by replacing the Feynman propagator with the retarded Green's function
\begin{equation}
G_R(x)
=
-\frac{\Theta(t)}{4\pi |\vec x|}
\delta(t-|\vec x|)=-\frac{\Theta(t)}{2\pi }
\delta\left(t^2-\vec x^{\,2}\right)\,.
\end{equation}
In the scalar theory, we obtain
\begin{equation}
\vec\phi_R(t,\vec x)
=
-\frac{\Theta(t^2-\vec x^{\,2})}{4\pi}
\sum_{i=1,2}
\frac{\vec h_i}{R_i}\stackrel{\vec{x}_{\perp}=0}{=}
-\frac{\Theta(t^2-\vec x^{\,2})}{4\pi\,L}\left[\frac{\vec{h}_1}{\sinh\left(\frac\eta 2-\bar{\eta}\right)}+
\frac{\vec{h}_2}{\sinh\left(\frac\eta 2+\bar{\eta}\right)}\right]\,.
\end{equation}
This decays exponentially as for the time-ordered prescription.  Using the improved stress tensor, in the regime \(1\ll\eta-2\bar{\eta}\ll \eta+2\bar{\eta}\), where the first defect dominates, we obtain
\begin{equation}
\langle T_{00}(t,x,\vec{0}_\perp)\rangle_{\mathcal D}
\simeq
\frac{\vec h_1^{\,2}}{96\pi^2L^4}
\csch^4\left(\frac{\eta}{2}-\bar\eta\right)
\simeq
\frac{\vec h_1^{\,2}}{6\pi^2L^4}
e^{-4\left(\frac\eta2-\bar\eta\right)} .
\end{equation}
Note that this decays faster than the previously obtained time-ordered expectation value. At the symmetric point \(\bar\eta=0\), instead,
\begin{equation}
\begin{split}
\langle T_{00}(t,\vec{0})\rangle_{\mathcal D}
&=
\frac{1}{96\pi^2t^4\sinh^4\frac\eta 2}
\left[\vec h_1^{\,2}+\vec h_2^{\,2}-2\vec{h}_1\cdot\vec{h}_2\cosh\eta\right]
\simeq
-\frac{\vec h_1\cdot\vec h_2}{6\pi^2t^4}e^{-\eta}
\,.
\end{split}
\end{equation}

In Maxwell theory the retarded gauge field is
\begin{equation}
    A_R^\mu=-\frac{e^2q}{4\pi}\sum_i \frac{\epsilon_i u_i^\mu}{R_i}=
-\frac{e^2q}{2\pi}\partial^\mu \bar{\eta}
+O(e^{-\eta})\,,
\end{equation}
which becomes pure gauge for large boosts as is well known~\cite{Collins:1989gx}. As a result the physically measurable electric field vanishes as $\eta\rightarrow\infty$,
\begin{equation}
    \langle F_{\mu\nu}(t,\vec{x})\rangle_{\mathcal{D}}=O(e^{-\eta})\,,
\end{equation}
in contrast to the time-ordered expectation value.

\section{The Spin Impurity Cusp}\label{sec_spin_imp}

In this section we consider spin impurity defects both in free theories and in the $O(3)$ CFT. These are defined by~\cite{PhysRevB.61.4041,sachdev1999quantum,vojta2000quantum,Liu:2021nck}:
\begin{equation}\label{eq_spin_imp_def}
    \mathcal{D}_s=\text{Tr}\left[P e^{\gamma\int d\tau \phi_a T^a}\right]\,,
\end{equation}
where the matrices $T_a$ furnish a spin $s$ irrep. of $SU(2)$. Physically,~\eqref{eq_spin_imp_def} represents a localized spin $s$ degree of freedom coupled to the bulk order parameter. 

For a free bulk, the spin impurity is equivalent to the non-local Bose-Kondo model~\cite{PhysRevB.61.4041,Vojta_review}. The coupling $\gamma$ admits an interesting phase diagram, displaying two fixed points for $d$ sufficiently near $4$ dimensions~\cite{Cuomo:2022xgw,Beccaria:2022bcr,Nahum:2022fqw,Weber:2022ada}. One fixed point is always perturbatively accessible in the $\varepsilon$-expansion for every fixed $s$. Additionally the full phase diagram can be studied perturbatively in a double-scaling limit $\varepsilon\rightarrow 0$ with $\varepsilon s=\text{fixed}$~\cite{Cuomo:2022xgw,Beccaria:2022bcr,Nahum:2022fqw}.

In the interacting $O(3)$ model, both perturbative arguments and numerical results~\cite{PhysRevLett.98.087203,PhysRevLett.99.027205}, including most notably recent fuzzy-sphere calculations~\cite{Sarma:2026lte}, provide strong evidence that spin impurities flow to nontrivial fixed points.\footnote{For $s=1/2$ the existence of a nontrivial fixed point is enforced by the $g$-theorem~\cite{Cuomo:2021rkm} and anomalies~\cite{Komargodski:2025jbu}.}  In particular, each $s$ yields a different DCFT. These fixed points can be studied perturbatively in the $\varepsilon$-expansion. Additionally, \cite{Cuomo:2022xgw} argued that, for every $d$, the $s\rightarrow \infty $ limit of the spin impurity can be related to a pinning field defect (up to global zero modes), with calculable corrections in $1/s$. 

Below we will study the cusp between two spin impurities, in free theory and in the interacting $O(3)$ CFT, both in the $\varepsilon$-expansion and in various large $s$ limit. Our results are new and may be of interest also beyond the large boost limit which is the focus of this work. We therefore discuss them more extensively than the previous examples.

\subsection{Perturbation Theory at Fixed \texorpdfstring{$s$}{s}}

Both for a free and an interacting bulk, the spin impurity admits a fixed point in $d=4-\varepsilon$ at small $\gamma^2 \sim \varepsilon$ for fixed $s$. To the first nontrivial order, the analysis is the same for both the free and the interacting theory, which can therefore be discussed together. The beta function reads
\begin{equation}\label{eq_spin_imp_beta}
\beta_{\gamma^2}=-\varepsilon\gamma^2+\frac{\gamma^4}{2\pi^2}+O\left(\gamma^6s\right)\,.
\end{equation}
This admits an IR stable fixed point at
\begin{equation} \label{eq:FixedPointPerFree}
\gamma^2_*=2\pi^2\varepsilon+O\left(\varepsilon^2\right)\,.
\end{equation}

The most general cusp configuration involves two representations $s_1$ and $s_2$ and a choice of the relative sign between the couplings, i.e. we consider
\begin{equation}
\langle
\left(P e^{-\gamma_1\int_0^\infty d\tau \phi_a T^a_1}\right)_{AB}\left(P e^{-\gamma_2\int_{-\infty}^0 d\tau \phi_a T^a_2}\right)_{CD}\rangle\,,
\end{equation}
where $T_1^a$ and $T_2^a$ are, respectively, in the spin $s_1$ and the spin $s_2$ irreps of $SU(2)$. The indices $A,B,\ldots$ arise as we consider open defects, since the cusp operator is degenerate at order $O(\gamma^0)$. As we shall see explicitly below, in general the various cusp operators can be arranged in different representations of the tensor product $s_1\otimes s_2^*=|s_1-s_2|\oplus\cdots\oplus|s_1+s_2|$.

To order $O(\gamma^2)$ the cusp anomalous dimension is easily determined to be
\begin{equation}
\begin{split}
    \Gamma_{AB,CD}^{(\gamma_1,s_1),(\gamma_2,s_2)}(\theta) &=\frac{\gamma^2_1 s_1(s_1+1)+\gamma_2^2s_2(s_2+1) }{8\pi^2}\delta_{AB}\delta_{CD}-\frac{\gamma_1\gamma_2}{4\pi^2}\frac{\pi-\theta}{\sin\theta}(T^a_1)_{AB}(T^a_2)_{CD}\,,\\
    &=
    \frac{\gamma^2_1 s_1(s_1+1)+\gamma_2^2s_2(s_2+1) }{8\pi^2}\mathds{1}_1\otimes\mathds{1}_2-\frac{\gamma_1\gamma_2}{4\pi^2}\frac{\pi-\theta}{\sin\theta}T^a_1\otimes T_2^{a\,*}
    \end{split}
\end{equation}
where in the second line we wrote the result in tensor product notation. Note that to this order the defect is classically conformal for arbitrary values of the couplings $\gamma_1$ and $\gamma_2$, hence we do not need to specialize to the fixed points. Using then
\begin{equation}
    T^a_1\otimes T_2^{a\,*}=\frac12\Big[\underbrace{T^a_1T^a_1}_{=s_1(s_1+1)\mathds{1}_1}\otimes\, \mathds{1}_2+\mathds{1}_1\otimes \underbrace{T_2^{a\,*}T_2^{a\,*}}_{=s_2(s_2+1)\mathds{1}_2}-\underbrace{(T^a_1\otimes \mathds{1}_2-\mathds{1}_1\otimes T_2^{a\,*})^2}_{=s'(s'+1)\mathds{1}_1\otimes \mathds{1}_2}\Big]\,,
\end{equation} 
we obtain the final result
\begin{equation}\label{eq_spin_imp_cusp_tree}
\begin{split}
    \Gamma_{(s')}^{(\gamma_1,s_1),(\gamma_2,s_2)}(\theta) &=\frac{\gamma^2_1 s_1(s_1+1)+\gamma_2^2s_2(s_2+1) }{8\pi^2}\\
    &-\frac{\gamma_1\gamma_2}{8\pi^2}\frac{\pi-\theta}{\sin\theta}\left[s_1(s_1+1)+s_2(s_2+1)-s'(s'+1)\right]\,,
    \end{split}
\end{equation}
where $s'$ is the representation in which the cusp operator transforms.

Several comments are in order. First, for $\gamma_1=\gamma_2$ the ground state on the cusp is in the smallest possible representation, $s'=|s_1-s_2|$, and in particular transforms in the trivial representation for $s_1=s_2$. For $\gamma_1=-\gamma_2$ instead the most favorable state transforms in the spin $s'=|s_1+s_2|$ representation.

Setting to zero one of the two couplings we obtain the dimension of the defect creation operator
\begin{equation}\label{eq_spin_creation}
    \Delta_{s,0}=\frac{\gamma^2 s(s+1)}{8\pi^2}\stackrel{\gamma=\gamma_*}{=}\frac{\varepsilon }{4}s(s+1)\,.
\end{equation}
Similarly, at $\theta=\pi$ the result~\eqref{eq_spin_imp_cusp_tree} gives the dimension of various defect changing operators
\begin{equation}
    \Delta_{(s')}^{(\gamma_1,s_1),(\gamma_2,s_2)} =\frac{\gamma^2_1 s_1(s_1+1)+\gamma_2^2s_2(s_2+1) }{8\pi^2}
    -\frac{\gamma_1\gamma_2}{8\pi^2}\left[s_1(s_1+1)+s_2(s_2+1)-s'(s'+1)\right]\,.
\end{equation}
We can also extract the normalization $C_D$ of the displacement operator two-point function setting $\gamma_1=\gamma_2$, $s_1=s_2$, $s'=0$ and expanding near $\theta=\pi$:
\begin{equation}
    \Gamma_{(0)}^{(\gamma,s),(\gamma,s)}(\theta) \simeq-\frac{C_D}{12}(\theta-\pi)^2+O\left((\theta-\pi)^4\right)\quad\implies\quad
    C_D=\frac{\gamma^2 s(s+1)}{2\pi^2}\stackrel{\gamma=\gamma_*}{=}\varepsilon s(s+1)\,.
\end{equation}

The Lorentzian cusp reads
\begin{equation}\label{eq_spin_imp_cusp_tree_boost}
\begin{split}
    \Gamma_{(s')}^{L\,(\gamma_1,s_1),(\gamma_2,s_2)}(y) &=\frac{\gamma^2_1 s_1(s_1+1)+\gamma_2^2s_2(s_2+1) }{8\pi^2}\\
    &-\frac{\gamma_1\gamma_2}{4\pi^2}\frac{y\log y}{y^2-1}\left[s_1(s_1+1)+s_2(s_2+1)-s'(s'+1)\right]\,.
    \end{split}
\end{equation}
The analysis of the large boost limit is identical to the pinning field defect~\eqref{eq_pinning_field} in the free scalar model. The only difference is that now the endpoint operators are degenerate, and hence the interaction depends on the \emph{channel} $s'$ similarly to the usual quark-antiquark potential.

It is finally interesting to consider the $\theta\rightarrow 0$ limit. As explained in~\cite{Cuomo:2024psk,Kravchuk:2024qoh,Diatlyk:2024zkk}, this can be understood by considering the EFT describing the fusion of defects. This is particularly clear on the cylinder, where $\theta$ is the angular separation on the defects and the cusp anomalous dimension is just the energy of the lowest spin $s$ state of the Hilbert space with two line insertions. We claim that, to order $O(\gamma^2)$, the fusion EFT of two spin impurities consists of a defect in the reducible $s_1\otimes s_2$ representation with couplings
\begin{equation}\label{eq_spin_couplings}
    \sum_{s= |s_1-s_2|}^{s_1+s_2} \frac{C_s}{\theta} P_s+\frac{\gamma_1-\gamma_2}{2}(T_1^a+T_2^a)\phi_a+\frac{\gamma_1+\gamma_2}{2}(T_1^a-T_2^a)\phi_a+O\left(\theta\right)\,,
\end{equation}
where $P_s$ is the projector in the spin $s$ subspace and we defined the Casimir energy coefficients
\begin{equation}
    C_{s'}=-\frac{\gamma_1\gamma_2}{8\pi}\left[s_1(s_1+1)+s_2(s_2+1)-s'(s'+1)\right]\,.
\end{equation}
It is instructive to rewrite~\eqref{eq_spin_couplings} decomposing the defect Hilbert space into irreducible representations. The combination $T_1^a+T_2^a$ is just the spin generator and thus acts diagonally on all the representations $s\subset s_1\otimes s_2$. The difference operator $T_1^a-T_2^a$ instead connects adjacent representations, and may therefore be written as a linear combination of the spin generator and the defect changing operators between spin impurities in the $s$ and the $s\pm 1$ representations. Explicitly, we normalize such defect changing operators as
\begin{equation}
    \sum_{a=1}^3 \mathcal O^a_{s,s+1}\mathcal O^a_{s+1,s}=P_s\,,
    \qquad
    \mathcal O^a_{s,s+1}=\left(\mathcal O^a_{s+1,s}\right)^\dagger \,.
\end{equation}
Note that by taking the trace this implies
\begin{equation}
    \sum_{a=1}^3 \mathcal O^a_{s+1,s}\mathcal O^a_{s,s+1}
    =
    \frac{2s+1}{2s+3}P_{s+1}\,.
\end{equation}
Further denoting
\begin{equation}
    T^a_s=P_s(T_1^a+T_2^a)P_s\,,
\end{equation}
we can rewrite~\eqref{eq_spin_couplings} as
\begin{multline}\label{eq_spin_couplings_2}
     \sum_{s= |s_1-s_2|}^{s_1+s_2} \left\{\frac{C_s}{\theta} P_s +\left[
    \frac{\gamma_1-\gamma_2}{2}
    +
    \frac{\gamma_1+\gamma_2}{2}
    \frac{s_1(s_1+1)-s_2(s_2+1)}{s(s+1)}
    \right]
    T^a_s\phi_a\right\} \\
    +\sum_{s=|s_1- s_2|}^{s_1+s_2-1}
    \frac{\gamma_1+\gamma_2}{2}\rho_s
    \left(
    \mathcal O^a_{s+1,s}
    +
    \mathcal O^a_{s,s+1}
    \right)\phi_a +O\left(\theta\right)\,,
\end{multline}
where
\begin{equation}
    \rho_s^2=
    \frac{
    \left[(s+1)^2-(s_1-s_2)^2\right]
    \left[(s_1+s_2+1)^2-(s+1)^2\right]
    }{
    (s+1)(2s+1)
    } \,.
\end{equation}
The first line of \eqref{eq_spin_couplings_2} describes a direct sum of ordinary spin-$s$ impurities with different zero-point masses $C_s/\theta$. The second line introduces defect changing operators connecting adjacent representations. For $\gamma_1=-\gamma_2$ the coefficients of the latter vanish and the result simplifies to a direct sum. Note that the defect changing operators are classically marginal; we expect that the corresponding couplings will acquire nontrivial beta functions at loop level. 

Using~\eqref{eq_spin_couplings_2}, to the order of interest, the small angle limit of the cusp anomalous dimension is finally given by~\cite{Cuomo:2024psk}
\begin{equation}
\Gamma_{(s')}^{(\gamma_1,s_1),(\gamma_2,s_2)}(\theta) =\frac{C_{s'}}{\theta}+\Delta^{(\rm fus)}_{s' ,0}+O(\theta)\,,
\end{equation}
where $\Delta^{(\rm fus)}_{s', 0}$ is the dimension of the defect creation operator in the spin $s'$ representation for the defect~\eqref{eq_spin_couplings_2} (neglecting the Casimir energies). Comparing with~\eqref{eq_spin_imp_cusp_tree} we find
\begin{equation}
\Delta^{(\rm fus)}_{s',0}
=
\frac{
\gamma_1^2 s_1(s_1+1)
+
\gamma_2^2 s_2(s_2+1)
+
\gamma_1\gamma_2
\left[
s_1(s_1+1)+s_2(s_2+1)-s'(s'+1)
\right]
}{8\pi^2}\,.
\end{equation}
Note that for $\gamma_1=-\gamma_2$, $\Delta^{(\rm fus)}_{s',0}$ agrees with~\eqref{eq_spin_creation} as expected.

We remark that in the $\theta\rightarrow 0$ limit the fusion is dominated by the spin impurity in the channel with the smallest Casimir energy. This is the one with $s'=|s_1-s_2|$ for $\gamma_1=\gamma_2$, whereas it is the one with $s'=|s_1+s_2|$ for $\gamma_1=-\gamma_2$. However, in perturbation theory we need to retain the whole sum in~\eqref{eq_spin_couplings_2}, since the Casimir energies are perturbative in the defect couplings, and thus the limits $\gamma_{1,2}\rightarrow 0$ and $\theta\rightarrow 0$ do not commute.

\subsection{The Large Spin Limit in the Interacting Theory}

In the interacting $O(3)$ model, spin impurities with $s\gg 1$ may be well approximated in terms of the pinning field defect data, which are often easier to extract. Intuitively, such a relation exists because for $s\rightarrow \infty$ the impurity \emph{classicalizes} and acts as a source. More precisely, considering for concreteness a bulk observable $\mO$ (potentially representing several insertions),~\cite{Cuomo:2022xgw} argued that for $s\rightarrow\infty$ we have
\begin{equation}\label{eq_large_s_int}
   \lim_{s\rightarrow\infty} \frac{1}{2s+1}\langle \mO\,\mathcal{D}_s\rangle=\frac{1}{4\pi}\int d^2\hat{m}\langle \mO \,\mathcal{D}_{h_*\hat{m}}^{\text{pinning}}\rangle\,,
\end{equation}
where the right-hand side is the correlator evaluated in the presence of the pinning field DCFT averaged over the magnetic field direction $\hat{m}$.~\eqref{eq_large_s_int} relates for instance the spin impurity partition function or the one-point function of the operator $\phi_a^2$ for $s\rightarrow\infty$ with the same observables in the presence of pinning field defect. A similar dictionary applies to defect correlators. See~\cite{Cuomo:2022xgw} for a discussion of $1/s$ corrections. Note that although the spin impurity acts as a classical source at large spin, the response of the bulk system remains strongly quantum.

The generalization of~\eqref{eq_large_s_int} to open segments is straightforward. The only difference is that now the choice of the endpoint state sets the boundary condition in the spin impurity saddle-point, hence freezing the global zero-mode and removing the average over the direction $\hat{n}$ from~\eqref{eq_large_s_int}.\footnote{Explicitly, using the coadjoint orbit representation of the defect action in the notation of~\cite{Cuomo:2022xgw}
\begin{equation}
    S_{imp}=\int d\tau\left( \bar{z}\dot{z}+\gamma s\,\bar{z}\frac{\sigma^a}{2}z\phi_a\right)\,,
\end{equation}
the endpoint operators for maximally charged states under the Cartan are $[z_+(\tau_f)]^{2s}$ and $[\bar{z}^+(\tau_i)]^{2s}$. These thus introduce a boundary term
\begin{equation}
    \delta S=-2s\log z_+(\tau_f)-2s\log \bar{z}^+(\tau_i)\,,
\end{equation}
which can be easily checked to lead to a constant fully polarized saddle-point for $z$.} We therefore conclude that, for instance, the defect creation operator dimensions simplify 
\begin{equation}
    \Delta_{s,0}=\Delta_{\vec{h}^*,0}
\Big\vert_{\text{pinning}}+O\left(\frac{1}{s}\right)\,.
\end{equation}
Similarly, the cusp obtained deforming a large spin impurity without additional insertions coincides with the cusp of two aligned pinning fields
\begin{equation}
    \lim_{s\rightarrow\infty}\Gamma^{(+,s),(+,s)}_{(0)}(\theta)= \Gamma_{\vec{h}^*,\vec{h}^*}(\theta)\Big\vert_{\text{pinning}}+O\left(\frac{1}{s}\right)\,.
\end{equation}
To identify the dictionary for general cusp angle, we note that in the $\varepsilon$-expansion the pinning field equivalence can be verified explicitly by working in the double scaling limit
\begin{equation}\label{eq_ds_int}
    \gamma^2\sim\varepsilon\rightarrow 0\quad\text{with}\quad
    \varepsilon s^2=\text{fixed}\,,
\end{equation}
where the effective pinning field description applies in the regime $\varepsilon s^2\gg 1$ upon the replacement $\gamma s\rightarrow h$~\cite{Cuomo:2022xgw}.\footnote{In the limit~\eqref{eq_ds_int} the theory remains perturbative but certain higher loop corrections get enhanced. For instance, the two-loop $\sim \lambda \gamma^4 s^2$ correction to the beta function~\eqref{eq_spin_imp_beta} modifies the fixed point coupling to
\begin{equation}
    \gamma^2_*=\frac{2\pi^2\varepsilon}{1+2\pi^2\varepsilon s^2/11}\left[1-\varepsilon^{1/2}\frac{2\pi^2\sqrt{\varepsilon s^2}}{11+2\pi^2\varepsilon s^2}
+O\left(\varepsilon\right)    
    \right]\,,
\end{equation}
giving $\gamma_*^2s^2\simeq 11$ for $\varepsilon s^2\gg 1$ in agreement with the fixed point value $(h^*)^2$~\eqref{eq_h_fix} for $N=3$.} Comparing the perturbative results~\eqref{eq_tree_level} and~\eqref{eq_spin_imp_cusp_tree} we see that they agree with the identification
\begin{equation}
\hat{m}_1\cdot\hat{m}_2\rightarrow\text{sgn}(\gamma_1\gamma_2)\frac{\vec{T}_1\cdot\vec{T}_2}{s_1 s_2}\simeq \text{sgn}(\gamma_1\gamma_2)\frac{s_1^2+s_2^2-(s')^2}{2 s_1 s_2}\,.
\end{equation}
Intuitively, the cusp state chooses the relative orientation between the semiclassical spin degrees of freedom. Generalizing, we conclude
\begin{equation}
    \Gamma^{(\pm,s_1),(\pm,s_2)}_{(s')}(\theta)\simeq  \Gamma_{\vec{h}_1^*,\vec{h}^*_2}(\theta)\Big\vert_{\text{pinning},\, \hat{m}_1\cdot\hat{m}_2\rightarrow\text{sgn}(\gamma_1\gamma_2)\frac{\vec{T}_1\cdot\vec{T}_2}{s_1 s_2}}\,,
\end{equation}
where this holds for $s_1\rightarrow\infty,\,s_2\rightarrow\infty$ with $s_1/s_2$ fixed.

Finally we comment that we only claim that the relations discussed in this section hold in the limit $s_i\rightarrow\infty$ with $\theta$ fixed. In particular, the large spin limit might not commute with the $\theta\rightarrow 0$ one, since the structure of the fusion of spin impurities is more intricate than for pinning field defects.

\subsection{The Large Spin Limit in the Free Theory}

In the free theory, the spin impurity, or equivalently the nonlocal Bose-Kondo model, admits the following interesting semiclassical limit:
\begin{equation}\label{eq_ds_free}
    \varepsilon\sim \gamma^2\quad \text{with} \quad \varepsilon s=\text{fixed}\,.
\end{equation}
Note that this is different from the one relevant for the interacting theory~\eqref{eq_ds_int}. It is convenient to define
\begin{equation}
    \alpha=\frac{\gamma^2 s}{4\pi^2}\,.
\end{equation}
In the limit~\eqref{eq_ds_free} the beta function can be written as~\cite{Beccaria:2022bcr,Cuomo:2022xgw,Nahum:2022fqw}:
\begin{equation}\label{eq_BETA}
\beta_{\alpha}=-\varepsilon\alpha+\frac{1}{s}\frac{2\alpha^2}{1+\pi^2\alpha^2}\,.
\end{equation}
For $\varepsilon s<1/\pi$ there are two fixed points given by
\begin{equation}\label{eq_spin_imp_free_fixed}
    \alpha_{1,2}=\frac{1\mp\sqrt{1-\pi ^2 s^2 \eps^2}}{\pi ^2 s \eps }+O\left(\frac{1}{s}\right)\,.
\end{equation}
The first fixed point is stable while the second is unstable. The fixed points merge at $\varepsilon s=1/\pi$ and for $\varepsilon s>1/\pi$ $\alpha$ runs to infinite coupling. Below we consider the simplest cusp configuration, which is obtained deforming the contour of a single spin $s$ impurity without any insertions, at the two fixed points~\eqref{eq_spin_imp_free_fixed}.\footnote{Similar semiclassical analyses of cusped Wilson-line states carrying large charge appeared, for instance, in \cite{Gromov:2012eu,Drukker:2012de,Sizov:2013joa,Bonomi:2025ctm,Bersini:2026tpq}.}

We first recall that the defect action can be represented within the coadjoint orbit approach in terms of a two-component complex vector $z$ such that $\bar{z}z=2$:
\begin{equation}\label{eq_s_action_free}
    S=\int d^dx\frac{1}{2}(\pd\phi_a)^2+s\int_{\mathcal{D}} d\tau\left[\bar{z}\dot{z}-\gamma_0\bar{z}\frac{\sigma^a}{2}z\phi_a\right]\,,
\end{equation}
where $\gamma_0$ is the bare coupling; we refer to~\cite{Cuomo:2022xgw} and app.~\ref{app_technical_spin} for details on the renormalization scheme we use.
Note that the action is invariant under gauge transformations $z\rightarrow e^{i\alpha(\tau)}z$, which make the defect field target space a sphere. Rescaling $\phi_a\rightarrow \sqrt{s}\phi_a$, we see that in the double-scaling limit~\eqref{eq_ds_free} the theory can be studied semiclassically in a saddle-point approximation. 

For our purposes, it is convenient to work on the cylinder. As reviewed before, the cusp anomalous dimension can be extracted from the partition function with two defect insertions at angular separation $\theta$. For comparison, we will also consider the defect creation operator dimension, which requires studying a single defect insertion on the cylinder. It is further convenient to integrate out the scalar using the cylinder propagator
    \begin{equation}\label{eq_cyl_prop}
G_{cyl}(x,y)=
\frac{1}{(d-2)\Omega_{d-1}[\Delta X^2(x,y)]^{\frac{d-2}{2}}}\,,\qquad
\Delta X^2(x,y)=2\cosh\tau_{xy}-2\hat{n}_x\cdot\hat{n}_y\,,
\end{equation}
and expand the defect fields around a constant background as
\begin{equation}
    z\simeq\left(\begin{array}{c}
         1-\frac{1}{\sqrt{2s}}\chi \\
         1+\frac{1}{\sqrt{2s}}\chi
    \end{array}\right)\,,
\end{equation}
where $\chi$ is a complex fluctuation and we neglected an overall global zero-mode rotating the field direction, since this is irrelevant for our purposes. We therefore obtain the following action
\begin{equation}\label{eq_free_defect_action_nonlocal}
\begin{aligned}
S=&-s\frac{\alpha_0}{2}\int_{\mathcal{D}} d\tau\int_{\mathcal{D}} d\tau'
\frac{1}{[\Delta X^2(x(\tau),x(\tau'))]^{\frac{d-2}{2}}}\\[0.3em]
&+\int_{-\infty}^\infty d\tau\,\left(\bar{\chi}\dot{\chi}\vert_0-\bar{\chi}\dot{\chi}\vert_\theta\right)-\frac{\alpha_0}{2}
\int_{\mathcal{D}} d\tau \int_{\mathcal{D}} d\tau'
\frac{\left(\bar{\chi}\chi'+\bar{\chi}'\chi-\bar{\chi}\chi-\bar{\chi}'\chi'\right)}{[\Delta X^2(x(\tau),x(\tau'))]^{\frac{d-2}{2}}} \\[0.3em]
&+O\left(\frac{1}{s}\right)\,,
\end{aligned}
\end{equation}
where for the cusp
\begin{equation}
    \int_{\mathcal{D}} d\tau\equiv\int_{-\infty}^\infty d\tau\Bigg\vert_{0}+\int_{-\infty}^\infty d\tau\Bigg\vert_{\theta}\,,
\end{equation}
while we set to zero all contributions at $\theta$ when considering a single defect insertion.

To extract the cusp anomalous dimension we must evaluate the cylinder partition function using~\eqref{eq_free_defect_action_nonlocal} and apply~\eqref{cusp_cyl}. At leading order, the partition function consists only of the first classical term in~\eqref{eq_free_defect_action_nonlocal}, and we immediately obtain the cusp anomalous dimension
\begin{equation}\label{eq_ds_cusp_0}
   \Gamma_{(0)}^{(+,s),(+,s)}(\theta)=s\,\alpha \left(1-\frac{\pi-\theta}{\sin\theta}\right)+O\left(1\right)\,,
\end{equation}
which coincides with the naive tree-level answer~\eqref{eq_spin_imp_cusp_tree}. Similarly, we find the defect creation operator dimension
\begin{equation}\label{eq_ds_dim_0}
    \Delta_{s,0}=\frac{\alpha s}{2}+O\left(1\right)\,,
\end{equation}
again in agreement with~\eqref{eq_spin_creation}. 

The $O(1)$ subleading corrections to the above results arise from the one-loop determinant over the fluctuations and the renormalization of the coupling. The calculation is described in detail in app.~\ref{app_technical_spin}. Below we report the results and analyze various limits of interest.

\begin{figure}[t]
    \centering
    \begin{subfigure}{0.49\textwidth}
        \centering
        \includegraphics[width=\linewidth]{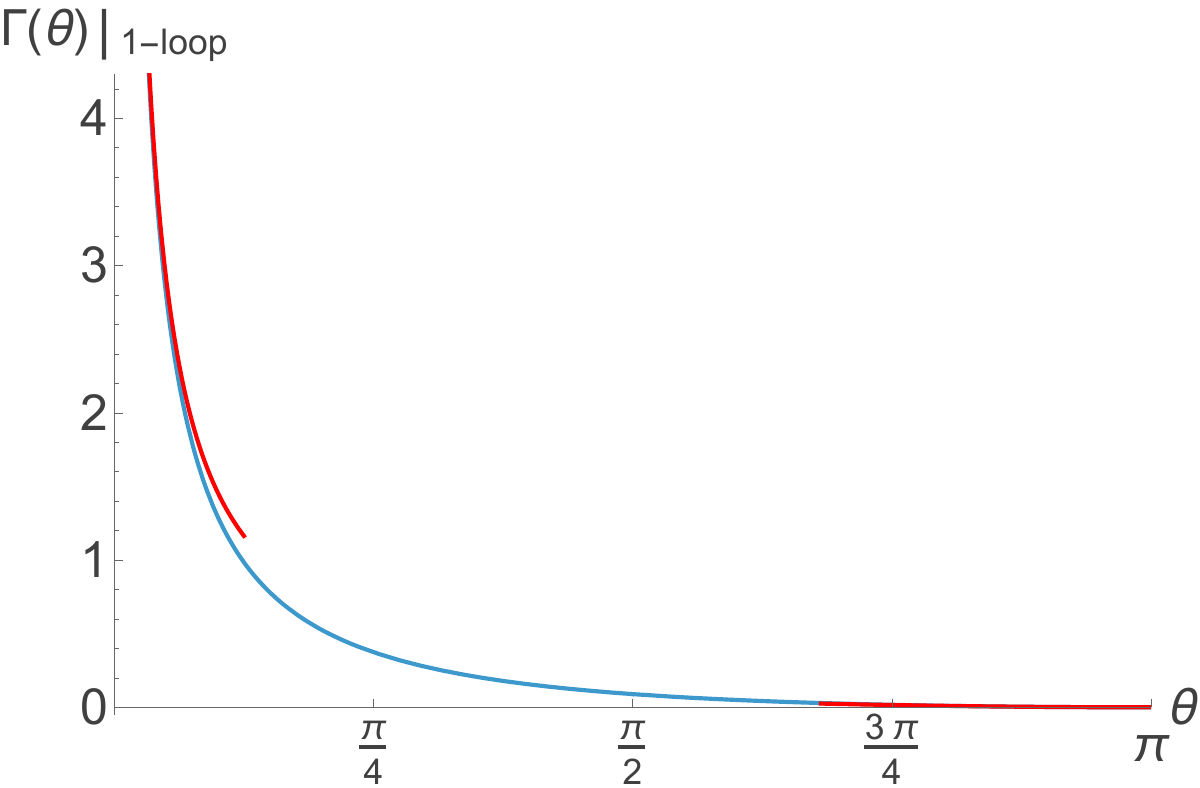}
        \caption{}
          \label{fig:SI_one_loop}
    \end{subfigure}
    \hfill
    \begin{subfigure}{0.49\textwidth}
        \centering
        \includegraphics[width=\linewidth]{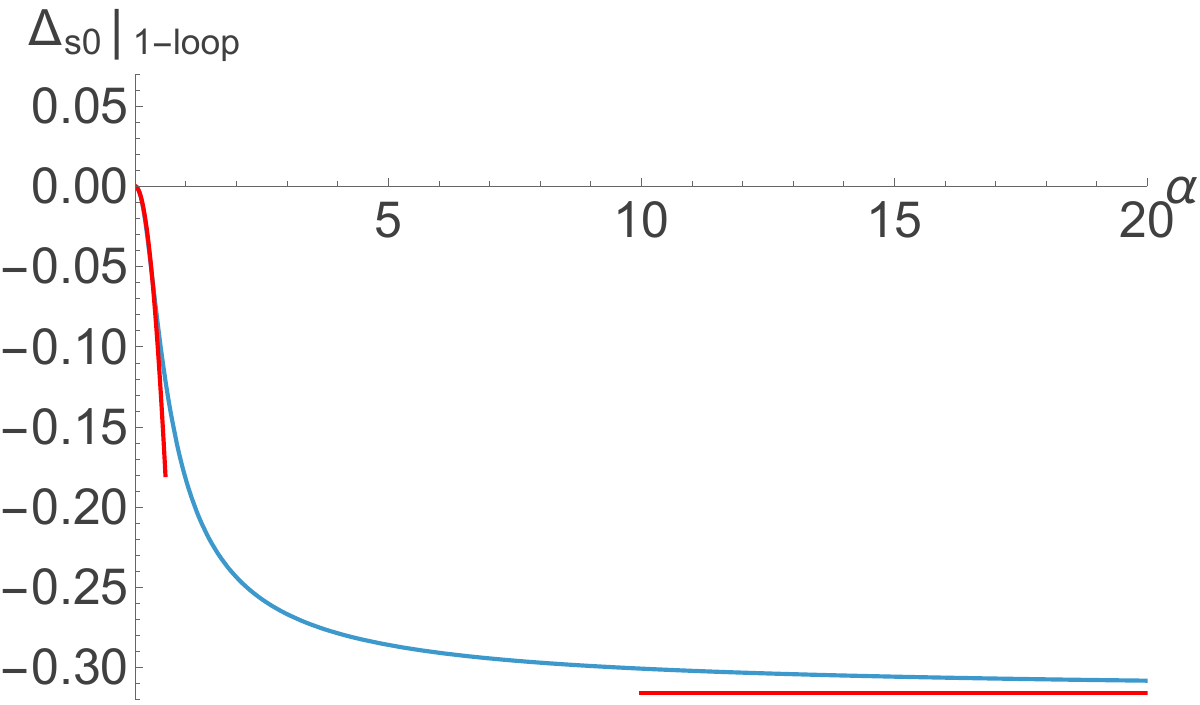}
        \caption{}
          \label{fig:SI_defect_creation}
    \end{subfigure}
    \caption{The one-loop contributions to: the cusp anomalous dimension $\Gamma^{(+,s)(+,s)}_{(0)}(\theta)\vert_{\text{1-loop}}=\Gamma^{(+,s)(+,s)}_{(0)}(\theta) -(s+1)\alpha\left(1-\frac{\pi-\theta}{\sin\theta}\right)$ at $\alpha=1/\pi$, fig.~\ref{fig:SI_one_loop}; and to the defect creation operator dimension $\Delta_{s,0}\vert_{\text{1-loop}}=\Delta_{s,0} -(s+1)\alpha/2$, fig.~\ref{fig:SI_defect_creation}. The blue curves are obtained from numerical integration while the red ones are the (semi-)analytic asymptotics reported in the main text.}
\end{figure}

Expressing $\varepsilon$ in terms of the fixed point coupling using~\eqref{eq_BETA}, we find that at both fixed points the cusp anomalous dimension can be written as:
\begin{equation}\label{eq_cusp_ds_1_loop}
\begin{split}
    \Gamma^{(+,s)(+,s)}_{(0)}(\theta) 
    &=(s+1)\alpha\left(1-\frac{\pi-\theta}{\sin\theta}\right)+\frac{2\alpha^2}{1+\pi^2\alpha^2}\left[1-\left.\frac{d f_{4-\varepsilon}(\cos\theta)}{d\varepsilon}\right\vert_{\varepsilon=0}\right]+\bar{I}^{(ss)}_{\alpha}(\theta)\\
    &+O\left(\frac{1}{s}\right)\,.
    \end{split}
\end{equation}
Here $\bar{I}^{(ss)}_{\alpha}(\theta)$ is the following function:
\begin{align}
\nonumber
    \bar{I}^{(ss)}_{\alpha}(\theta)
    % &=I^{(ss)}_{\alpha}(\theta)-I_{\alpha}^{(ss)}(\pi) \\
    &=2\int_0^\infty\frac{dk}{2\pi}\bar{G}^{(ss)}_{\theta,\alpha,\Lambda}(k)
    -\frac{\alpha^2\log\left(\Lambda^2 e^{2\gamma_E-2}\right)}{1+\pi^2\alpha^2}\left(\frac{\pi-\theta}{\sin\theta}-1\right)\,,
    \\[0.5em]
    \nonumber
\bar{G}^{(ss)}_{\theta,\alpha,\Lambda}(k)&=\log \left(\frac{k^2+\alpha ^2 \left[1-\frac{\pi -\theta}{\sin\theta}-\pi  k \coth (\pi  k)\right]^2
-\alpha ^2 \frac{\pi ^2 \sinh ^2((\pi -\theta ) k)}{\sin ^2(\theta ) \sinh ^2(\pi  k)}
}{k^2(1+\pi ^2 \alpha ^2)}\right)\\
&-\Theta\left(k-\Lambda\right)\frac{2\pi \alpha^2}{(1+\pi^2\alpha^2)k}\left(\frac{\pi-\theta}{\sin\theta}-1\right)\,,
\label{eq_app_Gss}
    \end{align}
where $\Lambda>0$ is arbitrary. Although we have not managed to find a more explicit expression, it can be easily checked that $\bar{I}^{(ss)}_{\alpha}(\theta)$ is independent of $\Lambda$ (for instance acting with $d/d\Lambda$ and commuting the derivative with the integral). For illustration, we plot the one-loop contribution to the cusp anomalous dimension at $\alpha=1/\pi$ in fig.~\ref{fig:SI_one_loop}.

The defect creation operator dimension reads
\begin{equation}\label{eq_dim_ds_1_loop}
\begin{split}
    \Delta_{s,0} &=(s+1)\frac{\alpha}{2}+\frac{\alpha^2}{1+\pi^2\alpha^2}+\bar{I}^{(s0)}_{\alpha}+O\left(\frac{1}{s}\right)\,,
    \end{split}
\end{equation}
where $\bar{I}^{(s0)}_{\alpha}$ is the following function:
\begin{align}
\nonumber
\bar{I}^{(s0)}_{\alpha}
% &=I^{(s0)}_{\alpha}-\frac12 I_{\alpha}^{(ss)}(\pi) 
&=\int_0^\infty\frac{dk}{2\pi}\bar{G}^{(s0)}_{\alpha,\Lambda}(k)
    +\frac{\alpha^2\log\left(\Lambda^2 e^{2\gamma_E-2}\right)}{2+2\pi^2\alpha^2}\,,
    \\[0.5em]
\bar{G}^{(s0)}_{\alpha,\Lambda}(k)&=\log \left(\frac{k^2+\alpha ^2 \left[\pi  k \coth (\pi  k)-1\right]^2}{k^2(1+\pi ^2 \alpha ^2)}\right)+\Theta\left(k-\Lambda\right)\frac{2\pi \alpha^2}{(1+\pi^2\alpha^2)k}\,.
    \end{align}
As before it can be checked that $\bar{I}^{(s0)}_{\alpha}$ is independent of $\Lambda$. At small and large coupling the asymptotics of the dimension can be extracted using
\begin{equation}\label{eq_Is0_asymp}
    \bar{I}^{(s0)}_{\alpha}=\begin{cases}
        -\frac{3}{2}\alpha^2+O\left(\alpha^4\right) &\alpha\ll 1 \\
-0.417116\ldots+O\left(\alpha^{-2}\right)&\alpha \gg 1\,.
           \end{cases}
\end{equation}
We plot the one-loop contribution to the defect creation operator dimension in fig.~\ref{fig:SI_defect_creation}.

Let us analyze different limits of the cusp anomalous dimension. First, we note that 
\begin{equation}
  \left.\frac{d^2\bar{I}^{(ss)}_{\alpha}(\theta)}{d\theta^2}\right\vert_{\theta=\pi}=\frac{11 \alpha ^2}{9(1+ \pi ^2 \alpha ^2)}\,,
\end{equation}
and
\begin{equation}
    \left.\frac{d f_{4-\varepsilon}(\cos\theta)}{d\varepsilon}\right\vert_{\varepsilon=0}=1+\frac{1}{18} (\pi -\theta )^2+O\left((\pi-\theta)^4\right)\,.
\end{equation}
Using the above, we obtain the second derivative of the cusp anomalous dimension at $\theta=\pi$,
\begin{equation}\label{eq_ds_cusp_pi_1_loop}
       \left.\frac{d^2 \Gamma^{(+,s)(+,s)}_{(0)}(\theta)}{d\theta^2}\right\vert_{\theta=\pi}=-\frac{C_D}{6}=-(s+1)\frac{\alpha}{3}+\frac{\alpha ^2}{1+\pi ^2 \alpha ^2}+O\left(\frac{1}{s}\right)\,,
\end{equation}
which is related to the normalization $C_D$ of the displacement two-point function.

\begin{figure}[t]
    \centering
    \begin{subfigure}{0.49\textwidth}
        \centering
        \includegraphics[width=\linewidth]{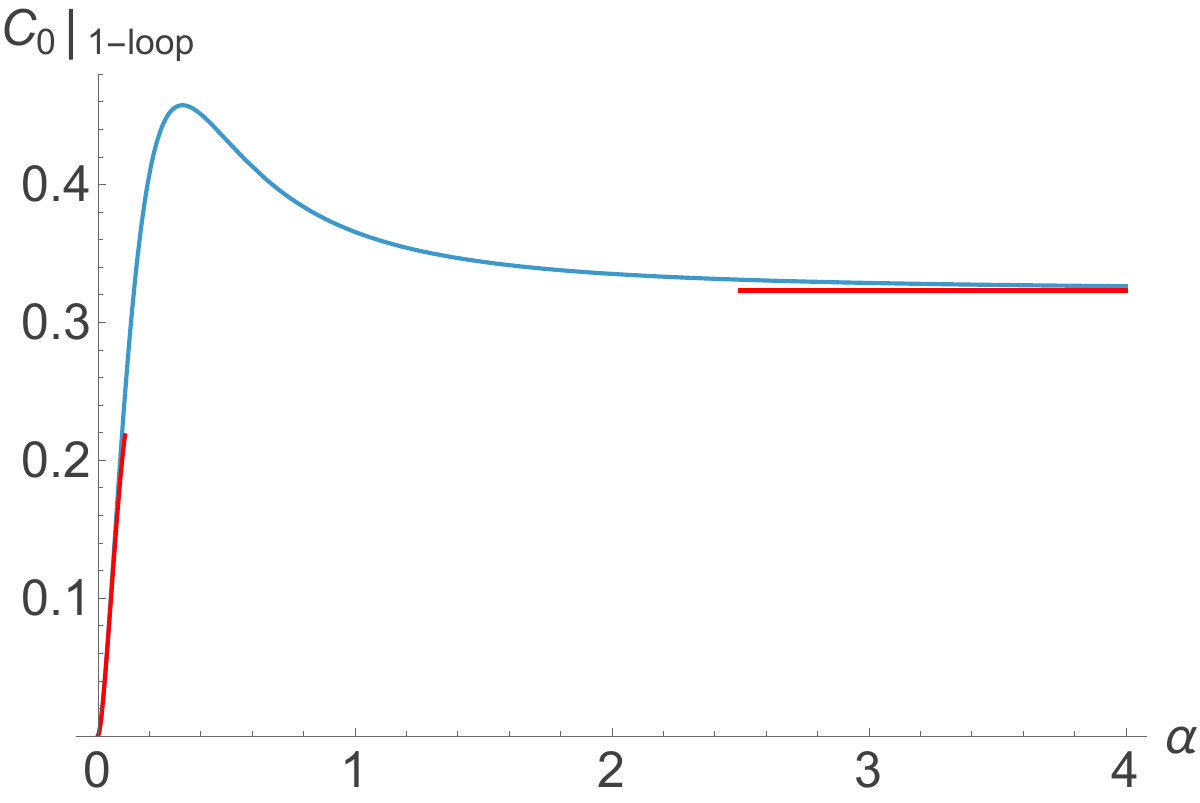}
        \caption{}
          \label{fig:SI_Casimir_energy}
    \end{subfigure}
    \hfill
    \begin{subfigure}{0.49\textwidth}
        \centering
        \includegraphics[width=\linewidth]{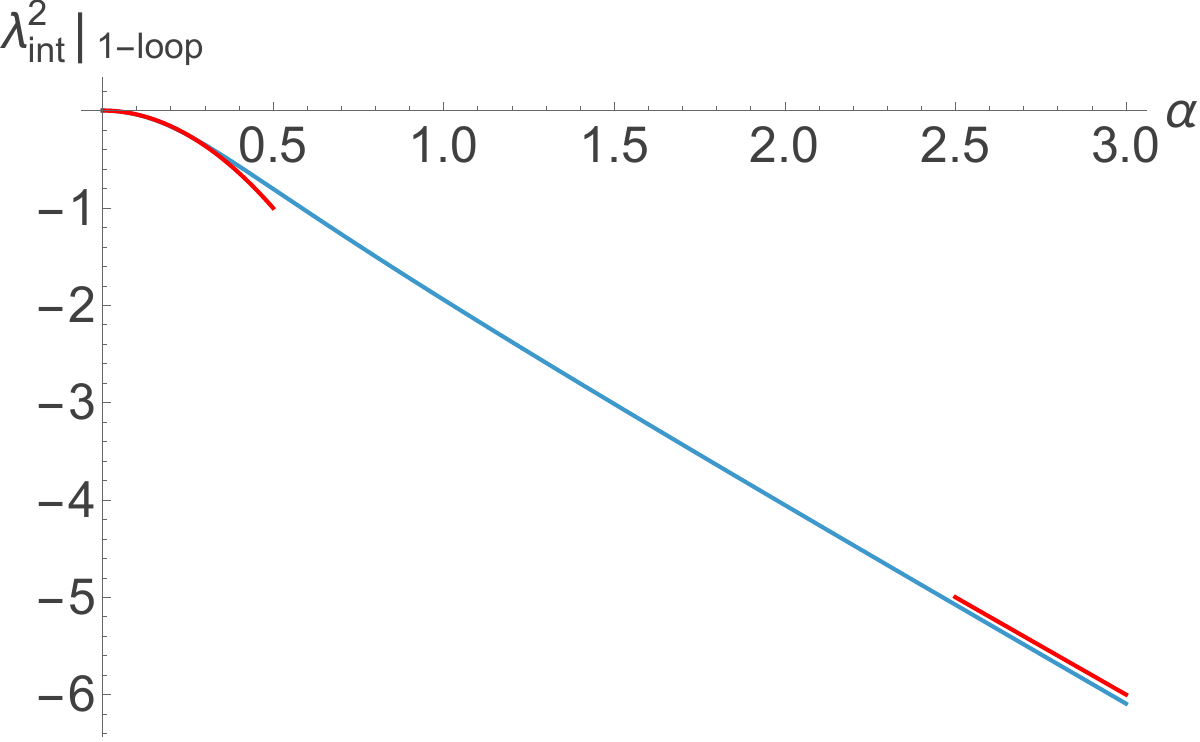}
        \caption{}
        \label{fig:SI_boost}
    \end{subfigure}
    \caption{The one-loop contributions to: the Casimir energy $C_0\vert_{\text{1-loop}}=C_0 +(s+1)\pi\alpha$, fig.~\ref{fig:SI_Casimir_energy}; and the large boost interaction coefficient $\lambda^2_{\rm int}\vert_{\text{1-loop}}=\lambda^2_{\rm int}- 2(s+1)\alpha$, fig.~\ref{fig:SI_boost}. The blue curves are obtained from numerical integration while the red ones are the (semi-)analytic small and large coupling asymptotics reported in the main text.}
\end{figure}

For $\theta\rightarrow 0$ the leading contribution in $\bar{I}^{(ss)}_{\alpha}(\theta)$ arises from the region $k\sim 1/\theta$. We find that this yields the following behavior
\begin{equation}
    \bar{I}^{(ss)}_{\alpha}(\theta)\simeq\frac{1}{\theta}\left[
    \frac{2 \pi  \alpha ^2 }{1+\pi ^2 \alpha ^2}
\left(\log \theta +1-\gamma_E\right)+I_C(\alpha)
    \right]+O\left(\log\theta\right)\,,
\end{equation}
where we defined
\begin{equation}\label{eq_spin_imp_IC}
\begin{split}
    I_C(\alpha)&=\int_0^\infty\frac{d q}{\pi}\left[
    \log \left(\frac{q^2+\pi ^2 \alpha ^2 \left[(q+1)^2-e^{-2 q}\right]}{q^2\left(1+\pi ^2 \alpha ^2\right) }\right)-\frac{2 \pi ^2 \alpha ^2}{(q+1)\left(1+\pi ^2 \alpha ^2\right) }
    \right] \\[0.5em]  
&=\begin{cases}
    \pi  \alpha ^2 \left[4 \log \left(\frac{1}{\pi ^2 \alpha ^2}\right)-2 \gamma_E +6-10 \log 2\right]+O\left(\alpha^4\right) & \alpha\ll 1 \\[0.35em]
    0.495111\ldots+O\left(\alpha^{-2}\right)
    & \alpha\gg 1    \,.
\end{cases}
    \end{split}
\end{equation}
Also using that
\begin{equation}
    \left.\frac{d f_{4-\varepsilon}(\cos\theta)}{d\varepsilon}\right\vert_{\varepsilon=0}=\frac{\pi  (\log \theta +\log 2)}{\theta }+O(1)\,,
\end{equation}
we find that the $\log \theta/\theta$ term cancels and the cusp anomalous dimension displays the expected $1/\theta$ singularity\footnote{The corrections are $O(\log\theta)$ due to the anomalous dimension of the defect perturbation at the fixed point, as found in~\cite{Cuomo:2024psk} for the pinning field defect.}
\begin{equation}\label{eq_ds_cusp_small_1_loop}
    \Gamma^{(+,s)(+,s)}_{(0)}(\theta)\simeq \frac{C_{0}}{\theta}+O\left(\log\theta\right)\,,
\end{equation}
where the Casimir energy is given by
\begin{equation}
    C_{0}=-(s+1)\pi\alpha+\frac{2 \pi  \alpha ^2(1-\gamma_E-\log 2) }{1+\pi ^2 \alpha ^2}+I_C(\alpha)+O\left(\frac{1}{s}\right)\,.
\end{equation}
We plot the one-loop contribution to the Casimir energy in fig.~\ref{fig:SI_Casimir_energy}. Note that, as follows from the small coupling expansion of $I_C(\alpha)$ in~\eqref{eq_spin_imp_IC}, the Casimir energy is non-analytic in the coupling for $\alpha\rightarrow 0$. A similar behavior occurs for fundamental Wilson lines in $\mathcal{N}=4$ SYM (and QCD), where it is due to the approximate degeneracy of the singlet and adjoint fusion channels in perturbation theory~\cite{Pineda:2007kz,Gromov:2016rrp}. For spin impurities, several fusion channels are likewise degenerate at $\alpha=0$, as seen in the previous sections: this is the origin of the non-analyticity. In gauge theories, the Casimir energy can be computed perturbatively using a non-relativistic EFT~\cite{Pineda:2007kz}; it would be interesting to develop an analogous analysis in the present setup.

Let us finally analyze the large boost limit. Noting that
\begin{equation}
   \left. \frac{\pi  \sinh ((\pi -\theta ) k)}{\sin (\theta ) \sinh (\pi  k)}\right\vert_{\theta=\pi-i\log y}\stackrel{y\rightarrow\infty}{\rightarrow}\frac{2 \pi \sin (k\log y) }{\sinh (\pi k)}\left(\frac{1}{ y}+\frac{1}{ y^3}+\ldots\right)\,,
\end{equation}
we see that we can neglect the last term inside the log in~\eqref{eq_app_Gss}. We then obtain the expansion
\begin{equation}
    2\bar{G}^{(ss)}_{\pi-i\log y,\alpha,\Lambda}(k)\stackrel{ y\rightarrow\infty}{\rightarrow}2\bar{G}^{(s0)}_{\alpha,\Lambda}(k)
    +\frac{ \log  y  }{ y }\bar{H}^{(ss)}_{\alpha,\Lambda}(k)+O\left(\frac{1}{ y^2}\right)\,,
\end{equation}
where we defined the following function
\begin{equation}
    \bar{H}^{(ss)}_{\alpha,\Lambda}(k)\equiv 8 \alpha ^2   \left[\frac{\pi  k \coth (\pi  k)-1}{k^2+\alpha ^2 (\pi  k \coth (\pi  k)-1)^2}-\frac{\pi \, \Theta (k-\Lambda )}{\left(1+\pi ^2 \alpha ^2\right) k}\right]\,.
\end{equation}
Using the above, the final result for the large boost cusp anomalous dimension takes the same form as in the pinning-field examples
\begin{equation}
\begin{split}
    \Gamma^{(+,s)(+,s)}_{(0)}(\pi-i \log y) &\stackrel{y\gg 1}{\simeq} 2\Delta_{s,0}-\frac{\lambda^2_{\rm int}}{ y^{\Delta_{\phi}}}\left[\log y-\psi^{(0)}(\Delta_\phi)-\gamma_E\right]\,,
    \end{split}
\end{equation}
where $\Delta_{\phi}=1-\varepsilon/2+O\left(\varepsilon^2\right)$ and
\begin{equation}
    \lambda^2_{\rm int}=2(s+1)\alpha-\bar{V}^{(ss)}_{\alpha} +O\left(\frac{1}{s}\right)\,.
\end{equation}
Here $\bar{V}^{(ss)}_{\alpha}$ is the following function:
\begin{equation}
\bar{V}^{(ss)}_{\alpha}=\int_0^\infty\frac{dk}{2\pi}\bar{H}^{(ss)}_{\alpha,\Lambda}(k)-
\frac{2\alpha^2\log\left(\Lambda^2 e^{2\gamma_E-2}\right)}{1+\pi^2\alpha^2}=\begin{cases}
4\alpha^2+O(\alpha^4) & \alpha\ll 1 \\
2\alpha +O(1) &\alpha\gg 1\,,
\end{cases}
\end{equation}
which can be checked to be independent of $\Lambda$ similarly to the other functions defined above. We plot the one-loop contribution to $\lambda_{\rm int}^2$ in fig.~\ref{fig:SI_boost}.

\section{Outlook}\label{sec:outlook}

In this work we studied the cusp anomalous dimensions in the large boost limit. Our main result is the general expression~\eqref{lorentzcusp}. We verified our prediction in several perturbative examples. Below, we would like to comment on some open questions for future research.

% In this work, we proposed two positivity conditions on the Lorentzian cusp anomalous dimension,
% \begin{equation}\label{eq_bound_Gamma_conclusions}
% \Gamma^L_{ab}(y)\geq 0
% \qquad\text{and}\qquad
% \Re,\Gamma^L_{ab}(-y-i\epsilon)\geq 0\,,
% \end{equation}
% for physical $y>0$. Both of these condition deserve further study. The first bound in~\eqref{eq_bound_Gamma_conclusions} follows from Rindler positivity. While Rindler positivity is rigorously established for Wightman functions of local operators, its extension to general defects requires an additional assumption; for defects admitting endpoint operators that connect them to the trivial line, we justified this assumption in sec.~\ref{subsec:rindler-positivity}, but we lack a general proof for arbitrary defects. The second bound in~\eqref{eq_bound_Gamma_conclusions} relies instead on the interpretation of the defect correlator as a physical amplitude for heavy external particles, and therefore applies whenever such a heavy-particle realization is available. Examples of this sort include Wilson lines, pinning field defects and spin impurities for instance, but a general CFT proof for arbitrary defects is missing.

In this work, we proposed two positivity conditions on the Lorentzian cusp anomalous dimension,
\begin{equation}\label{eq_bound_Gamma_conclusions}
\Gamma^L_{ab}(y)\geq 0
\qquad\text{and}\qquad
\Re\,\Gamma^L_{ab}(-y-i\epsilon)\geq 0\,,
\end{equation}
for physical $y>0$. Both conditions deserve further study. The first bound in~\eqref{eq_bound_Gamma_conclusions} follows from Rindler positivity. While Rindler positivity is rigorously established for Wightman functions of local operators, its extension to general defects requires an additional assumption. For defects admitting endpoint operators that connect them to the trivial line, we justified this assumption in sec.~\ref{subsec:rindler-positivity}, but a proof for general defects is still lacking. The second bound in~\eqref{eq_bound_Gamma_conclusions} relies instead on the interpretation of the defect correlator as a physical amplitude for heavy external particles, and therefore applies whenever such a heavy-particle realization is available. Examples include Wilson lines, pinning-field defects, and spin impurities, among others, but a general CFT derivation remains an open problem.

% In section~\ref{subsec_HEFT} we used physical arguments to derive a positivity bound on the cusp anomalous dimension~\eqref{eq_condition_Gamma}. Although the argument was discussed in detail for the localized magnetic defect in the Ising model, it can be easily generalized to a large class of defects in other theories, including Wilson lines and spin impurities for instance. In general one obtains the following condition 
% \begin{equation}\label{eq_bound_Gamma_conclusions}
%     \text{Re}\,\Gamma^L_{ab}(-y-i\epsilon)\geq 0\,,
% \end{equation}
% for physical $y>0$. \GC{changed here} As mentioned in sec. \ref{subsec:rindler-positivity}, it would be interesting to explore the conditions for the validity of~\eqref{eq_bound_Gamma_conclusions} in more detail, and provide a proof with CFT methods. We also note that in all our examples $\Gamma^L_{ab}(y)$ is positive for $y>0$ in agreement with the positivity bound \eqref{eq_rindler_gamma_bound} that we derived from Rindler positivity. 

It would be interesting to study the cusp anomalous dimension and its null limit in other examples. In particular, monodromy defects exist in all theories with internal symmetries and provide an important understudied setup for which our results do not apply. Additionally, it would be valuable to calculate the Lorentzian cusp anomalous dimension non-perturbatively in some (non-planar) examples, perhaps exploiting the recent advances on the fuzzy-sphere regularization of CFTs~\cite{Zhu:2022gjc,He:2026ong}.

It might also be interesting to study the cusp anomalous dimension and the fusion of spin impurities in the interacting $O(3)$ model at subleading orders, where one must disentangle the perturbative degeneracy among the different fusion channels. As discussed in sec.~\ref{sec_spin_imp}, this may require resumming infinitely many perturbative contributions, along the lines of~\cite{Pineda:2007kz}. We were informed about upcoming work in this direction \cite{Diatlyk:2026appear}.

Null lines enjoy additional symmetries compared to static ones~\cite{Braun:2003rp,Alday:2010ku}, at least kinematically~\cite{Erramilli:2025pfh}. For instance, a generic cusped configuration preserves only dilations and transverse rotations, but in the infinite rapidity limit it becomes also boost invariant as discussed at length in the main text. The symmetries of null cusps (and polygons) are known to have important implications for Wilson lines in gauge theories~\cite{Alday:2010ku}, where they are also related to amplitudes and local correlators~\cite{Alday:2010zy,Chen:2025ffl}. It would be interesting to analyze the implications of these symmetries for more general line defects. In particular, it is natural to ask if these can be used to systematically derive corrections to our main result~\eqref{lorentzcusp} with analytic bootstrap methods in terms of microscopic CFT data, perhaps along the lines of the argument in app.~\ref{app_spectral_representation}. 

The arguments in sec.~\ref{sec_Large_boost} imply that in the large boost limit the two defects meeting at the cusp are only weakly interacting with each other. We used this fact to argue for our main results, but we did not construct a complete EFT, with systematic power counting rules and manifest symmetries, that allows computing correlators and subleading corrections to the cusp anomalous dimension. As reviewed in app.~\ref{app_Large_S}, a related question arises in the analysis of multi-trace large spin operators, which are expected to admit a similar description~\cite{Alday:2007mf,Komargodski:2012ek,Fitzpatrick:2012yx}. Recent works~\cite{Fardelli:2024heb,Kravchuk:2024wmv,Fardelli:2025eun,Fardelli:2025fkn} proposed an \emph{emergent} higher-dimensional holographic EFT describing the large spin sector of CFTs. It might be possible to relate those works to our results and improve the EFT description of both boosted defects and large spin operators.\footnote{Perhaps such EFT could describe the off-shell propagation of heavy modes in terms of analytically continued non-relativistic states as suggested in~\cite{Lopes:2026erz}.}

Finally, as commented in the main text, our arguments suggest a transition in the behavior of the cusp anomalous dimension for Wilson lines not charged under one-form symmetries at non-perturbatively large values of the boost parameter. This might have implications for the various existing factorization theorems involving these defects \cite{Collins:1989gx}, such as those concerning local operators~\cite{Alday:2010zy,Chen:2025ffl} or Energy-Energy-Correlators in conformal theories \cite{Korchemsky:2019nzm,Moult:2025nhu}.

\section*{Acknowledgments}

We thank E.~Armanini, O.~Diatlyk, G.~Fardelli, R. Lanzetta, Y.-Z.~Li, M. Metlitski, A.~Monin, I.~Moult, L. Ricci, and A. Zhiboedov for useful discussions. We are especially grateful to Z.~Komargodski and Y. Wang for useful comments on a preliminary version of this manuscript.

\appendix

\section{Coordinate Transformations in Embedding Space}\label{app_coordinates}

In this section we study the coordinates~\eqref{eq_ds_alpha} and their analytic continuation~\eqref{eq_an_cont} using the conformal embedding formalism.

To set our conventions, let us briefly review the Lorentzian embedding space following~\cite{Kravchuk:2018htv,simmonsduffin2019lorentziancft}. 
The $d$-dimensional physical space can be embedded in the projective null cone of $\mathbb{R}^{d,2}$:
\begin{equation}\label{eq_app_projective_null}
    P\in \mathbb{R}^{d,2}\quad\text{s.t.}\quad P^2=0\quad\text{and}\quad P\sim \lambda P\,,
\end{equation}
where the last condition should be understood as a gauge redundancy. Taking the $d+2$ dimensional metric to be $\eta=\text{diag}(-1,-1,1,\ldots,1)$, we recover flat space specializing to the Poincar\'e section:
\begin{equation}
    P^A=(P^{-1},P^\mu,P^{d})=\left(\frac{1+x^2}{2},x^\mu,\frac{1-x^2}{2}\right)\,,
\end{equation}
which has a flat induced metric,
\begin{equation}
    ds^2=dP^A dP^B\eta_{AB}=dx^\mu dx_\mu\,.
\end{equation}
Rescaling $P^A$ we obtain other Weyl-equivalent manifolds. For instance, a patch of the Lorentzian cylinder $\mathbb{R}\times S^{d-1}$ can be embedded in $d+2$ dimensions as
\begin{equation}
    P^A=\left(\cos\tau,\sin \tau,\hat{m}^{i},\hat{m}^d\right)\,,
\end{equation}
where $\tau$ is the cylinder time coordinate and $\hat{m}=(\hat{m}^i,\hat{m}^d)$ is a $d$-dimensional unit vector parametrizing the sphere $S^{d-1}$.

On the embedding vector $P$ the conformal group $SO(d,2)$ acts linearly. Explicitly, defining the generators
\begin{equation}
    (L_{AB})^C_{\;D}=i\delta^C_{A}\eta_{BD}-i\eta_{AD}\delta^C_B\quad\implies\quad
[L_{AB},L_{CD}]=i\eta_{AD}L_{BC}+\ldots\,,
\end{equation}
we obtain the Lorentzian conformal algebra with the following identifications
\begin{equation}\label{eq_app_conformal_ids}
    M_{\mu\nu}=L_{\mu\nu}\,,\quad
P_\mu=L_{-1\mu}-L_{d\mu}\,,\quad
K_\mu=L_{-1\mu}+L_{d\mu}\,,\quad D=-L_{-1\,d}\,.
\end{equation}

We obtain AdS$_3\times S^{d-3}$ in~\eqref{eq_AdS3} setting
\begin{equation}
    P=\left(\frac{1+x^2+r^2-t^2}{2r},\frac{t}{r}\,,\frac{x}{r}\,,\hat{n}\,,\frac{1+t^2-x^2-r^2}{2r}\right)\,,
\end{equation}
where $\hat{n}$ is a $d-2$-dimensional unit vector. The coordinate change~\eqref{eq_alpha_coord} then gives
\begin{equation}\label{eq_P_alpha}
    P=\left(\cos \alpha \cosh \tau ,\sin\alpha  \cosh \gamma ,\sin\alpha  \sinh \gamma ,\hat{n},-\cos \alpha  \sinh \tau \right)\,.
\end{equation}
Comparing with~\eqref{eq_app_conformal_ids}, we see that dilations act as boosts in embedding space on the first and last coordinates and thus shift $\tau$. Similarly, $\gamma$ is manifestly the rapidity upon which $M_{01}$ acts linearly.

The analytic continuation~\eqref{eq_an_cont} leads to a complex vector, that can be written as
\begin{equation}\label{eq_app_P_wick}
    P\rightarrow \left(\frac{\bar{P}^{-1}-i\bar{P}^d}{\sqrt{2}},\frac{\bar{P}^{-1}+i\bar{P}^d}{\sqrt{2}},\frac{\bar{P}^{1}+i\bar{P}^0}{\sqrt{2}},\hat{n},\frac{\bar{P}^{1}-i\bar{P}^0}{\sqrt{2}}\right)\,,
\end{equation}
where $\bar{P}^A$ is a real Lorentzian embedding vector given by
\begin{equation}\label{eq_app_P_u_c}
\begin{split}
    \bar P=
    &\left(\cosh \sigma  \cosh \chi \cos u-\sinh \sigma \sinh \chi  \sin u\, ,\cosh \sigma \cosh \chi\sin u+
    \sinh \sigma  \sinh \chi   \cos u \,,\right.\\ 
    &
    \left.
    \cosh \sigma  \sinh \chi \cos u-\sinh \sigma  \cosh \chi \sin u\,,\hat{n}\,,\sinh \sigma  \cosh \chi \cos u+\cosh \sigma  \sinh \chi \sin u\right)\,.
    \end{split}
\end{equation}
It is now simple to see that
\begin{equation}
    i\pd_u=L_{-1\,0}-L_{1d}\,,\quad i\pd_{\chi}=L_{-1\, 1}+L_{0 d}\,.
\end{equation}
An elegant way to show this is to define the following null coordinates
\begin{equation}\label{eq_app_null_coord}
\begin{aligned}
\bar{P}^{\pm}_{-1\,1} &=\bar{P}^{-1}\pm \bar{P}^1=\frac{1}{\sqrt{2}}e^{\pm\chi}\sqrt{\cosh 2\sigma}\,\cos\left(u\pm\theta(\sigma)\right)\,,\\
\bar{P}^{\pm}_{0\,d} &=\bar{P}^{0}\pm \bar{P}^d=\frac{1}{\sqrt{2}}e^{\pm\chi}\sqrt{\cosh 2\sigma}\,\sin\left(u\pm\theta(\sigma)\right)\,,
\end{aligned}
\end{equation}
where
\begin{equation}
    \theta(\sigma)\equiv \arctan\left(\tanh\sigma\right)\,.
\end{equation}
The generator $L_{-1\,0}-L_{1d}$ generates $SO(2)$ rotations of the vectors $(\bar{P}^{\pm}_{-1\,1},\bar{P}^{\pm}_{0\,d}) $, therefore shifting $u$, while the generator $L_{-1\, 1}+L_{0 d}$ rescales the $\pm$ sectors oppositely, hence shifting $\chi$.

Finally, we identify the generator of translations in $u$ with the Euclidean twist generator. Note first that in the embedding-space conventions above we have,
\begin{equation}
    H_u\equiv -i\pd_u
    =-(L_{-1\,0}-L_{1d})
    =-\frac{P_0+K_0}{2}
    +\frac{P_1-K_1}{2}\,.
\end{equation}
It is natural then to consider the north-south quantization generated by $-(P_0+K_0)/2$. This arises naturally for instance if we embed Minkowski space in a patch of the Lorentzian cylinder and foliate spacetime according to the natural cylinder time coordinate. More precisely, we can use the following well known isomorphism between the Euclidean and the Lorentzian conformal algebra:
\begin{gather}
\nonumber
D^E=  -\frac{P_0+K_0}{2}\,, \quad
P_0^E=-\frac{P_0-K_0}{2}+i D\,,\quad
K_0^E=-\frac{P_0-K_0}{2}-i D\,,\\
P_i^E=-i\frac{P_i+K_i}{2}-M_{0i}\,,\quad
K_i^E=i\frac{P_i+K_i}{2}-M_{0i}\,,\quad
M_{0i}^E=-i\frac{K_i-P_i}{2}\,,\quad
M_{ij}^E= -iM_{ij}\,,
\end{gather}
where, given that the Lorentzian generators are unitary, the Euclidean generators satisfy the following Hermiticity conditions
\begin{equation}
    (D^E)^\dagger = D^E\,,\quad
    (P_\mu^E)^\dagger=K^E_\mu\,,\quad
    (M_{\mu\nu}^E)^\dagger = - M^E_{\mu\nu}\,.
\end{equation}
Using the above map, we see that $H_u$ measures indeed the twist spectrum
\begin{equation}\label{eq_app_Hu}
    H_u
    =
    D^E-iM_{01}^E\,.
\end{equation}

\section{Spectral Representation of the Large-Boost Interaction}
\label{app_spectral_representation}

In this appendix we interpret the defect interaction \eqref{eq_Gamma_L_int_2} with spectral methods. The argument relies on two assumptions.  First, we assume that the cusp can be obtained by bringing together two defects which are separately well defined.  Second, we assume that the leading connected interaction has an
extensive large-\(T\) limit of the form described below.

The first assumption states that we can express the cusp correlation function carving out a small circle of radius \(\epsilon\) around the junction, both in Euclidean and Lorentzian signature:
\begin{equation}
 \lim_{L\rightarrow \infty}\lim_{\epsilon\rightarrow0^+}
 \epsilon^{\Delta_{a0}+\Delta_{b0}}
 \left\langle
 \mathcal D_a(-L,-\epsilon)
 \mathcal D_b(\epsilon,L)
 \right\rangle \,.
 \label{eq:app_cusp_from_separated_defects}
\end{equation}
This regularization allows us to obtain the cusp by acting with a conformal transformation, and in particular boosts (or rotations in Euclidean), to a single defect segment. Note that this assumption fails for defects with no endpoint operators connecting them to the trivial line.

\begin{figure}
    \centering
    \includegraphics[width=0.9\linewidth]{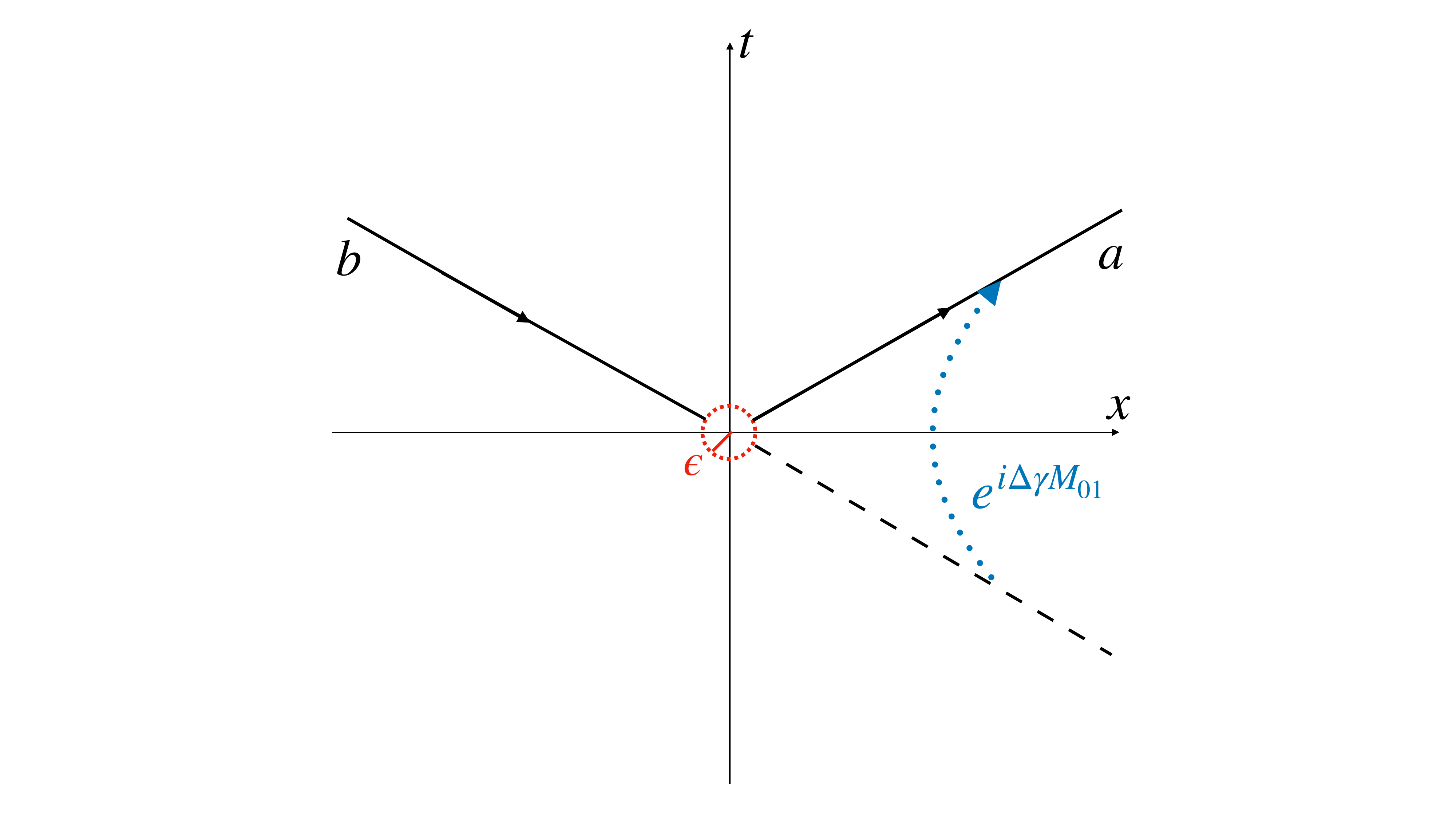}
    \caption{A depiction of the spectral decomposition of the cusp configuration. A small ball separating the defects (red) regulates the cusp as in \eqref{eq:app_cusp_from_separated_defects}, and the cusp is obtained via a boost (in blue) of the defect $a$ compared to $b$, leading to the spectral decomposition in \eqref{eq:app_ads3_completeness}.}
    \label{fig:spectral}
\end{figure}

After the Weyl transformation to the AdS$_3
\times S^{d-3}$ frame in \eqref{eq_AdS3}, we obtain a correlator of two defects with large length \(T\) in the \(\tau\)-direction, separated by \(\Delta\gamma>0\). The cusp anomalous dimension follows from the large $T$ limit of the matrix element
\begin{equation}
      \langle  \mathcal{D}_a(T,\Delta\gamma)\,
    \mathcal{D}_b(T,0)\rangle= \mathcal{Z}_{ab}\,e^{-\Gamma^L_{ab}(y)T}+\ldots\,,
\end{equation}
where $\mathcal{Z}_{ab}$ is a constant, we assumed a gap in the cusp spectrum and neglected terms which are further exponentially suppressed. 

The defect-defect correlator can be expressed by inserting a complete set of states between the defects and isolating the vacuum contribution:
\begin{equation}
 \langle  \mathcal{D}_a(T,\Delta\gamma)\,
    \mathcal{D}_b(T,0)\rangle= 
 \langle 0|\mathcal{D}_a(T,0)|0\rangle\langle 0|
    \mathcal{D}_b(T,0)|0\rangle\left[1+
     \mathcal{M}_{ab}^{(c)} (T,\Delta\gamma)\right]\,,
\end{equation}
where
\begin{equation}
  \langle 0|\mathcal{D}_a(T,0)|0\rangle\propto e^{-\Delta_{a0}T}\,,\qquad
  \mathcal{M}_{ab}^{(c)} (T,\Delta\gamma)=\frac{\langle  \mathcal{D}_a(T,\Delta\gamma)\,
    \mathcal{D}_b(T,0)\rangle_c}{\langle 0|\mathcal{D}_a(T,0)|0\rangle\langle 0|\mathcal{D}_b(T,0)|0\rangle}\,.
\end{equation}
As explained in the main text, on general grounds we expect that for $\Delta\gamma\gg 1$
\begin{equation}\label{eq_W_Vint}
     \mathcal{M}_{ab}^{(c)} (T,\Delta\gamma)=T\, V_{\rm int}(y)+\ldots\,,
\end{equation}
where $V_{\rm int}(y)=O(e^{-\Delta\gamma\,\tau_{\rm exch}})$.

To obtain $V_{\rm int}(y)$, we express the connected correlator by plugging a complete set of states diagonalizing dilations and boosts on the \(\alpha=\pi/4\) Cauchy surface, and obtain the spectral decomposition
\begin{equation}
 \mathcal{M}^{(c)}_{ab}(T,\Delta\gamma)
 =
 \int_{\mathbb R}\frac{dk}{2\pi}\,
 e^{ik\Delta \gamma}\hat{\rho}_{ab,T}(k)\,.
 \label{eq:app_ads3_completeness}
\end{equation}
The variable \(k\) is the eigenvalue of the generator of boosts, i.e. \(\gamma\)-translations.  For different defects \(a\) and \(b\), the spectral density in \eqref{eq:app_ads3_completeness} need not be
positive.

At large \(T\), translation invariance away from the endpoints allows us to resolve also the momentum \(p\) conjugate to \(\tau\):
\begin{equation}
 \hat{\rho}_{ab,T}(k)
 =
 \int_{-T/2}^{T/2}d\tau_1
 \int_{-T/2}^{T/2}d\tau_2
 \int_{\mathbb R}\frac{dp}{2\pi}\,
 e^{ip(\tau_1-\tau_2)}
 \rho_{ab,T}(p,k)\,.
 \label{eq:app_large_T_density}
\end{equation}
This form is immediate, for example, when a defect is constructed from an integrated local source,
\begin{equation}
 \mathcal D_a
 =
\exp\left(
 g_a\int d\tau\,\mathcal O_a(\tau)
 \right)\,.
\end{equation}
At every order in \(g_a\) and \(g_b\), the center-of-mass coordinate of the insertions factors out, while their relative coordinates remain in the matrix element defining the spectral density.  Additionally, by consistency with \eqref{eq_W_Vint} we expect the leading connected exchange $O(e^{-\Delta\gamma\tau_{\rm exch}})$ to admit a \(T\)-independent limit. This is our second assumption.\footnote{At subleading order we also expect contributions scaling as $\sim T^2,\, T^3\,,$ etc., since these are needed to exponentiate the interaction contribution. Physically, these higher powers of $T$ are associated with multi-particle exchanges, and we expect that they arise from the sum over multi-trace operators. Abusing notation, we neglect the $T$-dependence of the spectral density in what follows, implicitly focusing on the leading contribution to the interaction term.}  Given this, introducing \(\Delta\tau=\tau_1-\tau_2\), we obtain
\begin{equation}
 \mathcal{M}^{(c)}_{ab}(T,\Delta\gamma)=T
 \int_{-T}^{T}d\Delta\tau\left(1-\frac{|\Delta\tau|}{T}\right)
 \int_{\mathbb R^2}\frac{dp\,dk}{(2\pi)^2}\,
 e^{ip\Delta\tau+ik\Delta\gamma}\rho_{ab}(p,k)\,.
 \label{eq:app_extensive_spectral_density}
\end{equation}

We now would like to analytically continue the spectral density away from the real axis and deform the contour, so as to express the correlator in terms of the Wick-rotated coordinates \eqref{eq_an_cont}. To this aim, it is convenient to rewrite the original Fourier phase as
\begin{equation}
 e^{ip\Delta\tau+ik\Delta\gamma}
 =
 e^{i\mu\frac{\Delta\tau+\Delta\gamma}{2}
    +i\nu\frac{\Delta\gamma-\Delta\tau}{2}}\,,
    \quad
    \mu=p+k\,,\quad \nu=k-p\,,
 \label{eq:app_phase_continuation}
\end{equation}
and deform the contour of the $\mu$-integral. For \(\Delta\tau>-\Delta\gamma\), the phase decays exponentially for $\Im \mu>0$ and we can thus deform the contour into the upper half-plane. Conversely, for \(\Delta\tau<-\Delta\gamma\), we deform the contour into the lower half-plane $\Im \mu<0$. Since \(H_u\) measures the twist spectrum, which is discrete, the spectral density will have poles in the upper and lower half-plane corresponding to its discrete eigenvalues. We denote $h\equiv \Delta-J$ the eigenvalues of the collinear twist operator $H_u$. The momentum \(\nu\) conjugate to \(\chi\) remains continuous instead. We therefore obtain the following representation
\begin{multline}
\int_{\mathbb{R}^2}\frac{dp\,dk}{(2\pi)^2}\,
 e^{ip\Delta\tau+ik\Delta\gamma}\rho_{ab}(p,k)=
 \sum_{h>0}\int_{\mathbb R}\frac{d\nu}{2\pi}
 \bigg[
 \Theta(\Delta\tau+\Delta\gamma)\bar{\rho}_{ab,h}(\nu)
 e^{-\frac h2(\Delta\tau+\Delta\gamma)+\frac{i\nu}{2}(\Delta\gamma-\Delta\tau)}
 \\
 +
 \Theta(-\Delta\tau-\Delta\gamma)\bar{\rho}_{ba,h}(\nu)
 e^{\frac h2(\Delta\tau+\Delta\gamma)-\frac{i\nu}{2}(\Delta\gamma-\Delta\tau)}
 \bigg]\,,
 \label{eq:app_ordered_spectral_kernel}
\end{multline}
where the spectral densities $\bar{\rho}_{ab,h}(\nu)$ and $\bar{\rho}_{ba,h}(\nu)$ are obtained from the poles of $\rho_{ab}(p,k)$. Note that we could not have derived \eqref{eq:app_ordered_spectral_kernel} directly from the correlator in the space \eqref{eq_uc_metric}, since the defect operators are interlaced by the necessary Euclidean $\Im u$ time ordering. 

The physical meaning of \eqref{eq:app_ordered_spectral_kernel} is that the interaction between the defects is mediated by the propagation of \emph{particles} of mass $h$ and momentum $\nu$, which can be emitted and absorbed at any point on the defects $a$ and $b$. The factor $e^{-\frac h2(\Delta\tau+\Delta\gamma)+\frac{i\nu}{2}(\Delta\gamma-\Delta\tau)}$ is nothing but the propagation phase from $a$ to $b$, and vice versa $e^{\frac h2(\Delta\tau+\Delta\gamma)-\frac{i\nu}{2}(\Delta\gamma-\Delta\tau)}$ the phase associated with going from $b$ to $a$. Note that the time ordering in $\Im u$ specifies a direction in the propagation of the (off-shell) particle, so that the associated phase decays.

Taking the $\Delta\tau$-integral and the limit \(T\rightarrow\infty\), we finally find
\begin{equation}
V_{\rm int}(y)
 \simeq
 2\sum_{h>0}\int_{\mathbb R}\frac{d\nu}{2\pi}
 \left[
 \frac{e^{i\nu \Delta\gamma}\bar{\rho}_{ab,h}(\nu)}{h+i\nu}
 +
 \frac{e^{-i\nu \Delta\gamma}\bar{\rho}_{ba,h}(\nu)}{h+i\nu}
 \right]\,.
 \label{eq:app_integrated_spectral_kernel}
\end{equation}
To evaluate the \(\nu\) integral we deform the contour once again. Since \(\Delta\gamma>0\), the contour in the first term is deformed into the upper half-plane, while that in the second term is deformed into the lower half-plane. In particular, the denominator in the first ordering has a kinematical pole at
\begin{equation}
    \nu=ih\,.
\end{equation}
Additionally, we note that the double Wick rotation exchanges the \(u\)- and \(\chi\)-quantization axes. This operation exchanges the two collinear weights
\begin{equation}
    h_-=\Delta-J\,,
    \qquad
    h_+=\Delta+J\,.
\end{equation}
Consequently, the spectral density associated with the \(u\)-energy \(h_-\) has its first \(\chi\)-channel pole at \(\nu=
\pm ih_+\), while the density associated with the companion component of \(u\)-energy \(h_+\) has its first pole at \(\nu=\pm ih_-\).

At this point it is convenient to distinguish two cases. For \(J>0\), one has \(h_-<h_+\). In the term associated with the component of mass \(h_-\), the first upper-half-plane singularity is the kinematical pole $\nu=ih_-$, while the first pole of its spectral density lies farther away, at \(\nu=ih_+\). Conversely, for the component of mass \(h_+\), the spectral density already has a pole at \(\nu=ih_-\), whereas its kinematical pole is at \(\nu=ih_+\). The two mechanisms therefore give contributions with the same leading exponential behavior,
\begin{equation}
    V_{\rm int}(y)
    = C_{ab}
    e^{-(\Delta-J)\Delta\gamma}
    +\ldots,
    \qquad J>0\,,
\end{equation}
where
\begin{equation}
    C_{ab}
    =
    2\,\bar{\rho}_{ab,h_-}(ih_-)+
    \frac{i}{J}
    \underset{\nu=ih_-}{\operatorname{Res}}\,
    \bar{\rho}_{ab,h_+}(\nu)\,.
\end{equation}

For a scalar, instead, the two collinear weights coincide:
\begin{equation}
    h_-=h_+=\Delta\,.
\end{equation}
The first spectral pole then collides with the kinematical pole of the same integrand. We expand the spectral density in the vicinity of the pole as
\begin{equation}
    \bar{\rho}_{ab,\Delta}(\nu)
    =
    \frac{R_{ab}^{+}}{\nu-i\Delta}
    +F_{ab}^{+}
    +O(\nu-i\Delta)\,.
\end{equation}
The collision then produces a double pole, whose residue gives
\begin{equation}
    V_{\rm int}(y)
    =
    e^{-\Delta\Delta\gamma}
    \left[
        A_{ab}\Delta\gamma+B_{ab}
    \right]
    +\ldots\,,
\end{equation}
with
\begin{equation}
    A_{ab}=2iR_{ab}^{+}\,,
    \qquad
    B_{ab}
    =
    2F_{ab}^{+}
    -\frac{i}{\Delta}\,
    \underset{\nu=-i\Delta}{\operatorname{Res}}\,
    \bar{\rho}_{ba,\Delta}(\nu)\,.
\end{equation}
Here the last term in \(B_{ab}\) is the simple-pole contribution from the opposite ordering. This concludes our derivation of \eqref{eq_Gamma_L_int_2}.

Let us finally note that the states in the $u$-quantization, which were used to interpret the spectral decomposition in \eqref{eq:app_ordered_spectral_kernel}, correspond to local operators inserted at \(\Im u\to\pm\infty\) and \(\sigma=0\). In the original flat-space coordinates, these map to operator insertions on the null lines \(t=\mp x\), as follows from \eqref{eq_app_P_wick} and \eqref{eq_app_P_u_c}. This observation may allow our arguments to be reformulated in terms of the OPE of the defect endpoints, perhaps along the lines of \cite{Alday:2010ku}.

\section{Large Spin Operators Review}\label{app_Large_S}

Here we review the arguments of~\cite{Alday:2007mf} to demonstrate that large spin double-trace operators behave as composites of \emph{free} partons. Along the way, we clarify the role of the quantum-mechanical nature of the states, also emphasized from the bootstrap viewpoint in~\cite{Komargodski:2012ek,Fitzpatrick:2012yx}. We focus on $d=3$ spacetime dimensions for concreteness.

Let us consider the state on $\mathbb{R}\times S^{2}$ corresponding to a primary operator of the form $\sim \sum_k c_k \pd^{k}\phi \pd^{J-k}\phi$ with $J\gg 1$, where $\phi$ is a (gauge invariant) primary operator, that we take to be scalar for simplicity. At finite but large spin we expect this state to be well approximated by the product of free wavefunctions for the descendant states of the \emph{parton} operator $\phi$. Indeed the wave-function of a large spin descendant state on the cylinder reads 
\begin{equation}\label{eq_wave_functions_largeJ}
\langle 0|\phi(t,\theta,\phi)|\pd_z^{\ell}\phi\rangle\propto e^{-i(\Delta_{\phi}+\ell)\tau}e^{i\ell\phi}\sin^\ell\theta
\stackrel{\ell\gg 1}{\simeq}e^{-i\Delta_{\phi} \frac{\tau+\phi}{2}}e^{-i \ell\frac{\tau-\phi}{2}} e^{-\frac12\ell\left(\theta-\frac{\pi}{2}\right)^2}\,,
\end{equation}
where the cylinder metric is the natural one
\begin{equation}
ds^2_{\mathbb{R}\times S^2}=-d\tau^2+d\theta^2+\sin^2\theta d\phi^2\,.
\end{equation}
We see that for $\ell\sim J$ the wave-function is exponentially localized in a range $\delta\theta\sim 1/\sqrt{J}$ around $\theta=\pi/2$. Additionally, the fast phase factor $e^{-i \ell\frac{\tau-\phi}{2}}$ also implies that fluctuations in $\tau-\phi$ are $1/J$ suppressed. Since most of the components of the primary state $\sim \sum_k c_k \pd^{k}\phi \pd^{J-k}\phi$ have $k\sim J\gg 1$, this makes the individual partons localized and the state corresponding to this double-trace operator semiclassical.

We conclude therefore that in the infinite spin limit we can describe the states in terms of two classical trajectories given by (up to the obvious rotational zero-mode)
\begin{equation}\label{eq_classical}
\theta_1=\theta_2=\frac{\pi}{2}\,,\qquad
\phi_1=\tau\,,\qquad
\phi_2=\tau+\pi\,.
\end{equation}
Similarly to the null cusp, the trajectories~\eqref{eq_classical} are invariant under two important conformal symmetries. To identify these symmetries, it is convenient to consider the trajectories in embedding space:
\begin{align}
    P_1 =(\cos\tau,\sin\tau,\cos\tau,\sin\tau,1)\,,\qquad
 P_2 = (\cos\tau,\sin\tau,-\cos\tau,-\sin\tau,1)\,.
\end{align}
The first symmetry is obvious and amounts to a simultaneous rotation of $(P^{-1},P^0)$ and $(P^{1},P^2)$, which just shifts $\tau\rightarrow\tau +c$. This is isomorphic to the twist generator $D^E-i M_{01}^E$. The second symmetry is less intuitive in flat space, but can be identified as a simultaneous boost in the $(P^{-1},P^1)$ and $(P^{0},P^2)$ directions in embedding space. This acts by rescaling the coordinates and is an isometry due to the gauge redundancy $P\sim \lambda P$ of the projective null cone~\eqref{eq_app_projective_null}. Analogously to the boost symmetry of the null cusp, this isometry is slightly broken by the spread of the wave-function in the $\theta$ direction at finite spin. As in the case of the cusp, it is therefore convenient to analyze the double-trace state in a Weyl frame where this conformal symmetry becomes a manifest isometry.

It turns out that one such convenient frame is given by (two copies of) AdS$_3$, which will make clear that interactions die off very fast near the equator, in an appropriate sense.
Let us recall first that we can Weyl map the cylinder to two copies of AdS$_3$, connected at their asymptotic boundary, via
\begin{equation}
ds^2_{\mathbb{R}\times S^2} =\cos^2\theta\left(\frac{-d\tau^2+d\theta^2+\sin^2\theta d\phi^2}{\cos^2\theta}\right)\equiv \cos^2\theta ds^2_{\text{AdS}_3}\,,
\end{equation}
where
\begin{equation}\label{eq_AdS3_metric1}
ds^2_{\text{AdS}_3}=-\cosh^2\rho d\tau^2+\sinh^2\rho\, d\phi^2+d\rho^2\,,\qquad
\sinh\rho=\begin{cases}
\frac{1}{\cot\theta} & 0<\theta<\frac{\pi}{2}\\
-\frac{1}{\cot\theta} & \frac{\pi}{2}<\theta<\pi\,.
\end{cases}
\end{equation}
We see that the two particles are localized near the AdS boundary. Their extension into the bulk $\rho_0$ is determined by the above change of coordinates, which near $\theta=\pi/2$ reads
\begin{equation}\label{eq_rho_J}
\sin^2\theta=\tanh^2\rho\quad\implies\quad \delta\theta^2\simeq 4 e^{-2\rho_0}\quad\implies\quad
\rho_0\sim \frac12\log J\,.
\end{equation}
We see in particular that as $J\rightarrow\infty$, the wave-functions localize at the AdS boundary $\rho\rightarrow\infty$. This already hints at weak interactions between the two partons.

To make the above symmetries of the trajectories~\eqref{eq_classical} manifest,~\cite{Alday:2007mf} introduced the coordinates $(u,\chi,\sigma)$ discussed around~\eqref{eq_an_cont} in the main text. We recall that AdS$_3$ is the $SL(2,\mathbb{R})_L\times SL(2,\mathbb{R})_R$ homogeneous manifold, and hence given an $SL(2,\mathbb{R})$ representative parametrization $g$, we can obtain the metric from $ds^2=\text{Tr}(dg dg^{-1})$. The  relation between~\eqref{eq_AdS3_metric1} and the new coordinates is specified by
\begin{equation}
g=e^{i\sigma_2\frac{\tau+\phi-\pi/2}{2}}e^{\rho\sigma_3}e^{i\sigma_2\frac{\tau-\phi+\pi/2}{2}}=e^{i u\sigma_2}e^{\sigma\sigma_3}e^{\chi\sigma_1}\,.
\end{equation}
To leading order in the spin $J$ according to the scalings mentioned below~\eqref{eq_wave_functions_largeJ}, the relation simplifies to 
\begin{equation}\label{eq_rel_exp}
u\simeq \frac{\tau+\phi}{2}\,,\qquad
\chi\simeq \rho\,,\qquad
\sinh 2\sigma\simeq -(\tau-\phi)e^{2\rho}\,,
\end{equation}
Note that $(\tau-\phi)e^{2\rho}\sim O(1)$. The metric reads
\begin{equation}\label{eq_magic_metric}
ds^2_{\text{AdS}_3}=-du^2+d\chi^2-2\sinh 2\sigma du d\chi+d\sigma^2\,.
\end{equation}
As both the metric and the representative $g$ make clear, the shifts $u\rightarrow u+c_u$ and $\chi\rightarrow\chi+c_{\chi}$ are (left and right) isometries. In particular, the relation~\eqref{eq_app_Hu} derived in the previous section shows that the Hamiltonian $H_u$ measures the twist spectrum. Shifts in $\chi$ correspond instead to the second isometry mentioned at the beginning of this section. The metric~\eqref{eq_magic_metric} also explains geometrically why fluctuations in $(\tau-\phi)$ are $\sim 1/J$ suppressed in this geometry in terms of the warp factor $\sim e^{2\sigma} du d\chi$, which prevents the wave-function from spreading towards $\sigma\gg 1$ - as also~\eqref{eq_rel_exp} shows.

To compute the dimension of the double-trace operator, we finally note that $u$ and $\chi$ are exchanged under analytic continuation $u\rightarrow \pm i u$ and $\chi\rightarrow\mp i \chi$. This implies that when quantizing in the Euclidean $\chi$-direction the spectrum is the same as when we quantize in the natural $u$ time-coordinate. Since $H_u$ is associated to the twist, we conclude that the energy of the state is well approximated by the free partons' contributions up to terms which are exponentially suppressed by the twist $\tau_{\rm exch}$ of the lowest twist state that couples to $\phi$, that we assume to be positive. On general grounds, we expect the interaction potential to scale as $\sim e^{-2\tau_{\rm exch}\Delta\chi}\sim e^{-2\tau_{\rm exch}\rho_0}$, and thus using~\eqref{eq_rho_J} we conclude
\begin{equation}\label{eq_double_tr_energy}
\Delta\simeq E_{\rm free}-\frac{\lambda_{\rm int}}{J^{\tau_{\rm exch}}} \,,\qquad
E_{\rm free}=2\Delta_{\phi}+J\,.
\end{equation}
Note that to compute the coefficient $\lambda_{\rm int}$ in general we need to take into account both the spread of the partons' wave-functions as well as the detailed form of the potential. In many cases, more precise results can be obtained from first principles using the analytic bootstrap~\cite{Fitzpatrick:2012yx,Komargodski:2012ek} and the inversion formula~\cite{Caron-Huot:2017vep}. A generalization of this argument also predicts the universality of the dimensions of large transverse spin operators $\sim \pd_{\bot}^J\phi$ in defect CFTs~\cite{Lemos:2017vnx}

As discussed at length in~\cite{Alday:2007mf}, in gauge theories one may also consider single trace operators of the form $\sim \sum_k c_k\text{Tr}\left[D^k\Phi D^{J-k}\Phi\right]$. In this case the partons couple to the gauge-field, which creates a string in AdS$_3$. This leads to a result $\Delta-J\propto T\Delta\chi\sim T \log J$, where $T$ is the string tension.\footnote{Strictly speaking this result holds in perturbation theory, where $T\ll 1$, or in the planar limit. In non-planar models when $T\log J\gtrsim \Delta_{\mO}$, where $\Delta_{\mO}$ is the dimension of the lightest neutral single-trace operator, we expect string breaking and the onset of a different behavior described by multi-trace states.} This configuration can be related by analytic continuation to the cusp anomalous dimension~\cite{Alday:2010ku}.

Finally, it should be said that the above physical arguments fall short of providing an effective field theory framework for systematically computing the anomalous dimension of large spin operators. For instance, we did not provide a complete systematic recipe to write down the interactions between the partons compatibly with conformal invariance. This might be particularly useful for multi-trace states, which are harder to analyze within the bootstrap. Recent progress in this direction exploits a fictitious emergent AdS$_{4}$ dual~\cite{Fardelli:2024heb,Kravchuk:2024wmv,Fardelli:2025fkn,Fardelli:2025eun}. It would be interesting to see if these works could be related more explicitly to the AdS$_3$ frame~\eqref{eq_magic_metric} introduced in~\cite{Alday:2007mf}.

\section{Large Boost Limit of \texorpdfstring{$I_{12}(\cos\theta)$}{I12}}\label{app_I12}

Here we discuss the large boost limit of the function $I_{12}(\cos\theta)$ defined in \eqref{eq_I12}. We set
\begin{equation}
\theta=\pi-i\log y,
\qquad
\cos\theta=-\frac12\left(y+y^{-1}\right),
\end{equation}
and choose
\begin{equation}
    \hat n_1=(1,0,0,0)\,,\quad \hat{n}_2=(\cos \theta\,,\sin\theta,0,0)=\left(-\frac{y}{2}-\frac{1}{2 y},\frac{i y}{2}-\frac{i}{2 y},0,0\right)\,.
\end{equation}
We parameterize $S^3$ as
\begin{equation}
\hat n=
\left(
t,v,\sqrt{1-t^2-v^2}\cos\varphi,
\sqrt{1-t^2-v^2}\sin\varphi
\right),
\qquad
t^2+v^2\leq1.
\end{equation}
Defining \(q=-t+iv\), one has
\begin{equation}
d\Omega_3=dt\,dv\,d\varphi,
\qquad
\hat n\cdot\hat n_2
=\frac12\left(yq+y^{-1}\bar q\right),
\end{equation}
and therefore
\begin{equation}
I_{12}
=
2\pi\int_Ddt\,dv\,f_4^2(t)
f_4\left(\frac{yq+y^{-1}\bar q}{2}\right)^2,
\qquad
D=\{(t,v):t^2+v^2\leq1\}.
\end{equation}

To analyze the large $y$ limit, we use the method of matched asymptotic expansions. We introduce a matching scale \(y^{-1}\ll\epsilon\ll1\) and split the integral into the regions \(|q|>\epsilon\) and \(|q|<\epsilon\). In the
outer region,
\begin{equation}
f_4\left(\frac{yq+y^{-1}\bar q}{2}\right)
=
-\frac{2}{yq}\left[\log y+\log(-q)\right]+\cdots,
\end{equation}
so that
\begin{equation}
I_{\mathrm{out}}(\epsilon)
=
\frac{8\pi}{y^2}
\int_{D\setminus B_\epsilon}dt\,dv\,
\frac{f_4^2(t)}{q^2}
\left[\log y+\log(-q)\right]^2+\cdots.
\end{equation}
At finite \(\epsilon\), we find
\begin{align}
\int_{D\setminus B_\epsilon}dt\,dv\,
\frac{f_4^2(t)}{q^2}
&=
2\int_{-1}^1dt\,\sqrt{1-t^2}f_4^2(t)
-\frac{2}{\epsilon^2}
\int_{-\epsilon}^{\epsilon}dt\,
f_4^2(t)\sqrt{\epsilon^2-t^2}
\nonumber\\
&=
2\int_{-1}^1dt\,\sqrt{1-t^2}f_4^2(t) 
-\pi f_4(0)+O(\epsilon^2)\,.
\label{eq:A0-epsilon}
\end{align}
To evaluate the remaining integral, set
\begin{equation}
t=\cos\alpha\,,\quad
dt=-\sin\alpha\,d\alpha\,,
\end{equation}
with $\alpha\in[0,\pi]$. Since
\begin{equation}
f_4(\cos\alpha)=\frac{\pi-\alpha}{\sin\alpha},
\end{equation}
we obtain
\begin{align}
\int_{-1}^1dt\,\sqrt{1-t^2}\,f_4^2(t)=
\int_0^\pi d\alpha\,(\pi-\alpha)^2
=\frac{\pi^3}{3}\,.
\end{align}
Using then that $t=0$ corresponds to $\alpha=\pi/2$, and therefore
\begin{equation}
f_4(0)
=
\frac{\pi}{2}\,,
\end{equation}
we obtain
\begin{align}
I_{\mathrm{out}}(\epsilon)
=\frac{10 \pi ^4}{3 y^2}\log^2 y
\left[
1+O(\epsilon^2)\right]+O\left(\frac{\log y}{y^2}\right)\,.
\end{align}

In the inner region we rescale \(z=yq\) and expand for large $y$, obtaining
\begin{equation}
I_{\mathrm{in}}(\epsilon)
=
\frac{2\pi f_4^2(0)}{y^2}
\int_{|z|<y\epsilon}d^2z\,
f_4^2\left(\frac z2\right)+\cdots\,.
\end{equation}
To extract the large-$y$ behavior, set
\begin{equation}
R=y\epsilon
\end{equation}
and choose a fixed $R_0>2$. The integral over $|z|<R_0$ is finite and does not affect the large-$R$ logarithms. For $z=re^{i\phi}$ with $r\gg1$, one has
\begin{equation}
f_4\left(\frac z2\right)
=
-\frac{2}{z}\log(1-z)
+O\left(\frac{\log|z|}{|z|^2}\right).
\end{equation}
Therefore,
\begin{align}
\int_{|z|<R}d^2z\,f_4^2\left(\frac z2\right)
&=
4\int_0^{2\pi}d\phi\,e^{-2i\phi}
\int_{R_0}^{R}\frac{dr}{r}\,
\log^2(1-re^{i\phi})
+O(1)\,.
\end{align}
For $0<\phi<2\pi$, we have
\begin{equation}
\log(1-re^{i\phi})
=
\log r+i(\phi-\pi)+O(r^{-1})\,.
\end{equation}
The radial integral gives
\begin{align}
\int_{R_0}^{R}\frac{dr}{r}\,
\log^2(1-re^{i\phi})
=
\frac13\log^3R
+i(\phi-\pi)\log^2R-\frac12(\phi-\pi)^2\log R
+O(1)\,.
\end{align}
The cubic logarithm vanishes after angular integration because
\begin{equation}
\int_0^{2\pi}d\phi\,e^{-2i\phi}=0\,.
\end{equation}
The quadratic logarithm is determined by
\begin{equation}
\int_0^{2\pi}d\phi\,
e^{-2i\phi}i(\phi-\pi)=-\pi\,.
\end{equation}
It follows that
\begin{equation}
\int_{|z|<y\epsilon}d^2z\,
f_4^2\left(\frac z2\right)
=
-4\pi\log^2 y
+O\left(\log y\right).
\end{equation}
Using $f_4(0)=\pi/2$, we therefore obtain
\begin{align}
I_{\mathrm{in}}(\epsilon)
=
-\frac{2\pi^4}{y^2}
\log^2y+O\left(\frac{\log y}{y^2}\right)\,.
\end{align}

Putting everything together we arrive at
\begin{equation}
I_{12}\left(-\frac{y+y^{-1}}{2}\right)
=
\frac{4\pi^4}{3}\frac{\log^2y}{y^2}
+O\left(\frac{\log y}{y^2}\right)\,.
\end{equation}
We verified this asymptotic behaviour numerically.

\section{Technical Details of the Spin Impurity Cusp Calculation}\label{app_technical_spin}

In this appendix we describe in detail the calculations leading to~\eqref{eq_cusp_ds_1_loop} and~\eqref{eq_dim_ds_1_loop}. We follow the conventions in~\cite{Cuomo:2022xgw}. In particular we define the bare coupling as
\begin{equation}
   \alpha_0\equiv\frac{\gamma_0^2 s}{(d-2)\Omega_{d-1}}\,,
\end{equation}
which is renormalized according to
\begin{equation}\label{eq_alpha_ct}
\alpha_0=\frac{4\pi^2M^{\varepsilon}}{(2-\varepsilon)\Omega_{3-\varepsilon}}\left(\alpha+
\frac{\delta\alpha}{\varepsilon}
\right)\,,\qquad
\delta\alpha=
\frac{2\alpha  \arctan(\pi  \alpha)}{\pi s}\,.
\end{equation}
Here $M$ is the sliding scale. The beta-function~\eqref{eq_BETA} is derived by demanding that the bare coupling is independent of $M$.

To compute the subleading corrections to~\eqref{eq_ds_cusp_0} and~\eqref{eq_ds_dim_0} we proceed similarly to appendix B of~\cite{Cuomo:2022xgw}. We go to momentum space and introduce a vector notation as
\begin{equation}
    \chi^a(\tau)=\int\frac{dk}{2\pi}e^{-ik\tau}\chi_k^a\,.\qquad
    (\chi^1,\chi^2)=(\chi\vert_0,\chi\vert_\theta)\,.
\end{equation}
Then we see that the quadratic action becomes
\begin{equation}
    S^{(2)}=\int\frac{dk}{2\pi}\bar{\chi}_k^a M^{ab}(k)\chi^b_k\,,
\end{equation}
where the matrix M is given by
\begin{equation}\label{eq_app_M_spin}
    M(k) =-i\,k\,\sigma^3-\alpha_0\left[\Sigma_{\eps}(k)-\Sigma_{\eps}(0)-F_{\varepsilon,\theta}(0)\right]\mathds{1}-\alpha_0 F_{\varepsilon,\theta}(k)\sigma^1\,,
    \end{equation}
for two defect insertions, and it reduces to a single diagonal component for a single defect. The functions $\Sigma_{\varepsilon}(k)$ and $F_{\varepsilon,\theta}(k)$ are defined as
\begin{align}
    \Sigma_{\varepsilon}(k)&=
    \int d\tau \frac{e^{ik\tau}}{(2\cosh\tau-2)^{1-\frac{\varepsilon}{2}}} \,,\\ 
     F_{\eps,\theta}(k)&=   \int d\tau \frac{e^{ik\tau}}{(2\cosh\tau-2\cos\theta)^{1-\frac{\varepsilon}{2}}}\,.
\end{align}
Diagrammatically, $\Sigma_{\varepsilon}(k)$ is a self-energy contribution from a scalar propagator from one defect to itself, while $F_{\varepsilon,\theta}(k)$ arises from the exchange diagram connecting the two defects. The self-energy contribution can be evaluated explicitly
\begin{equation}
\begin{split}
    \Sigma_{\varepsilon}(k) &=\Gamma (\varepsilon -1) \left[\frac{\Gamma \left(1+i k-\frac{\varepsilon }{2}\right)}{\Gamma \left(i k+\frac{\varepsilon }{2}\right)}+\frac{\Gamma \left(1-i k-\frac{\varepsilon }{2}\right)}{\Gamma \left(\frac{\varepsilon }{2}-i k\right)}\right]\\[0.7em] 
    &=\begin{cases}
       \displaystyle
       % -\frac{  2^{1-\frac{\varepsilon }{2}}\pi \sec \left(\frac{\pi  \varepsilon }{2}\right) }{\Gamma (2-\varepsilon )}
       c_{\varepsilon}| k| ^{1-\varepsilon }\left[1+O\left(\frac{1}{k^2}\right)\right]
        & k\rightarrow\infty \\[0.5em]
           -\pi  k \coth (\pi  k) & \varepsilon=0\,,
    \end{cases}\qquad
    c_{\varepsilon}= -\frac{  \pi \sec \left(\frac{\pi  \varepsilon }{2}\right) }{\Gamma (2-\varepsilon )}\,,
    \end{split}
\end{equation}
where we reported both the large $|k|$ asymptotic behavior and the form at $\varepsilon=0$ for future purposes. The exchange diagram is more complicated, but admits a simple closed form for $\varepsilon=0$:
\begin{equation}
    F_{0,\theta}(k)=\frac{\pi  \sinh ((\pi -\theta ) k)}{\sin (\theta ) \sinh (\pi  k)}\,.
\end{equation}
Additionally, at $k=0$ $F_{\varepsilon,\theta}(k)$ coincides with the previously defined function~\eqref{eq_fd_def}
\begin{equation}
    F_{\varepsilon,\theta}(0)= f_{4-\varepsilon}(\cos\theta)\,.
\end{equation}

In writing the quadratic action~\eqref{eq_app_M_spin} we have been careless about ordering issues, which would require introducing a small point-splitting regulator between $\bar{\chi}$ and $\chi$~\cite{Cuomo:2022xgw}. We checked that the effect of this regulator is simply to shift $s\rightarrow s+1$ in the leading contributions~\eqref{eq_ds_cusp_0} and~\eqref{eq_ds_dim_0} as in appendix B of~\cite{Cuomo:2022xgw}. For simplicity of presentation therefore, in what follows we implement this change by hand and neglect this subtlety.

Performing the path-integral for the fluctuations and dividing by the partition function of the free defect (i.e. with $\alpha=0$) for convenience, we find
\begin{equation}
 \frac{1}{T}\log\frac{Z_{ss}(\theta)}{Z_{ss}\vert_{\alpha=0}}=(s+1)\alpha_0\left[\Sigma_{\varepsilon}(0)+F_{\varepsilon,\theta}(0)  \right]-\int\frac{dk}{2\pi}\text{Tr}\log \frac{M(k)}{M_{\alpha=0}(k)}+O\left(\frac{1}{s}\right)\,,
\end{equation}
where $T$ is the defect extension.
Diagonalizing the matrix $M(k)$ we can write explicitly the one-loop determinant as 
\begin{equation}\label{eq_app_one_loop_det}
    \int\frac{dk}{2\pi}\text{Tr}\log \frac{M(k)}{M_{\alpha=0}(k)}=2\int_0^\infty\frac{dk}{2\pi}\log\left(1+\alpha_0^2\frac{\left[\Sigma_{\eps}(k)-\Sigma_{\eps}(0)-F_{\eps,\theta}(0)\right]^2-F^2_{\eps,\theta}(k)}{k^2}\right) \,.
\end{equation}

To compute~\eqref{eq_app_one_loop_det}, we need to isolate its divergent part at $\varepsilon=0$. This arises from the first two terms in the Laurent expansion of the integrand
\begin{multline}\label{eq_app_Laurent}
\log\left(1+\alpha_0^2\frac{\left[\Sigma_{\eps}(k)-\Sigma_{\eps}(0)-F_{\eps,\theta}(0)\right]^2-F^2_{\eps,\theta}(k)}{k^2}\right)\\
=\log\left(1+\alpha_0^2c_{\varepsilon}^2 k^{-2\varepsilon}\right)-2\alpha_0^2 c_{\varepsilon}\left[\Sigma_{\eps}(0)+F_{\eps,\theta}(0)\right]\frac{k^{-1+\varepsilon}}{k^{2\varepsilon}+\alpha_0^2 c_{\varepsilon}^2}+O\left(\frac{1}{k^2}\right)\,.
    \end{multline}
To compute~\eqref{eq_app_one_loop_det}, we separate the integral into two regions $k\in (0,\Lambda)\cup (\Lambda,\infty)$ where $\Lambda>0$ is arbitrary. From the region $k\in (\Lambda,\infty)$ we then subtract the first two terms of the Laurent expansion~\eqref{eq_app_Laurent}, to obtain a manifestly finite integral which may be evaluated directly in $d=4$. Using then
\begin{align}
    2\int_\Lambda^{\infty} \frac{dk}{2\pi}
\log\left(1+\alpha_0^2c_{\varepsilon}^2 k^{-2\varepsilon}\right)&=-\frac{\Lambda}{\pi} \log\left(1+\alpha^2_0c_{\varepsilon}^2\right)+O\left(\varepsilon\right)\,,\\[0.4em]
2\int_\Lambda^{\infty} \frac{dk}{2\pi}
\frac{-2\alpha^2_0 c_{\varepsilon}\,k^{-1+\varepsilon}}{k^{2\varepsilon}+\alpha^2_0 c_{\varepsilon}^2}&=
\frac{2\alpha_0}{-\pi\varepsilon}\arctan\left(c_{\varepsilon}\alpha_0\Lambda^{-\varepsilon}\right)
\,,
\end{align}
and expanding the bare coupling with~\eqref{eq_alpha_ct},
we recast the one-loop determinant as
\begin{equation}\label{eq_app_1loop_det_ss}
\begin{split}
    \int\frac{dk}{2\pi}\text{Tr}\log \frac{M(k)}{M_{\alpha=0}(k)}&=
    \frac{2 \alpha  \arctan(\pi  \alpha )}{\pi  \varepsilon }\left[\Sigma_{\varepsilon}(0)+F_{\varepsilon,\theta}(0)\right]
    +I^{(ss)}_{\alpha}(\theta)
    \\[0.5em]
    &+\log \left( \pi  M^2e^{\gamma_E }\right)\left[\frac{\alpha ^2 }{1+\pi ^2 \alpha ^2}+\frac{\alpha  \arctan(\pi  \alpha )}{\pi } 
    \right]
    \left[\Sigma_{0}(0)+F_{0,\theta}(0)\right]\\[0.5em]
    &+O\left(\varepsilon,\frac{1}{s}\right)\,.
    \end{split}
\end{equation}
Here $\gamma_E$ is the Euler--Mascheroni constant and we defined
\begin{align}\nonumber
    I^{(ss)}_{\alpha}(\theta)&=2\int_{0}^{\infty} \frac{dk}{2\pi} G^{(ss)}_{\theta,\alpha,\Lambda}(k)
    -\frac{\Lambda  \log \left(1+\pi ^2 \alpha ^2\right)}{\pi }
    -\frac{\alpha ^2 \log \left(\Lambda ^2 e^{2 \gamma_E -2}\right)}{1+\pi^2\alpha^2}
    \left[\Sigma_{0}(0)+F_{0,\theta}(0)\right]\,,
    \\[0.4em]
    \nonumber
    G^{(ss)}_{\theta,\alpha,\Lambda}(k)&=
\log\left(1+\alpha^2\frac{\left[\Sigma_{0}(k)-\Sigma_{0}(0)-F_{0,\theta}(0)\right]^2-F^2_{0,\theta}(k)}{k^2}\right)\\[0.4em]
&-\Theta(k-\Lambda)\left\{\log\left(1+\alpha^2\pi^2 \right)+ \frac{2\pi \alpha^2}{(1+\alpha^2\pi^2)k}\left[\Sigma_{0}(0)+F_{0,\theta}(0)\right]\right\}\,.
    \end{align}
The function $I^{(ss)}_{\alpha}(\theta)$ can be straightforwardly evaluated numerically for arbitrary values of $\alpha$ and $\theta$; we checked that the result is independent of the arbitrary scale $\Lambda$. 

We obtain the result for a single defect insertion dividing by $2$ and setting the exchange contribution $F_{\varepsilon,\theta}(k)$ to $ 0$:
\begin{equation}
 \frac{1}{T}\log\frac{Z_{s0}}{Z_{s0}\vert_{\alpha=0}}=(s+1)\frac{\alpha_0}{2}\Sigma_{\varepsilon}(0) -\int\frac{dk}{2\pi}\text{Tr}\log \frac{M^{11}(k)}{M^{11}_{\alpha=0}(k)}+O\left(\frac{1}{s}\right)\,,
\end{equation} 
\begin{equation}\label{eq_app_1loop_det_s0}
\begin{split}
    \int\frac{dk}{2\pi}\text{Tr}\log \frac{M^{11}(k)}{M^{11}_{\alpha=0}(k)}&=
    \frac{\alpha  \arctan(\pi  \alpha )}{\pi  \varepsilon }\Sigma_{\varepsilon}(0)
    +I^{(s0)}_{\alpha}
    \\[0.5em]
    &+\log \left( \pi  M^2e^{\gamma_E }\right)\left[\frac{\alpha ^2 }{2+2\pi ^2 \alpha ^2}+\frac{\alpha  \arctan(\pi  \alpha ) }{2\pi }
    \right]
   \Sigma_{0}(0)\\
   &+O\left(\varepsilon,\frac{1}{s}\right)\,,
    \end{split}
\end{equation}
where
\begin{align}\nonumber
    I^{(s0)}_{\alpha}&=\int_{0}^{\infty} \frac{dk}{2\pi} G^{(s0)}_{\alpha,\Lambda}(k)
    -\frac{\Lambda  \log \left(1+\pi ^2 \alpha ^2\right)}{2\pi }
    -\frac{\alpha ^2 \log \left(\Lambda ^2 e^{2\gamma_E-2}\right)}{2+2\pi^2\alpha^2}
    \Sigma_{0}(0)\,,\\[0.5em] 
    \nonumber
    G^{(s0)}_{\alpha,\Lambda}(k)&=
\log\left(1+\alpha^2\frac{\left[\Sigma_{0}(k)-\Sigma_{0}(0)\right]^2}{k^2}\right)-\Theta(k-\Lambda)\left[\log\left(1+\alpha^2\pi^2 \right)+ \frac{2\pi \alpha^2}{(1+\alpha^2\pi^2)k}\Sigma_{0}(0)\right]\,.
    \end{align}

Putting everything together and using the relation~\eqref{eq_alpha_ct} between the bare and the renormalized coupling, we find the following final results for the partition functions
\begin{align}\nonumber
 \frac{1}{T}   \log\frac{Z_{ss}(\theta)}{Z_{ss}\vert_{\alpha=0}}&=s\alpha\left[\Sigma_{\varepsilon}(0)+F_{\varepsilon,\theta}(0)\right]\left[1+\frac{
 \varepsilon}{2} \log \left(\pi   M^2e^{\gamma_E}\right)
+O\left(\varepsilon^2\right)\right]\\
\label{eq_app_Zss_ds}
&+
\alpha\left[1-\frac{\alpha  \log \left(\pi   M^2e^{\gamma_E}\right)}{1+\pi ^2 \alpha ^2}\right]\left[\Sigma_{0}(0)+F_{0,\theta}(0)\right]-I^{(ss)}_{\alpha}(\theta)+O\left(\varepsilon,\frac{1}{s}\right)\,,\\
\nonumber
  \frac{1}{T}  \log\frac{Z_{s0}}{Z_{s0}\vert_{\alpha=0}}&=\frac{s\alpha}{2}\Sigma_{\varepsilon}(0)\left[1+\frac{
 \varepsilon}{2} \log \left(\pi   M^2e^{\gamma_E}\right)
+O\left(\varepsilon^2\right)\right]\\
\label{eq_app_Zs0_ds}
&+
\frac{\alpha}{2}\left[1-\frac{\alpha  \log \left(\pi   M^2e^{\gamma_E}\right)}{1+\pi ^2 \alpha ^2}\right]\Sigma_{0}(0)-I^{(s0)}_{\alpha}+O\left(\varepsilon,\frac{1}{s}\right)\,.
\end{align}
Note that the $1/\varepsilon$ poles canceled with the counterterm and the results are both finite for $\varepsilon\rightarrow 0$. Additionally, noting that at both fixed points we can express $\varepsilon$ from~\eqref{eq_spin_imp_beta} as
\begin{equation}\label{eq_eps_spin_imp}
    \varepsilon=\frac{2\alpha}{s(1+\pi^2\alpha^2)}+O\left(\frac{1}{s^2}\right)\,,
\end{equation}
it is easy to check that the results are independent of the sliding scale at the fixed points, providing a nontrivial consistency check of the calculation.

From~\eqref{eq_app_Zss_ds} and~\eqref{eq_app_Zs0_ds} we obtain the cusp anomalous dimension and the defect creation operator dimension from
\begin{align}
    \Gamma^{(+,s)(+,s)}_{(0)}(\theta) &=-\frac{1}{T}\log\frac{Z_{ss}(\theta)}{Z_{ss}(\pi)}=-\frac{1}{T}\log\frac{Z_{ss}(\theta)}{Z_{ss}\vert_{\alpha=0}}+\frac{1}{T}\log\frac{Z_{ss}(\pi)}{Z_{ss}\vert_{\alpha=0}}\,,\\[0.4em]
    \Delta_{s,0} &=-\frac{1}{T}\log\frac{Z_{s0}}{\sqrt{Z_{ss}(\pi)}} =-\frac{1}{T}\log\frac{Z_{s0}}{Z_{s0}\vert_{\alpha=0}} +\frac{1}{2T}\log\frac{Z_{ss}(\pi)}{Z_{ss}\vert_{\alpha=0}}\,,
\end{align}
and using~\eqref{eq_eps_spin_imp}. The results are given in~\eqref{eq_cusp_ds_1_loop} and~\eqref{eq_dim_ds_1_loop} in the main text.

\bibliography{Biblio}
\bibliographystyle{JHEP.bst}
\end{document}